%% file: paper.tex
\documentclass[a4paper, twocolumn]{revtex4-2}
\pdfoutput=1
\usepackage[dvipsnames]{xcolor}
\usepackage{float, verbatim, amsmath, amssymb, ulem, booktabs, graphicx, dsfont}
\usepackage[colorlinks=true,     linkcolor=blue, citecolor=blue, filecolor=blue, urlcolor=blue]{hyperref}

\newcommand{\e}{\mathrm{e}}

\begin{document}

\title{Can a Quantum Computer Simulate Nuclear Magnetic Resonance Spectra Better than a Classical One?}
\author{Keith R.\ Fratus}
\author{Nicklas Enenkel}
\author{Sebastian Zanker}
\author{Jan-Michael Reiner}
\author{Michael Marthaler}
\author{Peter Schmitteckert}
\affiliation{HQS Quantum Simulations GmbH, Rintheimer Straße 23, 76131 Karlsruhe, Germany}

\input{Sections/abstract}

\maketitle

\input{Sections/intro}

\input{Sections/nmr}

\input{Sections/solver}

\input{Sections/exact}

\input{Sections/larger}

\input{Sections/improve}

\input{Sections/conclusion}

\input{Sections/acknowledgment}

\bibliographystyle{unsrtnat}
\bibliography{Sections/refs}

\newpage
~\newpage

\appendix

\input{Sections/appSF}
\input{Sections/appCalc}

\input{Sections/appParam}

\input{Sections/appExtraData}

\end{document}

%% file: Sections/abstract.tex
\begin{abstract}

The simulation of the spectra measured in nuclear magnetic resonance (NMR) spectroscopy experiments is a computationally non-trivial problem which, due to its natural interpretation as a quantum spin problem, maps in a straightforward way to a quantum computer. As such, it represents a problem for which such a device may provide some practical advantage over traditional computing methods. In order to understand the extent to which such problems may indeed provide examples of useful quantum advantage, it is important to understand the limitations of existing classical simulation methods. In this work, we benchmark our own classical solver designed to study such problems. This solver uses a clustering approximation to achieve a resource scaling which is linear in the total number of nuclear spins in a given molecule, for a fixed cluster size. The success of such an approximation would present a stark repudiation to the common claim that such problems require an exponential scaling of resources, the very claim which makes simulating an NMR spectra a candidate for quantum advantage. Our benchmarking results indicate that our approximation performs well throughout, and even somewhat beyond, the more typical experimental regimes. We discuss what implications this may have for future efforts to demonstrate quantum advantage in the context of NMR.

\end{abstract}

%% file: Sections/intro.tex
\section{Introduction}
\label{sec:introduction}

While advances continue to be made in the field of quantum computing hardware \cite{Kjaergaard2020, Krantz2019, Wendin2017, Noiri2022, Flamini_2019, Henriet2020, Lanyon2011, Bruzewicz2019, Gambetta2012, Kueng2016, Preskill2018}, demonstrating concrete examples of useful quantum advantage remains a challenge \cite{Montanaro2016, Shor1994, Shor1997, Fauseweh2024, OMalley2016, Kandala2017, Monroe2021, Zhang2017, Cerezo2021, liu2023, Cerezo2021_barren, Fauseweh2021, Barends2015, Lloyd1996, Blekos2024, Biamonte2017, Garcia2022, Bauer2020, McArdle2020, harrow2009quantum, berry2014high, kiani2022quantum}. Considerable effort has been invested in searching for problems in which a solution provided by a quantum computer presents some advantage over a solution provided by a classical computer (including the extreme case in which a classical computer cannot reasonably provide a solution at all). One such problem which has been proposed as a candidate for demonstrating quantum advantage is the prediction of the spectra measured in nuclear magnetic resonance (NMR) spectroscopy experiments. Such a problem is seen as a natural candidate for quantum computation, as the Hamiltonian describing a molecule studied through NMR takes the form of a typical quantum spin system, the degrees of freedom of which map in a straightforward fashion to the qubits of a quantum computer. Previous work \cite{Sels2020, Seetharam2023, khedri2024impactnoisesimulationnmr, PRXQuantum.3.030345, burov2024quantumutilitynmrquantum} has investigated the manner in which such a problem can be solved through the use of a quantum computer.

Of course, understanding whether such a problem is truly capable of demonstrating useful quantum advantage requires understanding the limits of classical computation in the context of computing NMR spectra. In \cite{KUPROV2007241, KUPROV201131, EDWARDS2014107, KUPROV2016124, kuprovtutorial2018, Hogben2011, VESHTORT2006248} the extent to which various classical approximation methods can tackle such a problem has been explored. In this work, we propose our own classical method for simulating NMR spectra, and examine its performance when applied to a selection of real-world molecules. This solver makes use of a cluster approximation which is designed so that for a fixed cluster size its resource consumption is inherently linear in the total number of nuclear spins in the molecule. We find that this approximation provides accurate results in the experimental regimes which are typical for standard NMR experiments. Our benchmarking analysis does not reveal any molecules which require a cluster size of more than a dozen or so spins in order to achieve reasonably accurate spectra. These results suggest that the problem of computing accurate NMR spectra in the standard experimental regimes is tractable with resources that scale linearly with the number of nuclear spins. We also consider regimes with magnetic fields and linewidths that are somewhat lower than what is typical in standard NMR spectrometers, and find good performance here as well. We use the convergence of our cluster approximation applied to these molecules to gain insight into the types of problems for which we may hope to see some degree of quantum advantage in the context of NMR spectroscopy.

The outline of this paper is as follows: In Section~\ref{sec:nmr} we briefly discuss the basics of NMR spectroscopy, and in Section~\ref{sec:solver} we explain the details of our solver. In Section~\ref{sec:exact} we discuss the performance of our solver when applied to a collection of molecules for which we possess an exact solution, and in Section~\ref{sec:larger} we consider this performance for molecules for which we do not have an exact solution. In Section~\ref{sec:improve} we discuss the necessary extensions to our basic solver when handling certain special molecules. We conclude in Section~\ref{sec:conclusion}. In Appendix~\ref{sec:appSF} we explain how the particular spectral function we compute is related to quantities measured in an NMR experiment. In Appendix~\ref{sec:appCalc} we discuss how we compute certain accuracy metrics based on the output of our solver. In Appendix~\ref{sec:appParam} we list where we obtain the parameters which define the molecules we study in this work. In Appendix~\ref{sec:appExtraData} we display some additional supporting data which was omitted from the main text.

%% file: Sections/nmr.tex
\section{Nuclear Magnetic Resonance}
\label{sec:nmr}

Nuclear magnetic resonance spectroscopy is an important tool in analytical chemistry, used to identify molecules and to elucidate their structure, among other applications \cite{Bluemich2018, Reif2021, molecules25204597, metabo12080678, Clayden2012, Levitt2008, Grootveld2019, keeler2005understanding, claridge2009high}. In broad terms, NMR spectroscopy involves placing a sample of a given molecule in a large, constant magnetic field, perturbing it with an electromagnetic pulse, and measuring the response. In this section we provide a very brief discussion of the quantity which is measured in a typical NMR experiment, but in enough detail to elucidate the basic calculations which must be performed by our solver. We assume that the molecules are in a liquid solution, which simplifies some aspects of the analysis due to motional averaging.

The central object describing the physics of our molecule placed in an NMR spectrometer is the spin Hamiltonian (in radians per second) given according to
\begin{equation}
\hat{\cal{H}} = -\sum_l \gamma_l \left( 1 + \delta_l \right) B^z  \hat{I}^z_l + 2 \pi\sum_{k < l} J_{kl} \mathbf{\hat I}_k \cdot \mathbf{\hat I}_l 
\end{equation}
The objects $\hat{I}^\alpha=\hat{S}^\alpha / \hbar$ with $\alpha \in \{x, y, z\}$ correspond to the usual SU(2) spin operators for a spin-S particle, which represent atomic nuclei present in the molecule. Only atomic nuclei with non-zero nuclear spin (``active nuclei'') will be relevant for an NMR experiment, with focus often being placed on nuclei with spin-1/2, such as $^{1}$H or $^{31}$P nuclei. The molecule is subjected to a magnetic field which we take to be in the Z direction, denoted by $B^z$, and each nucleus interacts with this field through a Zeeman interaction described by the gyromagnetic ratio $\gamma$. The different atomic nuclei can also couple to each other, parameterized by the terms $J_{kl}$. While each $J_{kl}$ could in principle be a tensor quantity, the motional averaging occurring in solution results in a single scalar quantity, thus resulting in an isotropic Heisenberg coupling. This term is generally a result of indirect spin-spin coupling mediated by electrons in chemical bonds.

The key fact which underpins the utility of NMR spectroscopy in general is that shielding of the external magnetic field by electrons in chemical bonds causes variations in the local magnetic field experienced by each nucleus in the molecule. In the case of molecules in solution, this can be characterized by the single scalar parameter $\delta$, the so-called chemical shift, which effectively modifies the gyromagnetic ratio of a given nucleus. The fact that different nuclei in a given molecule precess at slightly different Larmor frequencies,
\begin{equation}
\omega_{l} = \gamma_l \left( 1 + \delta_l \right) B^z,
\end{equation}
allows for the identification of different chemical groups present in the molecule. For each species of active nuclei, a standard reference compound is used to define $\delta = 0$. For the case of $^{1}$H nuclei, for example, we make use of the convention that the chemical shifts of the hydrogen nuclei in the molecule tetramethylsilane (TMS) are defined as zero (these hydrogen nuclei in TMS all possess identical chemical shifts). This therefore sets the value of $\gamma$ for hydrogen nuclei in general, and allows for the assignment of chemical shift values for the hydrogen nuclei in any other given molecule.

When a sample is placed in an NMR spectrometer, the magnetic field of the device results in some net magnetization of the constituent nuclei. In a standard 1D NMR experiment, a brief electromagnetic pulse flips one species of active nuclei into the XY plane, which therefore begin to precess, creating a signal that can be recorded by measuring the induced magnetic field along the X or Y axis.  In this work, we will focus exclusively on proton NMR, in which $^{1}$H nuclei are the active nuclei which are flipped into the XY plane. In Appendix \ref{sec:appSF}, we show that the measured X and Y components of this magnetic field directly lead to the so-called spectral function, which, up to proportionality, is defined as,
\begin{equation}
\label{eqn:spectral}
C\left ( \omega \right ) \propto \eta \sum_{n,m} \frac{\langle E_{n}| \hat{M}^{-} | E_{m} \rangle  \langle E_{m} | \hat{M}^{+} |  E_{n} \rangle  }{\eta^{2} + [\omega - (E_{n} - E_{m})]^{2}}
\end{equation}
where we have defined
\begin{equation}
\hat{M}^{\pm} \equiv \sum_{i} \gamma_{i} \left ( \hat{I}^{x}_{i} \pm i \hat{I}^{y}_{i} \right ) \equiv \sum_{i} \gamma_{i} \hat{I}^{\pm}_{i} \,.
\end{equation}
For the applicability of Eq.~\eqref{eqn:spectral} see Appendix~\ref{sec:appSF}. Note that we are typically not interested in the overall normalization of the spectral function, and since we are considering proton NMR, we will always normalize this function so that its integral is equal to the number of hydrogen nuclei in the molecule.

This (real and positive) quantity consists of a collection of Lorentzian peaks, centered at the transition frequencies of the system. The width of these peaks is controlled by the ``broadening'' $\eta$ (sometimes also known as the ``line width''), which is formally a convergence factor, but can also be used to model a limited spectrometer resolution (which results from several experimental factors, including the decay of the measured signal due to decoherence). The amplitudes of these peaks are determined by the matrix elements of the $\hat{M}^{\pm}$ operators, which are restricted by the fact that the NMR Hamiltonian conserves the total spin along the Z axis,
\begin{equation}
\hat{I}^{z} = \sum_{i} \hat{I}^{z}_{i}.
\end{equation}
In the non-interacting case, it is a straightforward exercise to verify that there is one Lorentzian peak corresponding to each flipped nucleus, centered at the Larmor frequency of that nucleus. This quantity is therefore extremely useful in determining the chemical shifts of the molecule in question. In the presence of interactions, these peaks will typically split apart into additional smaller peaks, as the interaction terms delocalize the eigenstates of the system and allow for additional transitions, at slightly shifted frequencies. Therefore, it is generally desirable to operate NMR spectrometers at the largest possible magnetic field, since higher magnetic fields will enhance the effect of the Zeeman interaction compared with both the Heisenberg couplings and the broadening scale, leading to an NMR spectrum which can more easily identify the chemical shifts of the constituent nuclei.

The parameter regimes in which typical proton NMR experiments operate result in a natural frequency $\nu_{\text{ref}} = \omega_{\text{ref}}/2\pi$ of a few hundred MHz for the $^{1}$H nuclei in TMS, although smaller benchtop devices typically operate at lower frequencies, e.g. 80 MHz. The chemical shifts in a typical molecule examined in proton NMR usually range from around $10^{-6}$ to $10^{-5}$ (often quoted as 1 to 10 ``parts per million,'' or ``ppm'' for short), resulting in peaks in the spectral function which are separated by anywhere from a few hundred to a few thousand Hz, while the coupling terms are typically on the order of tens of Hz. A broadening of a few Hz is usually reasonable. In our analysis in this paper, we will study three magnetic field strengths, corresponding to natural frequencies of 400, 80, and 20 MHz, which, for the purposes of this work, we will refer to as high, low, and very low field. We will also consider two different broadening values, corresponding to Lorentzian peaks with a full-width at half maximum of 1 Hz, and also 0.1 Hz, which we refer to as high and low broadening. Note that the parameter regimes corresponding to very low field or low broadening are somewhat atypical with respect to actual experiments and merely serve as a stress test for our solver.

%% file: Sections/solver.tex
\section{The Spin-Dependent Cluster Solver}
\label{sec:solver}

To make a prediction for the spectral function which will be measured in an NMR experiment, we must compute the quantity in equation (\ref{eqn:spectral}). A brute-force solution to this problem would require either performing a time evolution of the Hamiltonian dynamics and then a transformation into the frequency domain, or an exact diagonalization of the Hamiltonian, which can be used to compute the sum in equation (\ref{eqn:spectral}) directly. The conservation of total spin along the Z axis allows the Hamiltonian to be written in a block-diagonal form, such that the largest block has size $\mathcal{D} \times \mathcal{D}$, with a reduced Hilbert space dimension of
\begin{equation}
\mathcal{D} \sim {\textstyle\binom{N}{N / 2}} \;\sim \mathcal{O}\left((2^{N}/\sqrt{N})\right) \,,
\end{equation}
where $N$ is the number of active nuclei in the molecule. If we were to attempt an exact diagonalization of the Hamiltonian on a classical computer, we would therefore anticipate that the required memory would scale as $\mathcal{O}\left(2^{2N}/N\right)$, while the required computational time would scale as $\mathcal{O}\left(2^{3N}/N^{3/2}\right)$. For specific molecules with highly symmetric coupling, one can additionally exploit a local SU(2) symmetry to reduce the effective system size \cite{10.1063/1.1700949}. For example, a freely rotating methyl group can be replaced by a decomposition into a spin-3/2 particle and two spin-1/2 particles, effectively decreasing the system size by one spin. However, the combinatorial effort associated with this reduction increases the complexity of the problem, and for small molecules (with less than around twelve nuclei), this can lead to an overall run time which is in fact longer. Furthermore, for molecules beyond a size of approximately twenty spins, neither of these symmetries is sufficient to obtain an exact solution in a reasonable amount of computational time.

To circumvent this issue, our solver is designed to compute the ``spin-resolved'' spectral function
\begin{equation}
C_{i}\left ( \omega \right ) = \gamma_{i} \sum_{j} \gamma_{j} \eta \sum_{n,m} \frac{\langle E_{n}| \hat{I}^{-}_{i} | E_{m} \rangle  \langle E_{m} | \hat{I}^{+}_{j} |  E_{n} \rangle  }{\eta^{2} + [\omega - (E_{n} - E_{m})]^{2}}
\end{equation}
defined so that
\begin{equation}
C \left ( \omega \right ) = \sum_{i} C_{i}\left ( \omega \right ).
\end{equation}
Dividing up the computation in this way, where each $C_{i}$ is computed separately, is advantageous because it allows us to take advantage of the key approximation utilized in the NMR solver, which is to construct a cluster $\Gamma_i$ of spins which are most likely to influence the physics of the chosen $C_{i}$. The solver diagonalizes the reduced Hamiltonian of this cluster, disregarding the effects from spins outside of the cluster, resulting in an approximation to the spin-resolved spectral function,
\begin{equation}
C_{i}\left ( \omega \right ) \approx \gamma_{i} \sum_{j} \gamma_{j} \eta \sum_{n,m} \frac{\langle E^{(\Gamma_i)}_{n}| \hat{I}^{-}_{i} | E^{(\Gamma_i)}_{m} \rangle  \langle E^{(\Gamma_i)}_{m} | \hat{I}^{+}_{j} |  E^{(\Gamma_i)}_{n} \rangle  }{\eta^{2} + [\omega - (E^{(\Gamma_i)}_{n} - E^{(\Gamma_i)}_{m})]^{2}}
\end{equation}
which uses the approximate energy eigenstates $| E^{(\Gamma_i)}_{m} \rangle$ corresponding to the given cluster. This approximation becomes exact in the limit where the cluster size coincides with the number of active nuclei. Note that the clusters corresponding to different spins can overlap, which is essential for obtaining accurate results. Furthermore, at fixed cluster size, this approach leads to a linear scaling with total molecular system size, though additional runtime improvements can be made by diagonalizing the Hamiltonian for spins located in identical clusters only once, and also by exploiting the symmetry considerations discussed previously, for each cluster Hamiltonian independently.

There are several heuristic justifications that one could provide for the potential success of such a clusterization method. In the time domain, if one imagines information (in the form of quantum correlations) spreading throughout the spin system after the pulse is applied, we expect that this information should only travel so far before the signal has decayed due to decoherence \cite{lieb1972finite}. Thinking in the frequency domain, we expect that the effect of a small coupling term will be to slightly perturb the energies and eigenvectors of the Hamiltonian, leading to the locations and amplitudes of the peaks in the spectral function to be slightly altered, which may not be visible above the broadening scale. In any case, the ultimate validity of such an idea will be born out by its potential success in accurately computing these spectral functions.

To determine precisely which spins should be included in a cluster, we must rank them in order of importance. We expect the effect of a coupling $J_{ij}$ on the spectral function to be unimportant if $J_{ij}$ is small compared with the difference between the Larmor frequencies of those two spins. One metric for determining the relevance of a given spin when computing the quantity $C_{i}$ is thus given by the quantity
\begin{equation}
\frac{J_{ij}^{2}}{\left | \gamma_{i}\left ( 1 + \delta_{i} \right )B^{z} - \gamma_{j}\left ( 1 + \delta_{j} \right )B^{z} \right | + \mathcal{E}}
\label{cluster_metric}
\end{equation}
where $\mathcal{E}$ is a small term which prevents the metric from diverging in the case that two Larmor frequencies are identical. For the broadening values we study here, we find it sufficient to simply set $\mathcal{E}$ to a constant value of 0.1 radians per second, as altering this value does not make a noticeable difference in the resulting spectral functions. The user of the solver can, however, adjust this value if desired.

We emphasize that such a metric will have value zero for all spins that are not directly coupled, which may render it somewhat less useful for ordering the importance of spins in molecules which possess fewer direct couplings. However, as we will soon see, even this relatively simple metric results in a clustering approximation which is highly effective at computing the spectra of almost all of the molecules we study, and struggles only in a few special cases. For these special molecules, straightforward considerations regarding their structure lead us to a generalized clustering scheme which allows us to once again obtain accurate predictions, which we will discuss in more detail later.

%% file: Sections/exact.tex
\begin{figure*}[t]
   \centering
   \includegraphics[width=0.32\textwidth]{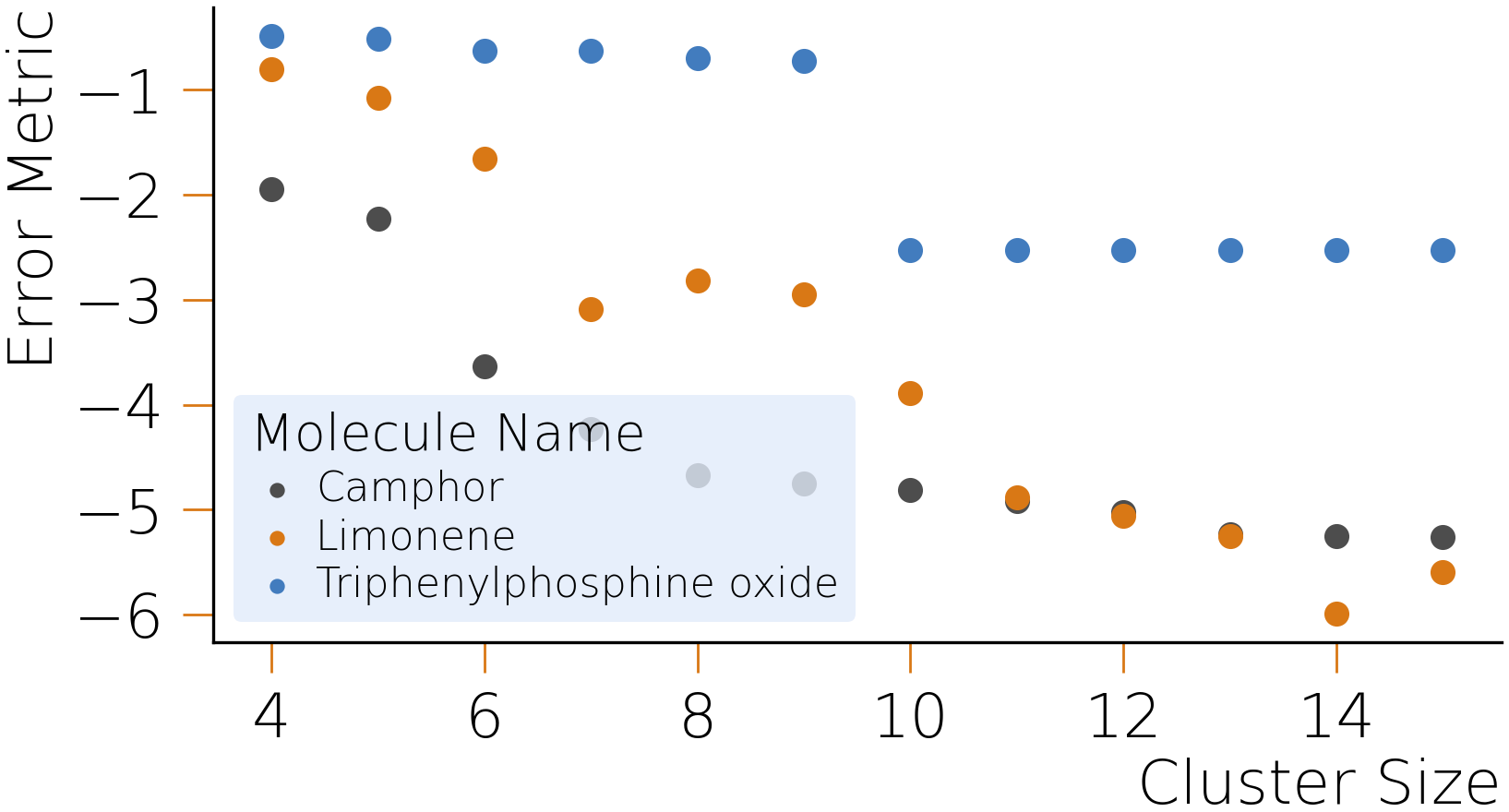}
   \includegraphics[width=0.32\textwidth]{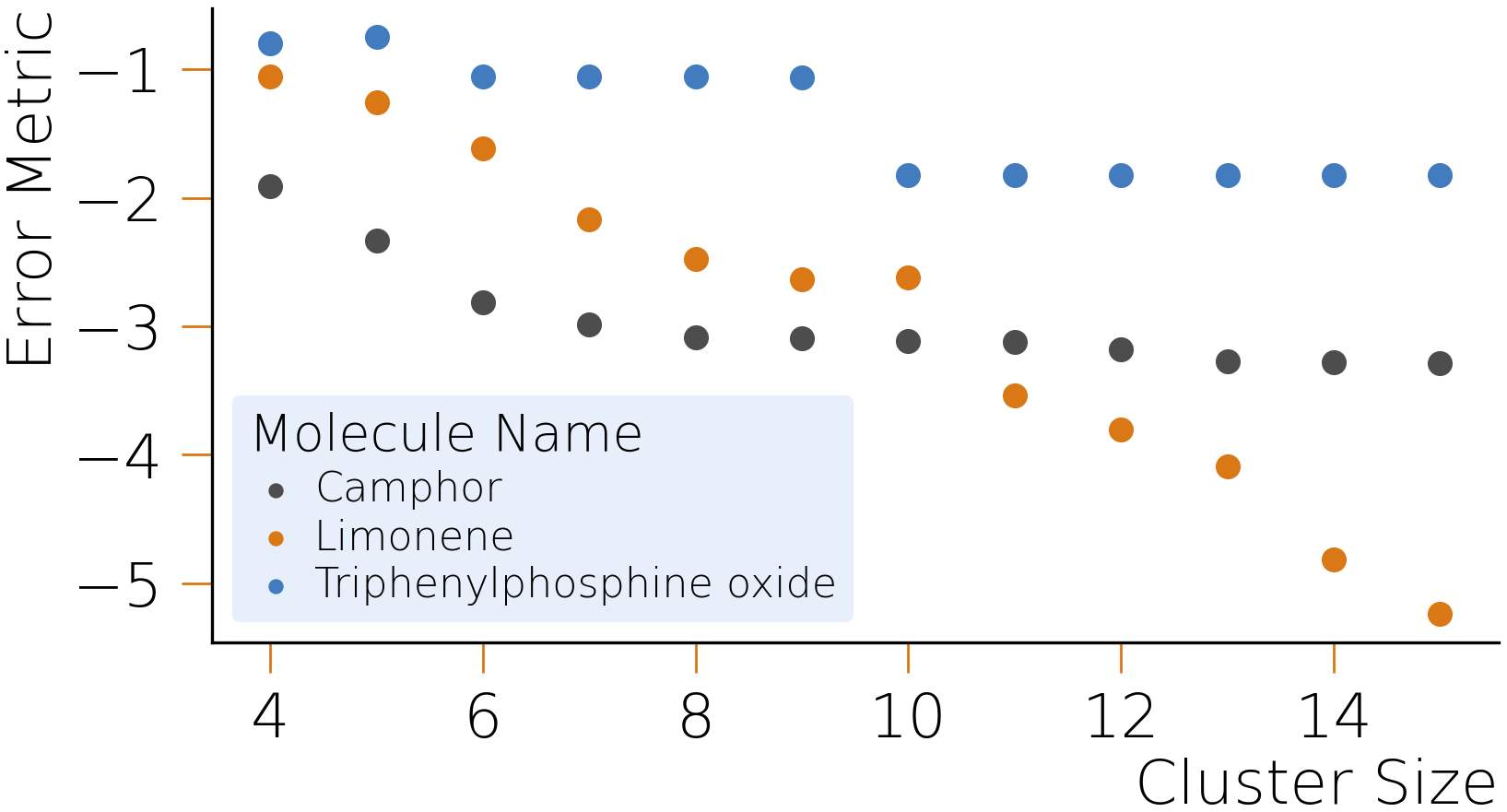}
   \includegraphics[width=0.32\textwidth]{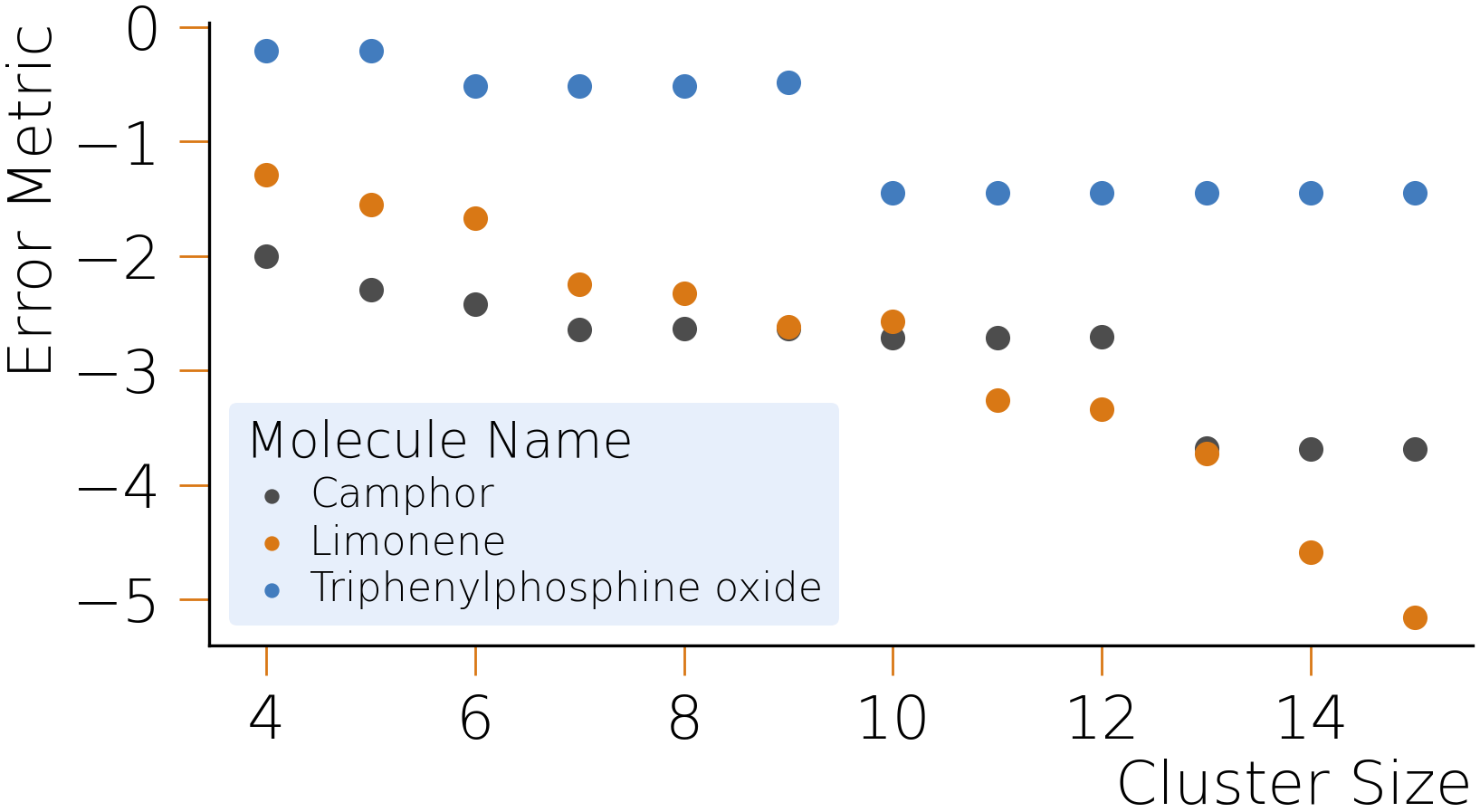}
   \caption{The convergence for selected molecules with 16 nuclear spins for the case of high broadening, for high, low, and very low field.}
   \label{fig:exact_high_broad}
\end{figure*}

\begin{figure*}
   \centering
   \includegraphics[width=0.32\textwidth]{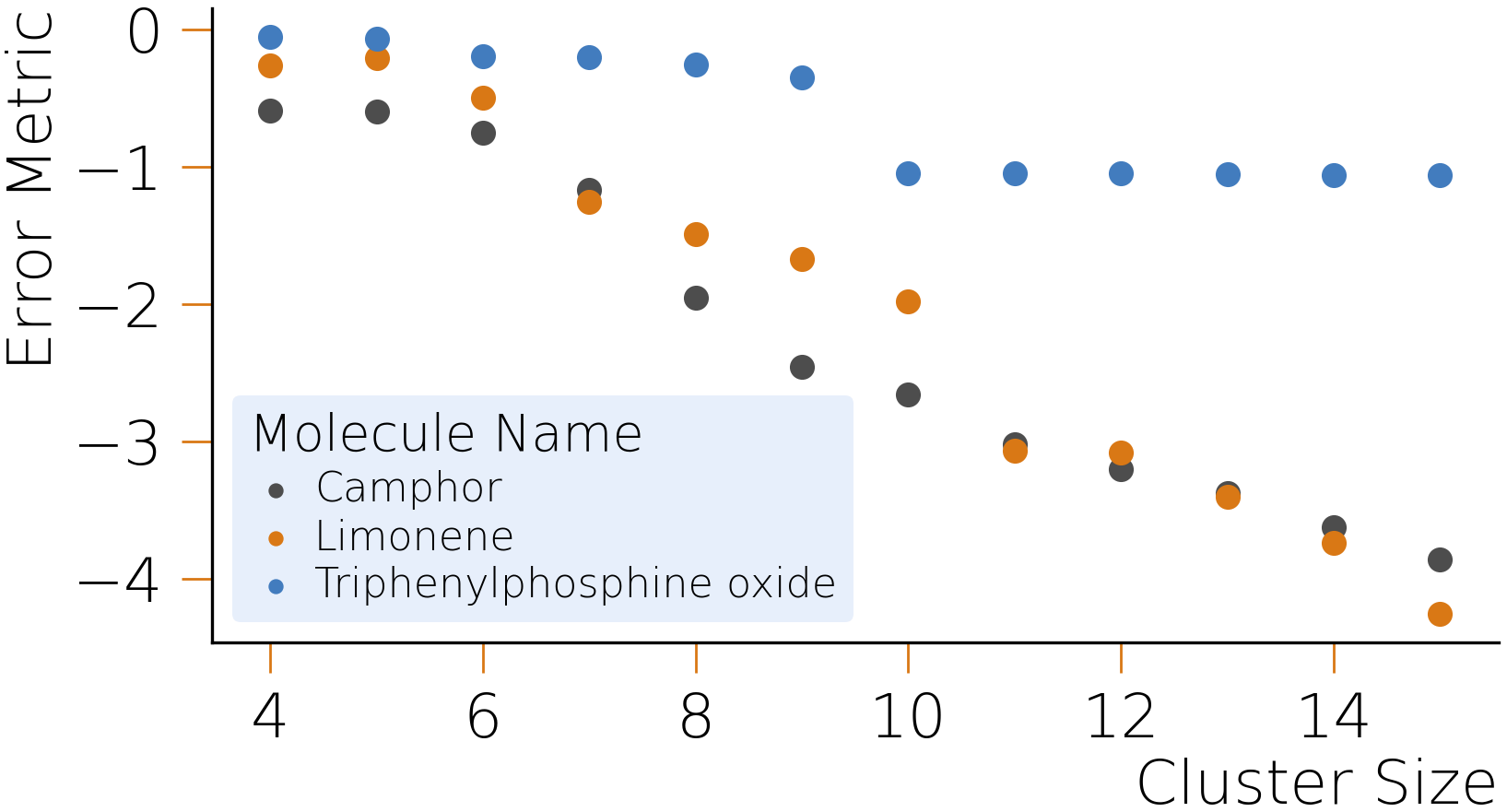}
   \includegraphics[width=0.32\textwidth]{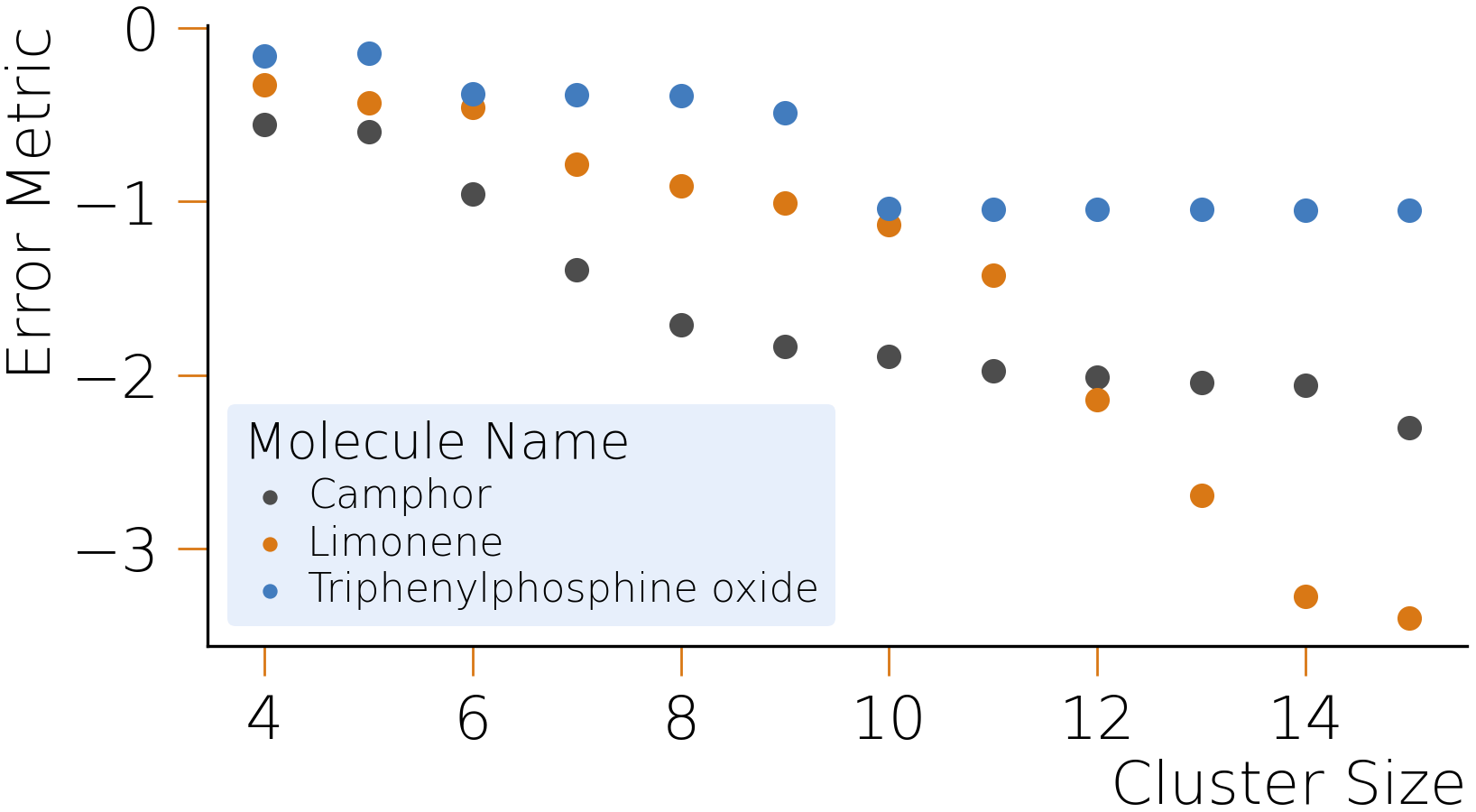}
   \includegraphics[width=0.32\textwidth]{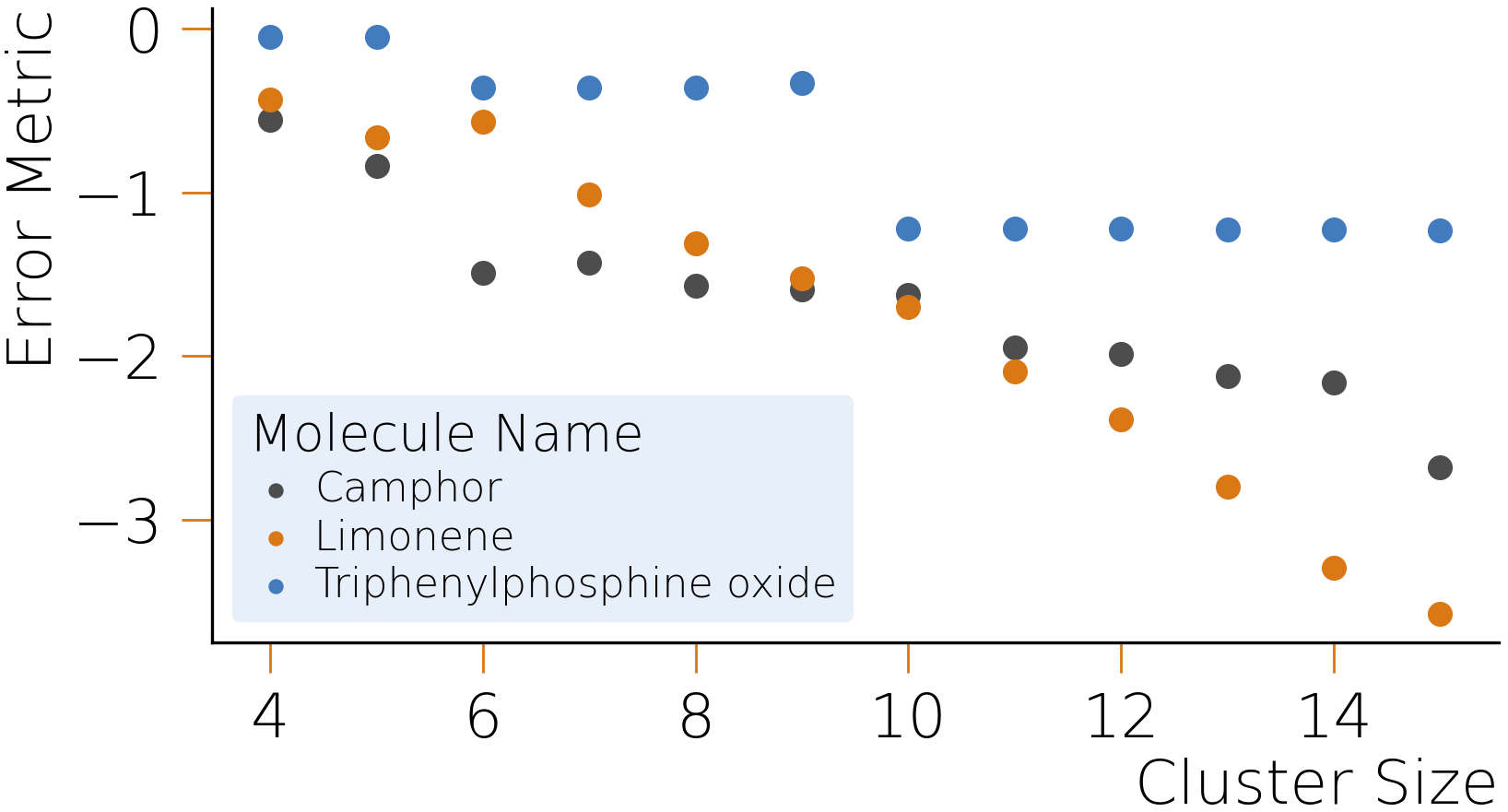}
   \caption{The convergence for selected molecules with 16 nuclear spins for the case of low broadening, for high, low, and very low field.}
   \label{fig:exact_low_broad}
\end{figure*}

\section{Exactly Solvable Examples}
\label{sec:exact}

We begin by discussing the performance of our solver for a selection of molecules whose exact solution is accessible to us, both in terms of accuracy and computational resources. In this section and the next, the couplings and chemical shifts that we use come from a combination of experimental databases, the Spinach Library \cite{Hogben2011}, and our own internal quantum chemical calculations (similar to the approach outlined in \cite{Grimme2017, Yesiltepe2018, Willoughby2014, Jonas2022}). In Appendix \ref{sec:appParam}, we provide a list of the molecules we study, and explain where we obtain the parameters for each of them.

To gauge the accuracy of our solver, we need a metric for describing the closeness of two functions. For this purpose, we will make use of the cosine similarity,
\begin{equation}
\cos \theta_{ab} \equiv \frac{\int_{-\infty}^{+\infty}C_{a}\left ( \omega \right )C_{b}\left ( \omega \right )d\omega}{\sqrt{\int_{-\infty}^{+\infty}C^{2}_{a}\left ( \omega \right )d\omega}~\sqrt{\int_{-\infty}^{+\infty}C^{2}_{b}\left ( \omega \right )d\omega}},
\end{equation}
which essentially measures the ``angle'' between the two functions $C_{a}$ and $C_{b}$, as the integrals can be understood as inner products on the space of square-integrable functions. An important feature of this metric is that it does not change if one or both of the functions is scaled uniformly. This is a desirable property for our metric to have, since the essential information contained in an NMR spectrum does not change if it is scaled by a constant factor. In Appendix \ref{sec:appCalc}, we discuss exactly how we compute this quantity from the numerical output of the solver. Since we will be mostly interested in small deviations between spectral functions, the specific error metric we will present is in fact the (base ten) logarithm of one minus the cosine similarity,
\begin{equation}
\epsilon_{ab} \equiv \log \left (1 - \cos \theta_{ab} \right ).
\end{equation}

\begin{figure*}
   \includegraphics[width=0.49\textwidth]{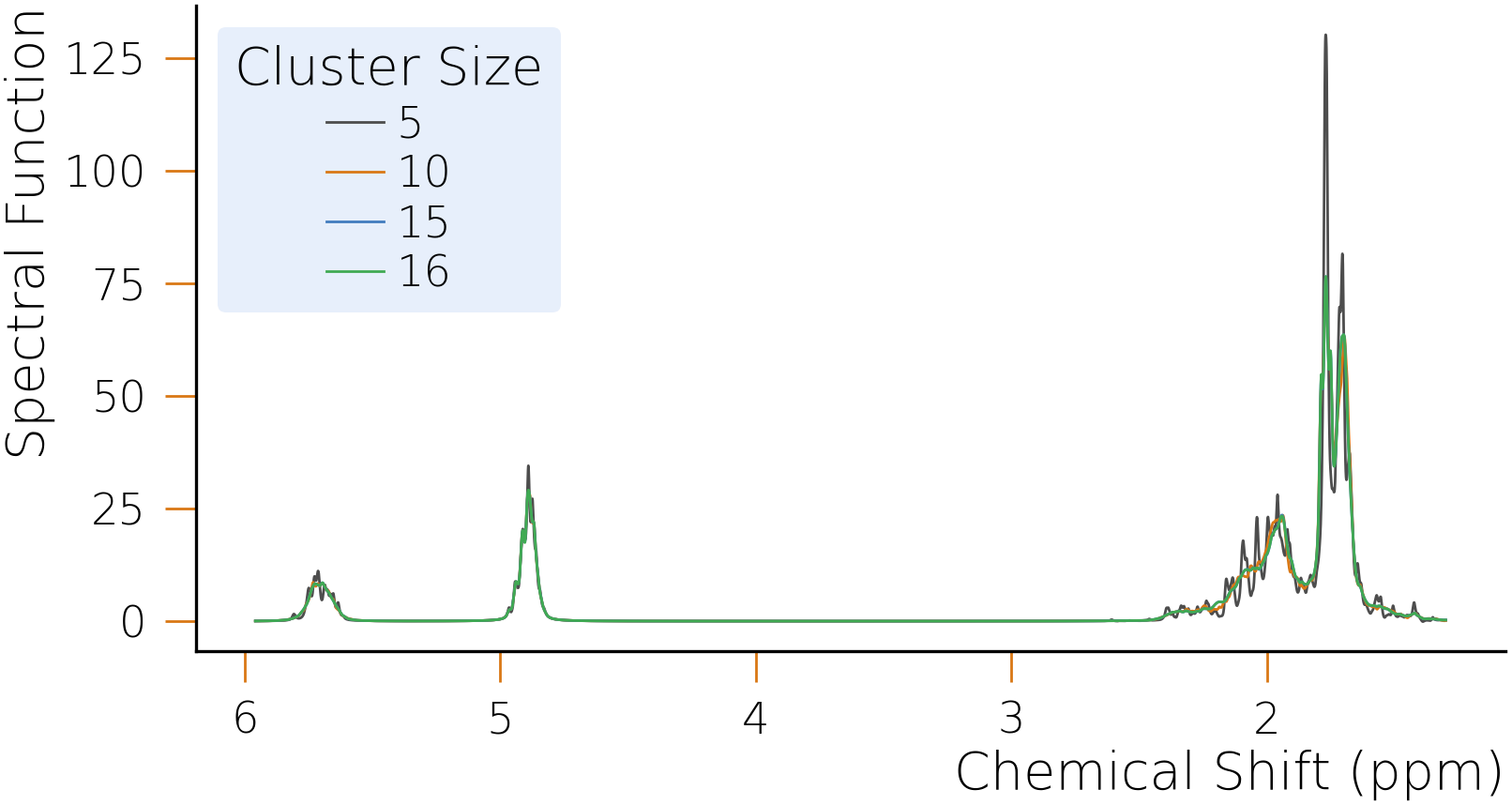}
   \includegraphics[width=0.49\textwidth]{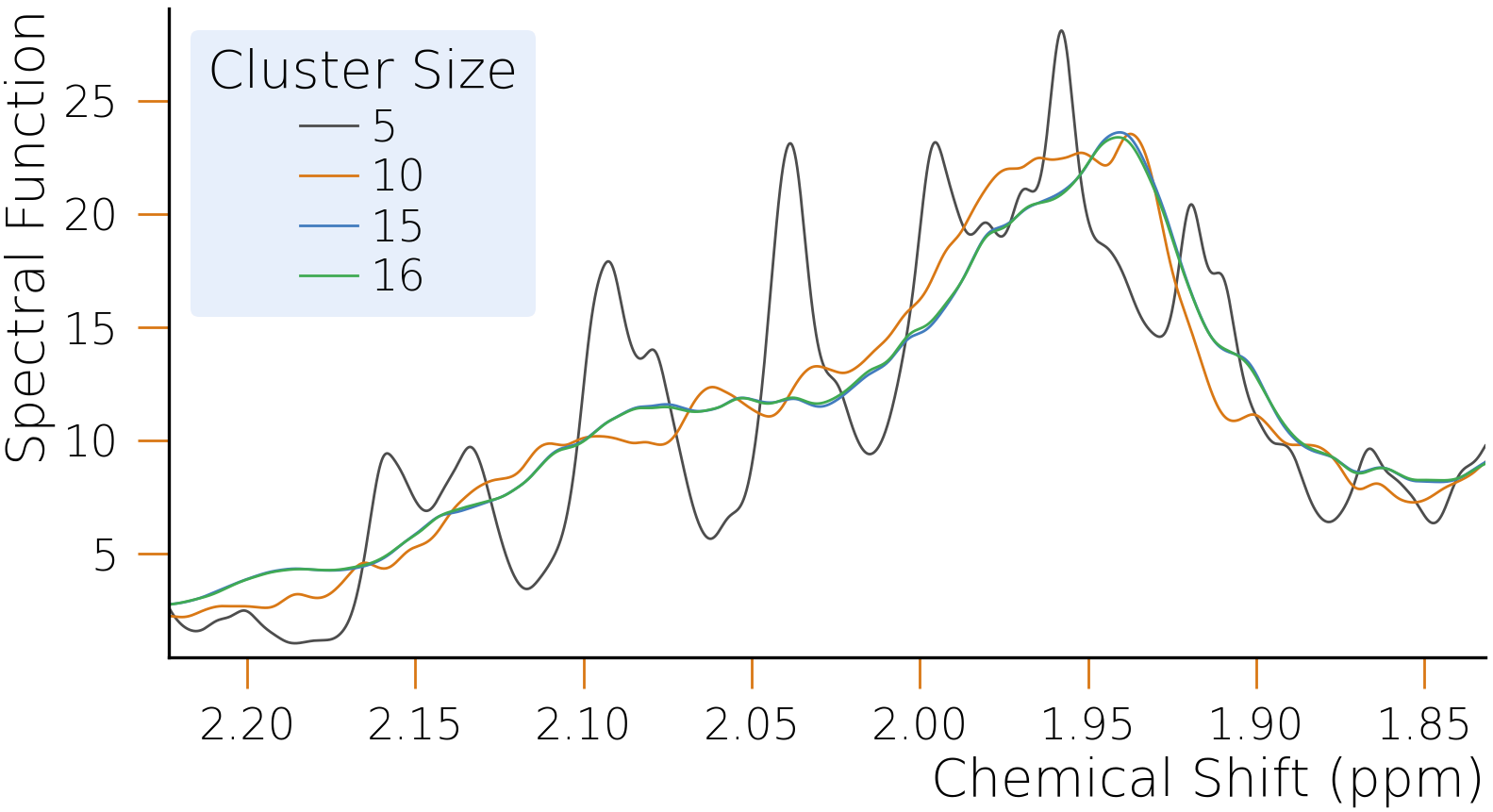}
   \caption{The convergence of the spectral function of Limonene as a function of cluster size, for low field and high broadening. The inverted horizontal axis is conventional in the chemistry literature.}
   \label{fig:limo} 
\end{figure*}

To lend credence to the utility of this metric as a means of comparing spectral functions, and to give an intuitive sense for what different values of this metric mean, we display in Figure \ref{fig:limo} the spectral function of the molecule Limonene at high broadening and low field, computed for several different maximum cluster sizes, in each case comparing against the exact result (which for this molecule corresponds to a maximum cluster size of sixteen). In these plots, rather than plotting against the frequency directly, we plot against the quantity
\begin{equation}
\Delta = \frac{\omega - \omega_{\text{ref}}}{\omega_{\text{ref}}} = \frac{\nu - \nu_{\text{ref}}}{\nu_{\text{ref}}},
\end{equation}
where $\omega_{\text{ref}}$ is, as usual, the Larmor frequency of TMS. In terms of this shifted and scaled variable, the spectral function has peaks located at the values of the chemical shifts $\delta$, rather than the Larmor frequencies themselves, which makes the identification of chemical groups simpler, and is the convention in the NMR literature (this is sometimes referred to as ``plotting against the chemical shift'').

As we increase the maximum cluster size from five to ten, and then again to fifteen, we see that the spectral function goes from relatively poor agreement to fairly good agreement, and then to almost perfect agreement (and in fact in the last case, we must zoom in significantly in order to see any deviation). These maximum cluster sizes correspond to error values of roughly -1.26, -2.62, and -5.24, respectively. This type of relationship between the numerical value of the error metric and the qualitative behaviour of the spectral functions, while admittedly somewhat subjective, is generally reflected across all of the molecules we have studied.

With this understanding of the error metric in mind, we present results for the convergence of three selected molecules as a function of maximum cluster size, all of which contain sixteen nuclear spins. We display the high broadening case in Figure \ref{fig:exact_high_broad} and the low broadening case in Figure \ref{fig:exact_low_broad}. As we would expect, the convergence generally improves for higher field and higher broadening, with the broadening seemingly having the more pronounced effect on accuracy. We highlight the fact that there are no sudden jumps in accuracy, either up or down, as a function of maximum cluster size, and that the error of the approximation is (almost always) decreasing. This fact will be important for attempting to benchmark the performance of the solver when applied to molecules for which we do not have an exact solution. 

\begin{figure}[H]
   \centering
   \includegraphics[width=0.45\textwidth]{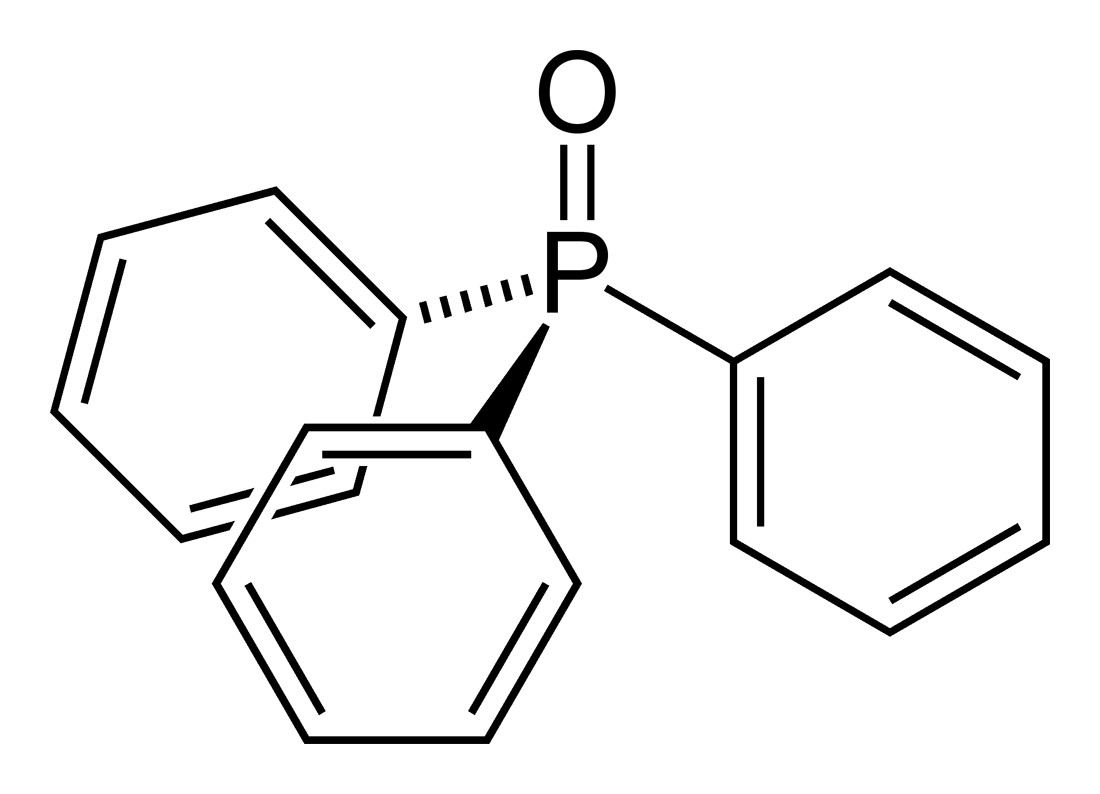}
   \caption{The molecule Triphenylphosphine oxide, with one phosphorous atom and fifteen hydrogen atoms.}
   \label{fig:tppo_mol}
\end{figure}

However, it is true that the convergence for one of the molecules, Triphenylphosphine oxide (TPPO), is a bit more stubborn than the other two. This is especially egregious in the case of low broadening. An explanation for this behaviour can be found by examining the structure of the TPPO molecule, which can be seen in Figure \ref{fig:tppo_mol}. The active nuclei in this molecule are the $^{1}$H nuclei contained within the three benzene rings, along with a $^{31}$P nucleus connecting them, all of which leads to a highly symmetric spin Hamiltonian. However, as the three outermost $^{1}$H nuclei are not directly coupled to the $^{31}$P nucleus, the clusters chosen for these nuclei do not, by default, contain the $^{31}$P nucleus, since it is always ranked as the least important nucleus, and thus these clusters lack crucial information about the global structure of this highly symmetric molecule. It is therefore not surprising that the spectra which are computed using the basic clustering scheme do not converge well. We will have more to say about this case (as well as a related one) in a later section. There we will propose a refined clustering method that does lead to good convergence behaviour for such unusual cases, where a lack of direct couplings leads to clusters that conflict with the symmetry of the molecule.

In Appendix \ref{sec:appExtraData} we provide convergence data for additional selected molecules with nuclei counts of eight, nine, ten, and twelve. All of these molecules display strong convergence behaviour.

\begin{figure}
   \centering
   \includegraphics[width=0.45\textwidth]{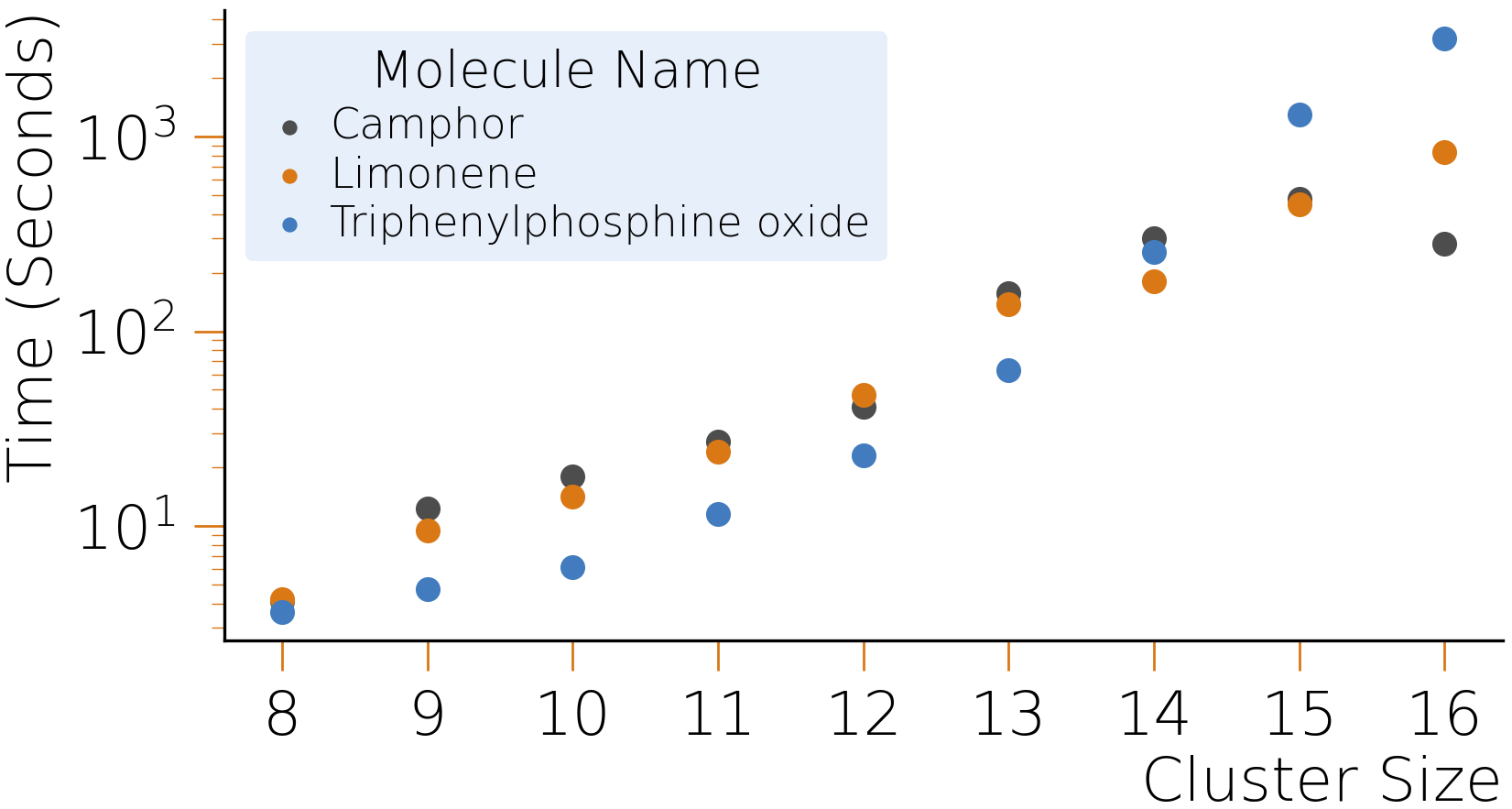}
   \caption{The computational time (on a logarithmic scale) required for selected molecules with 16 nuclear spins, for the case of very low field and low broadening.}
   \label{fig:timing}
\end{figure}

\begin{figure}
   \centering
   \includegraphics[width=0.45\textwidth]{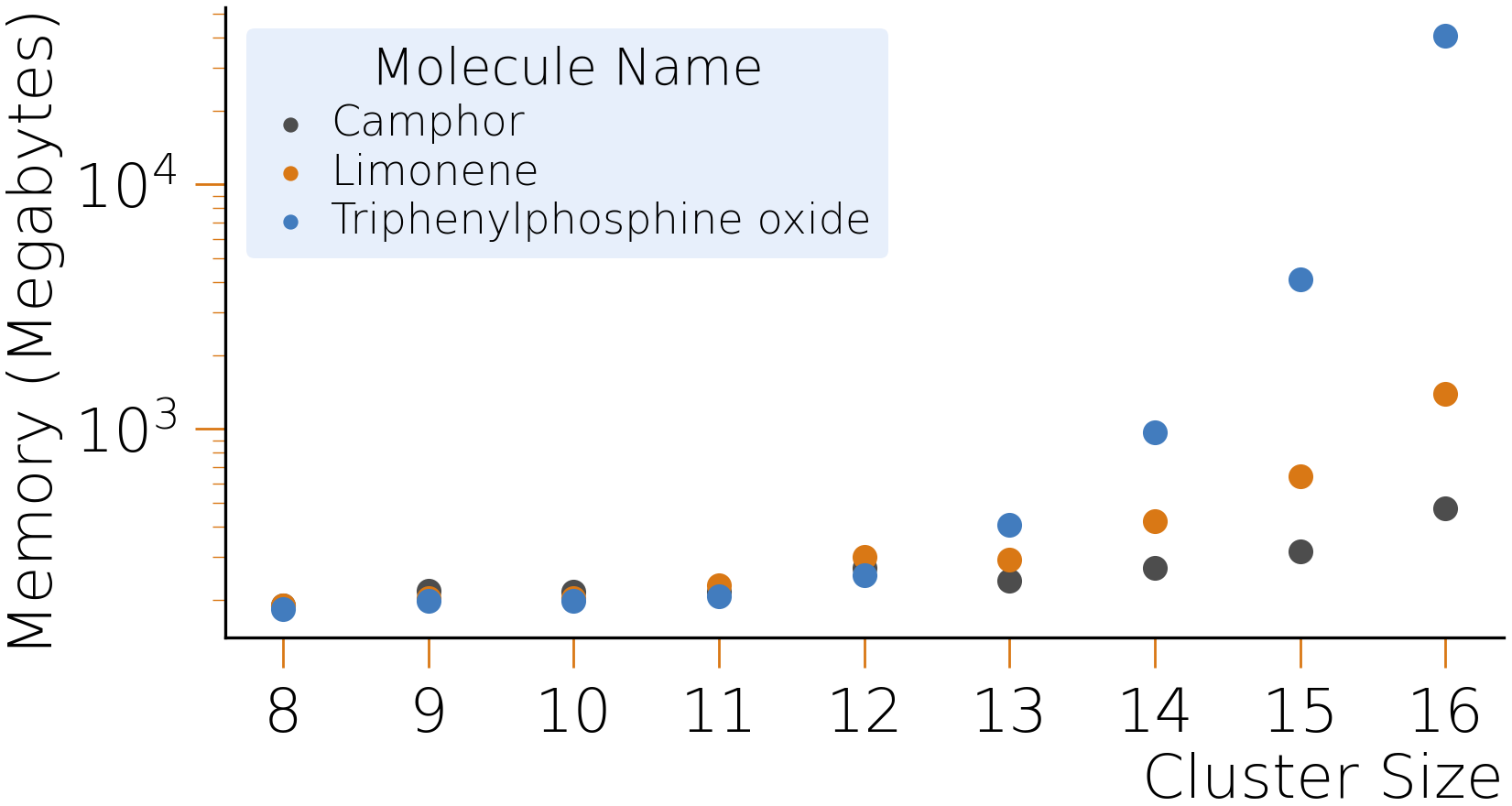}
   \caption{The memory (on a logarithmic scale) required for selected molecules with 16 nuclear spins, for the case of very low field and low broadening.}
   \label{fig:mem}
\end{figure}

As for the required computational resources, in Figure \ref{fig:timing} we display the required computational time to solve for the spectrum of the three molecules with sixteen nuclear spins, as a function of maximum cluster size, for the case of very low field and low broadening (other parameter regimes display similar behaviour). In Figure \ref{fig:mem} we display a similar plot for the required memory. As anticipated, the required computational time and memory both grow exponentially, since we perform an exact diagonalization for each cluster. However, even the most computationally intensive calculation requires approximately one hour of computational time, and in many cases, exceptionally good accuracy is reached already at cluster sizes which only require seconds or minutes of computational time. For fixed cluster size, the scaling is linear with total system size, since each additional spin adds an additional cluster to be diagonalized (this time can be sub-linear if multiple spins correspond to the same cluster, reducing the number of diagonalizations which must be performed). We note that the varying run times and memory requirements for molecules of the same total size can be traced back to the number of magnetically equivalent spins in each molecule, as the solver automatically exploits the local SU(2) symmetry where applicable. This data was collected on an Intel Xeon W-2145 processor at 3.70 GHz using 8 threads.

%% file: Sections/larger.tex
\section{Larger Molecules}
\label{sec:larger}

Here we discuss the performance of the solver when applied to larger molecules. Since we do not possess an exact solution for these molecules, the spectral functions for smaller cluster sizes are instead compared against the spectral function for the largest cluster size we can reasonably simulate (referred to as the ``reference'' size). As we examine the spectral functions corresponding to smaller cluster sizes, if we eventually reach a point beyond which we no longer see any significant differences between the smaller size and the reference size, then we will consider the solutions to have converged. While this does not represent a rigorous proof that we have converged to the correct solution, it is nevertheless plausible, given that we have not found any examples in the exactly solvable case in which the accuracy of the solution suddenly jumps from one approximate calculation to the next after having reached a plateau.

\begin{figure*}
   \centering
   \includegraphics[width=0.32\textwidth]{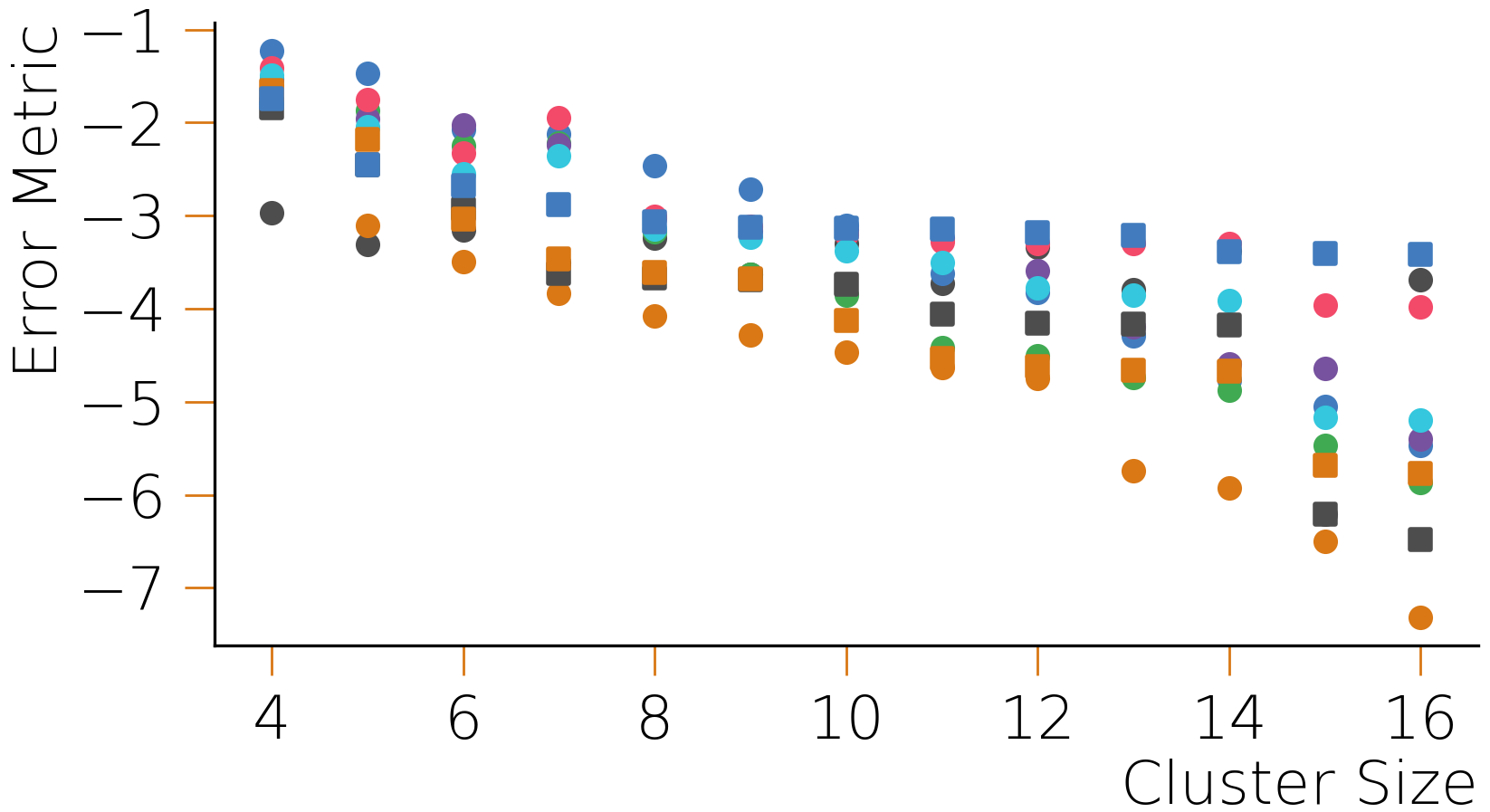}
   \includegraphics[width=0.32\textwidth]{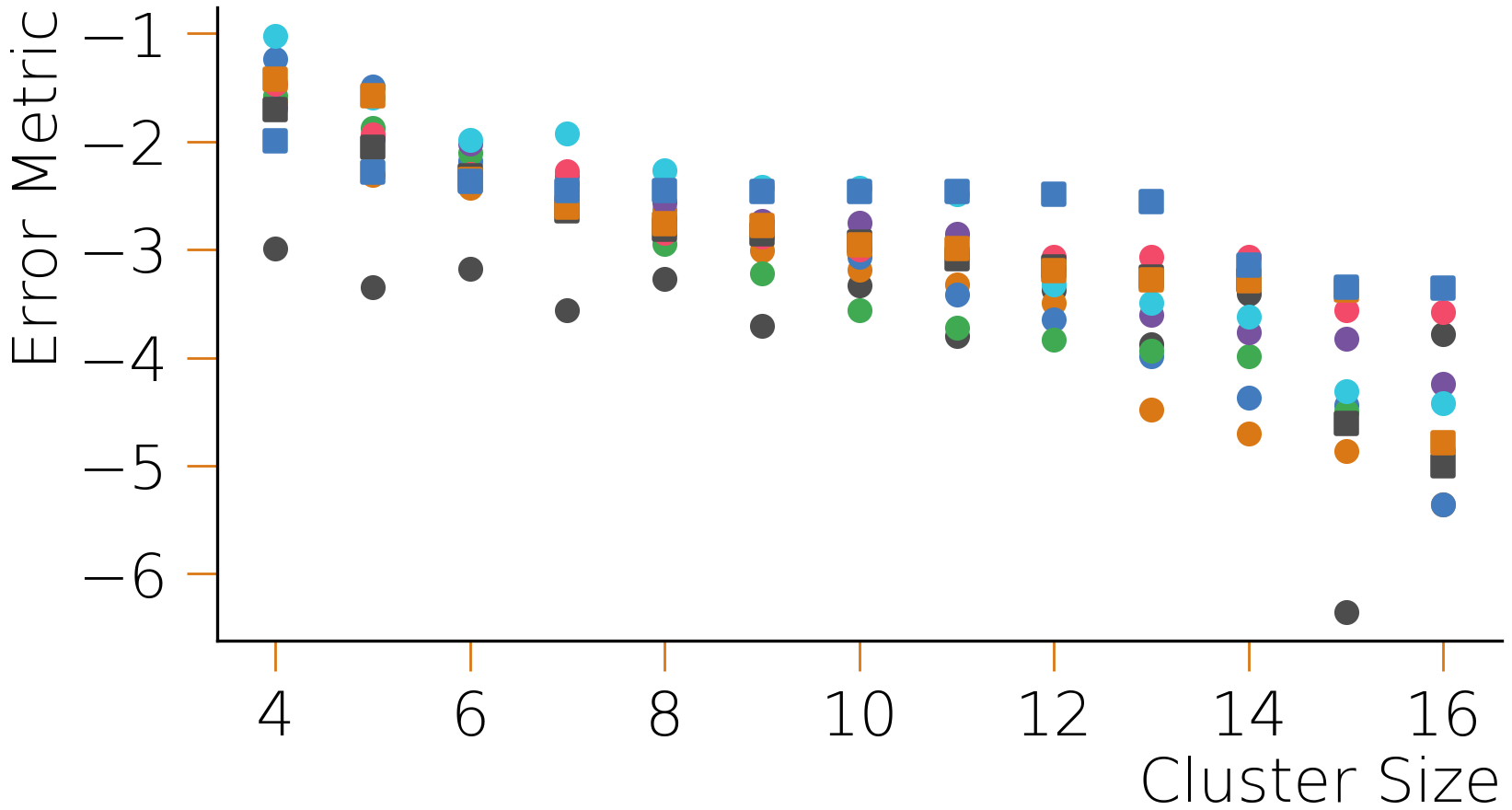}
   \includegraphics[width=0.32\textwidth]{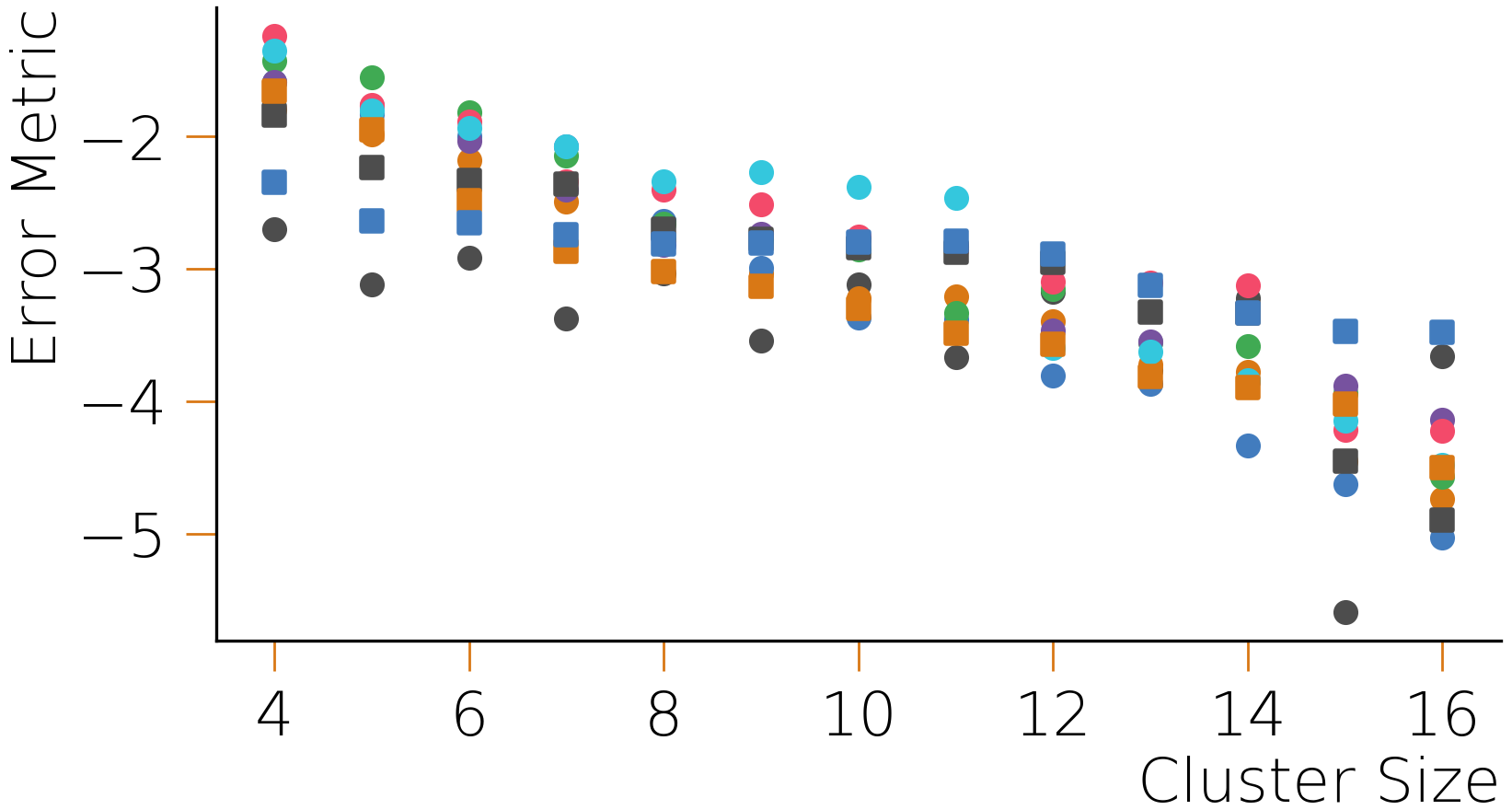}
   \caption{The convergence for selected molecules with more than 16 nuclear spins for the case of high broadening, for high, low, and very low field.}
   \label{fig:larger_high_broad}
\end{figure*}

\begin{figure*}
   \centering
   \includegraphics[width=0.32\textwidth]{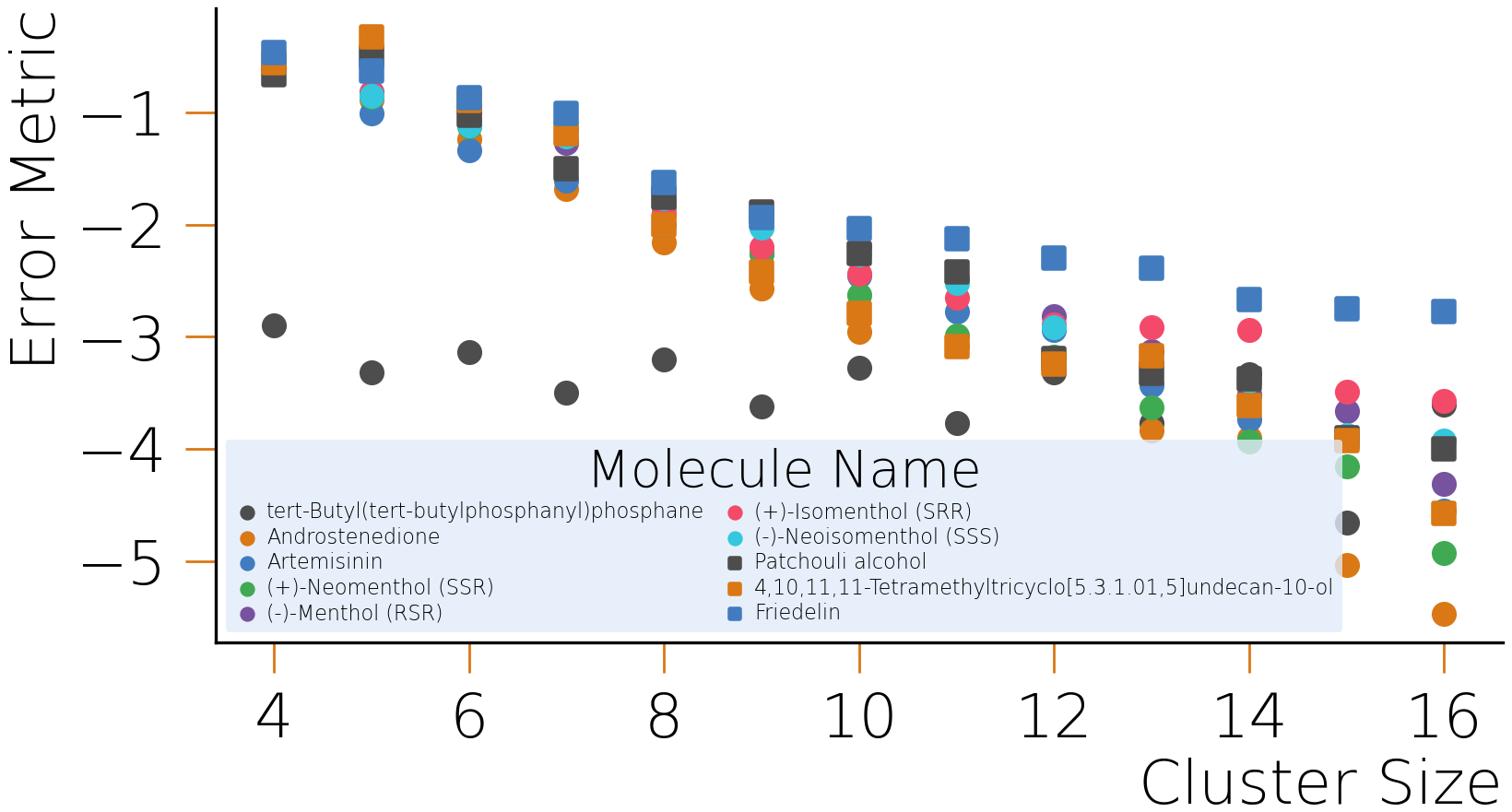}
   \includegraphics[width=0.32\textwidth]{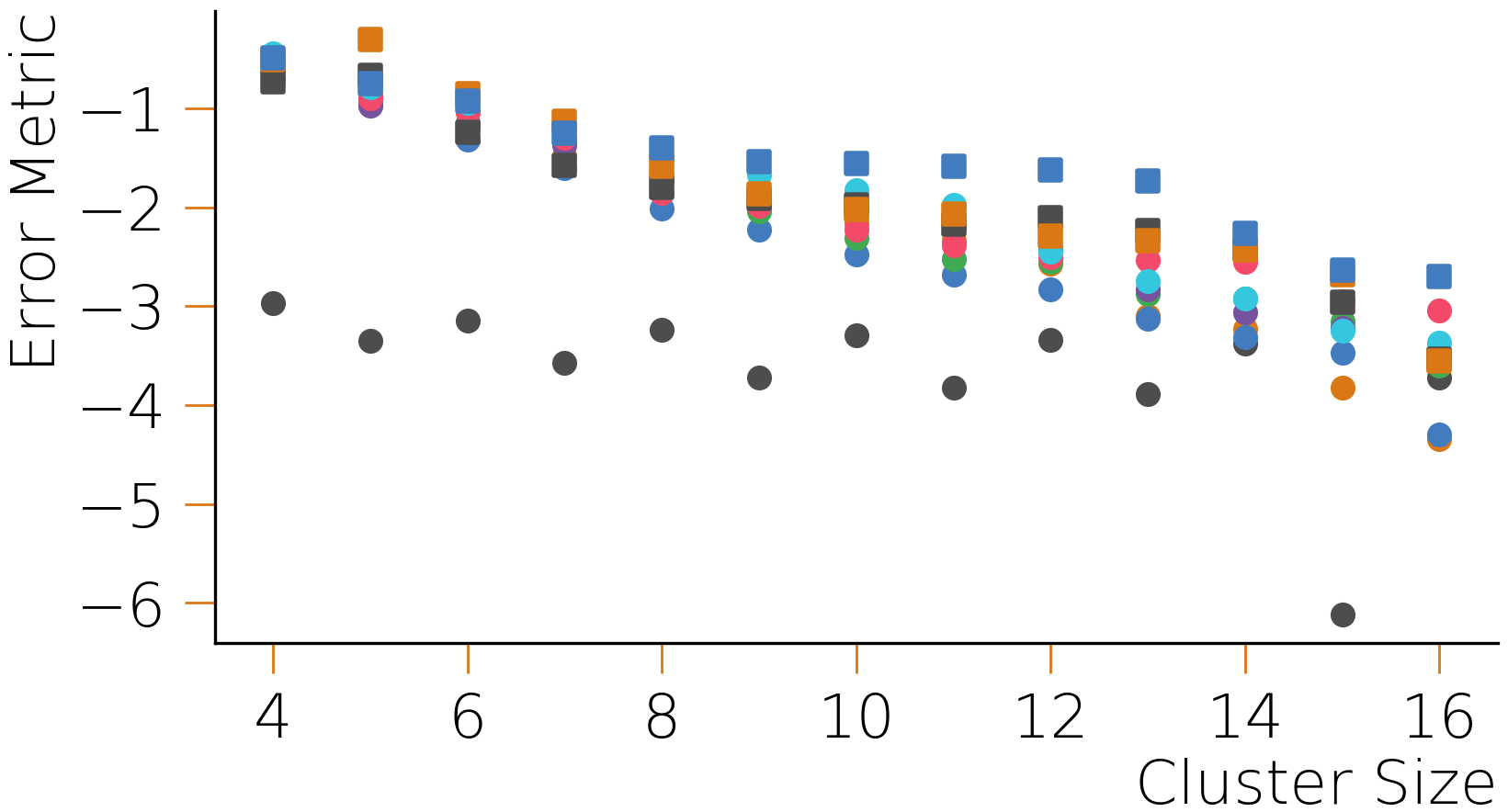}
   \includegraphics[width=0.32\textwidth]{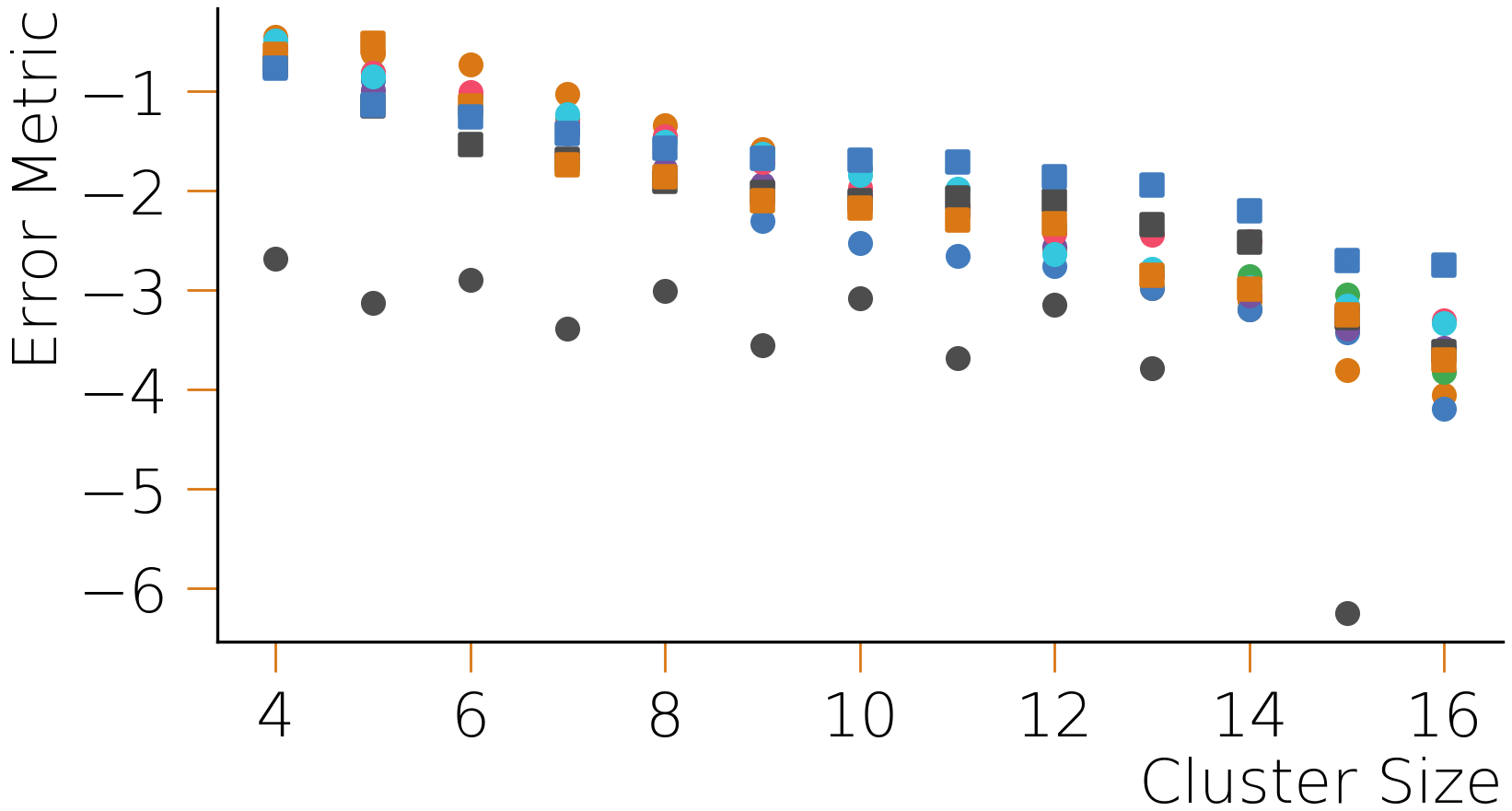}
   \caption{The convergence for selected molecules with more than 16 nuclear spins for the case of low broadening, for high, low, and very low field.}
   \label{fig:larger_low_broad}
\end{figure*}

In Figures \ref{fig:larger_high_broad} and \ref{fig:larger_low_broad} we display the results of our analysis for a selection of ten molecules of various size, all of them with a reference cluster size of seventeen. Once again, we see that the convergence is better for higher field and higher broadening, and the agreement between a cluster size of sixteen and a cluster size of seventeen is exceptionally good in all cases. According to our error metric, the molecule with the worst convergence would appear to be Friedelin at low broadening, with error values of -2.77, -2.70, and -2.75 for high, low, and very low field, respectively. We also notice that the error values for the molecule tert-Butyl(tert-butylphosphanyl)phosphane (also named 1,2-di-tert-butyl-diphosphane, referred to here as ``Diphosphane'') seem to progress in a somewhat erratic fashion from one cluster size to the next, for several parameter regimes. Even though these error values for Diphosphane are in principle very small, this erratic behaviour, along with the fact that Diphosphane is a highly symmetric molecule (similar to TPPO), is cause for suspicion. We briefly investigate the spectral functions of both of these molecules.

\subsection{Friedelin}

\begin{figure}[H]
   \centering
   \includegraphics[width=0.45\textwidth]{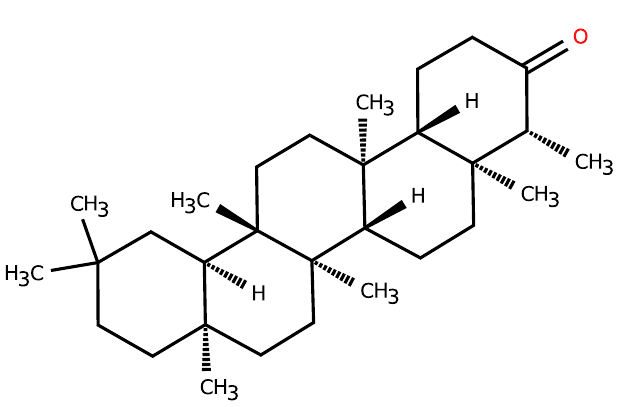}
   \caption{The molecule Friedelin, with fifty hydrogen atoms.}
   \label{fig:fried_mol}
\end{figure}

\begin{figure*}
   \centering
   \includegraphics[width=0.49\textwidth]{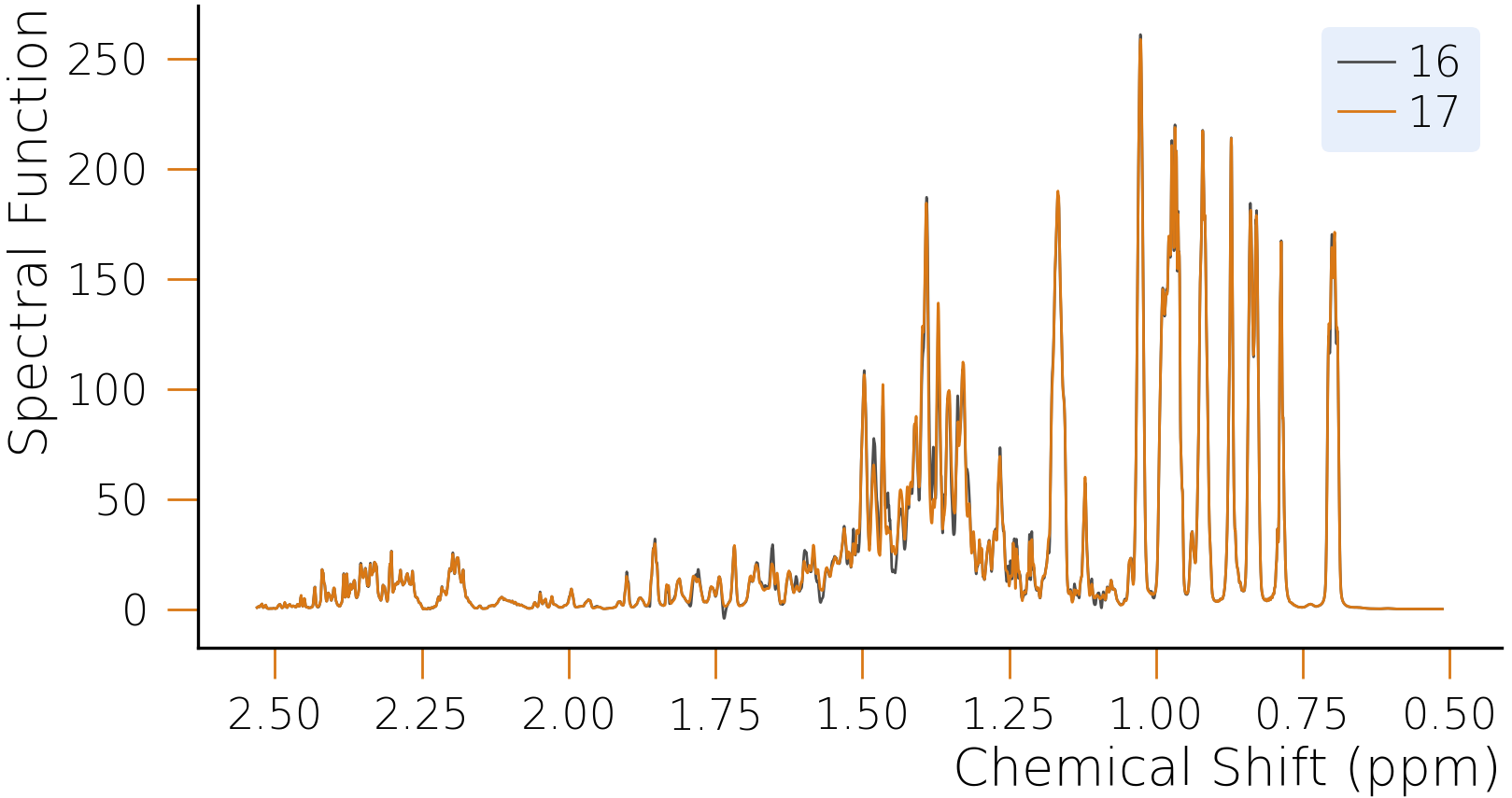}
   \includegraphics[width=0.49\textwidth]{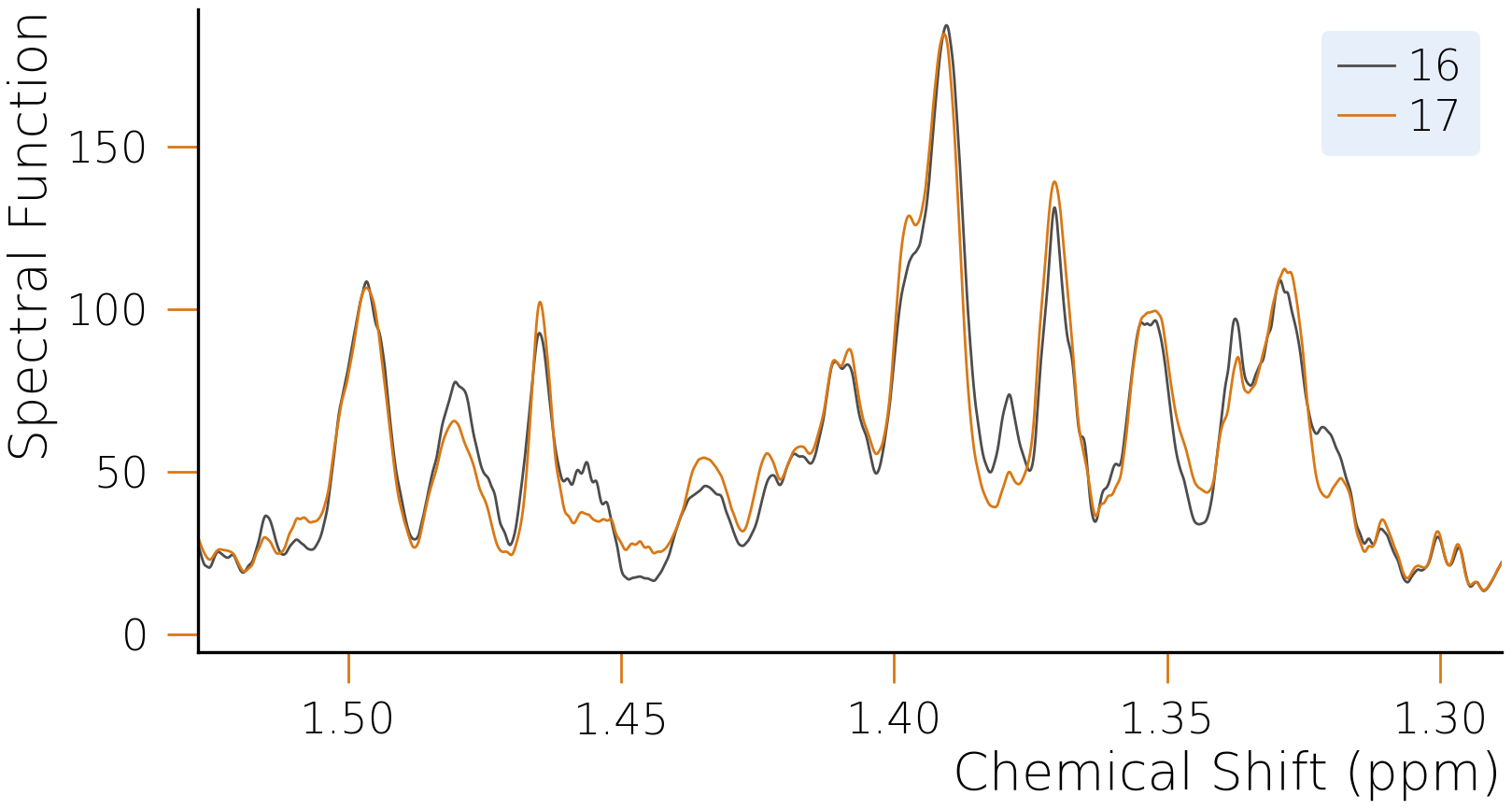}
   \caption{The spectral function for Friedelin for the case of low field and low broadening.}
   \label{fig:fried}
\end{figure*}

Friedelin is a naturally occurring triterpenoid whose rigid friedelane skeleton contains five fused rings and a single carbonyl group at C-3, while the remaining thirty carbon atoms are fully saturated. Because every carbon atom (apart from the quaternary ring junctions) carries at least one hydrogen, the molecule hosts 50 distinct $^{1}$H nuclei. Simulating a molecule with 50 active nuclei pushes the boundary of what is possible with classical cluster approximations, and thus Friedelin serves as a useful “large molecule” benchmark for our NMR solver. In Figure \ref{fig:fried_mol} we show a schematic representation of the structure of the Friedelin molecule.

In Figure \ref{fig:fried} we display the spectral function of Friedelin for the case of low field and low broadening, which, according to our error metric, should represent the worst convergence of any spectral function. While there are some small details which are not perfectly captured, the majority of the larger features appear to be resolved, to the extent that it is not clear whether a more precise simulation would add to the practical interpretation of the NMR spectrum. For the case of Friedelin, this implies that our solver is capable of reaching the level of accuracy necessary for industrial applications, especially since this level of broadening is already somewhat experimentally unrealistic.

\subsection{Diphosphane}

\begin{figure*}
   \centering
   \includegraphics[width=0.49\textwidth]{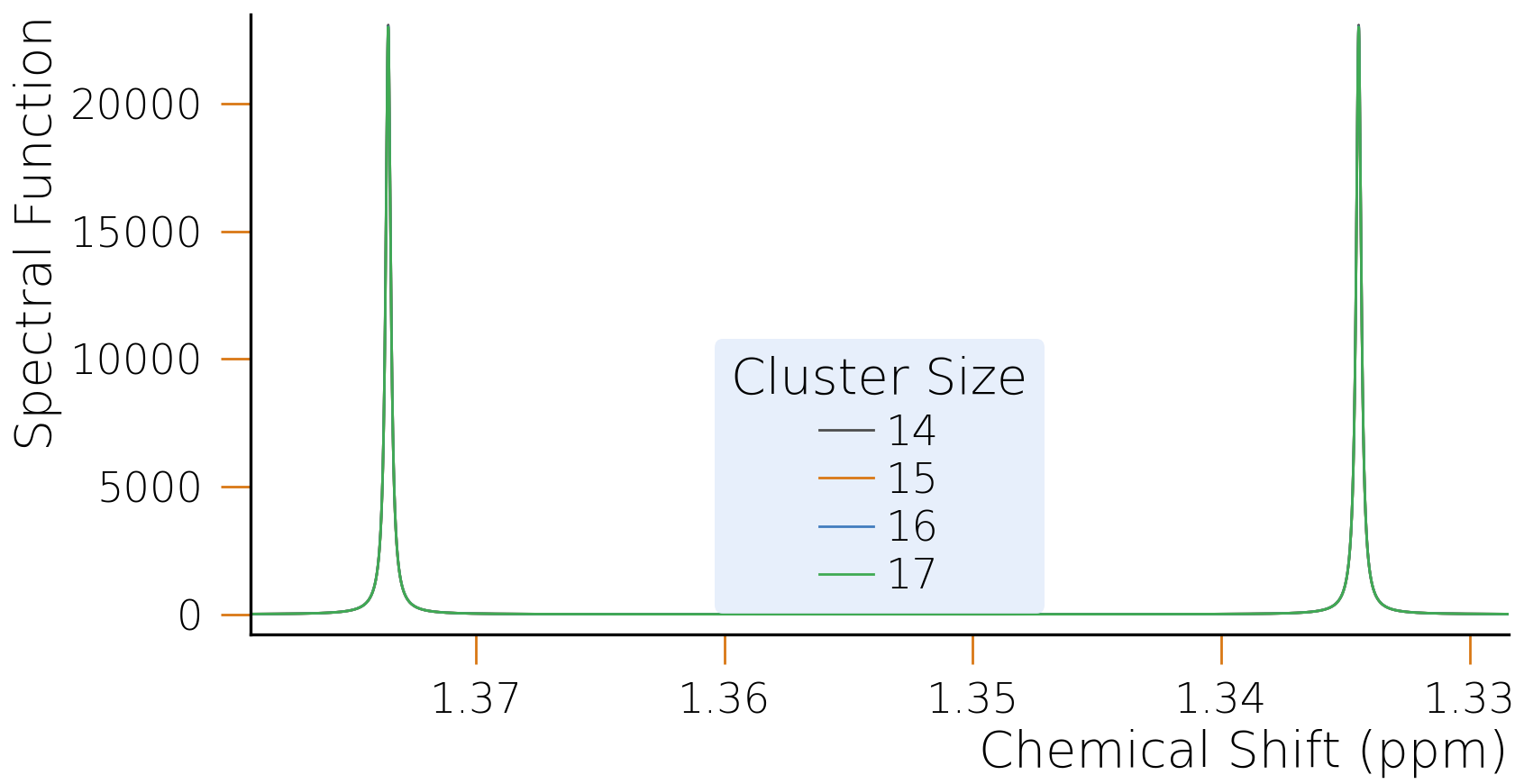}
   \includegraphics[width=0.49\textwidth]{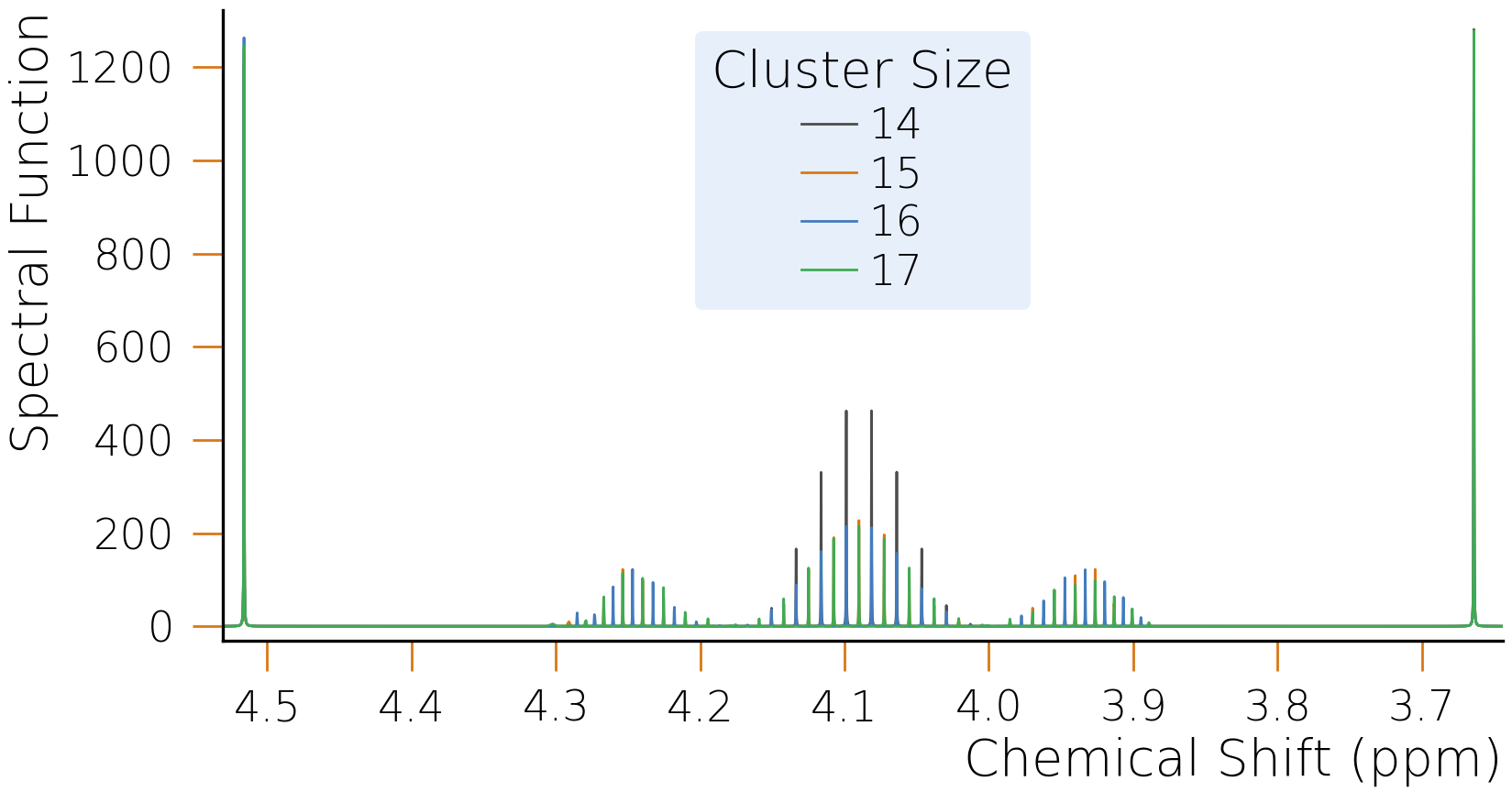}
   \caption{The spectral function for Diphosphane, for the case of high field and low broadening using the simple clustering method. For the improved clustering see Fig.~\ref{fig:diphosphane_spectra_new_clustering}.}
   \label{fig:diph}
\end{figure*}

The Diphosphane molecule is a single-bond analogue of hydrazine in which each phosphorus bears one tert-butyl group and one hydride (tBu–PH–PH–tBu). In Figure \ref{fig:diph} we display the spectral function of Diphosphane at the two portions of the proton NMR spectrum where there is a non-zero signal. To understand the behaviour seen in the convergence plots, we examine the spectral functions for cluster sizes fourteen through seventeen (the largest cluster sizes in the convergence study), specifically for the case of high field and low broadening. We notice two large peaks at low chemical shift, which do not seem to change as a function of cluster size, and then a collection of narrow features at higher chemical shift. We note that the location of these features seems to oscillate as a function of cluster size, rather than converge (this behaviour in fact also occurs at lower cluster sizes). 

\begin{figure}[H]
   \centering
   \includegraphics[width=0.45\textwidth]{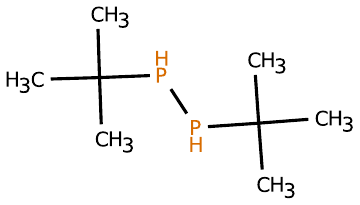}
   \caption{The molecule Diphosphane, with two phosphorous atoms and eighteen hydrogen atoms.}
   \label{fig:diph_mol}
\end{figure}

To understand the origin of this behaviour, in Figure \ref{fig:diph_mol} we show a schematic representation of the Diphosphane molecule. This molecule consists of two central phosphorous atoms, each of which is coupled to three Methyl groups that together contain nine hydrogen atoms. Additionally, each phosphorous atom has a hydrogen atom that is strongly coupled to it (leading to a total of twenty hydrogen atoms). The key feature of the NMR Hamiltonian for this molecule is that none of the hydrogen nuclei couple to each other directly. Rather, each hydrogen nucleus couples only to the nearest phosphorous nucleus. The Diphosphane molecule therefore represents another case in which a lack of direct couplings may lead to clustering which conflicts with the symmetry of the molecule, thereby leading to inaccurate results.

\begin{figure*}[t]
   \centering
   \includegraphics[width=0.49\textwidth]{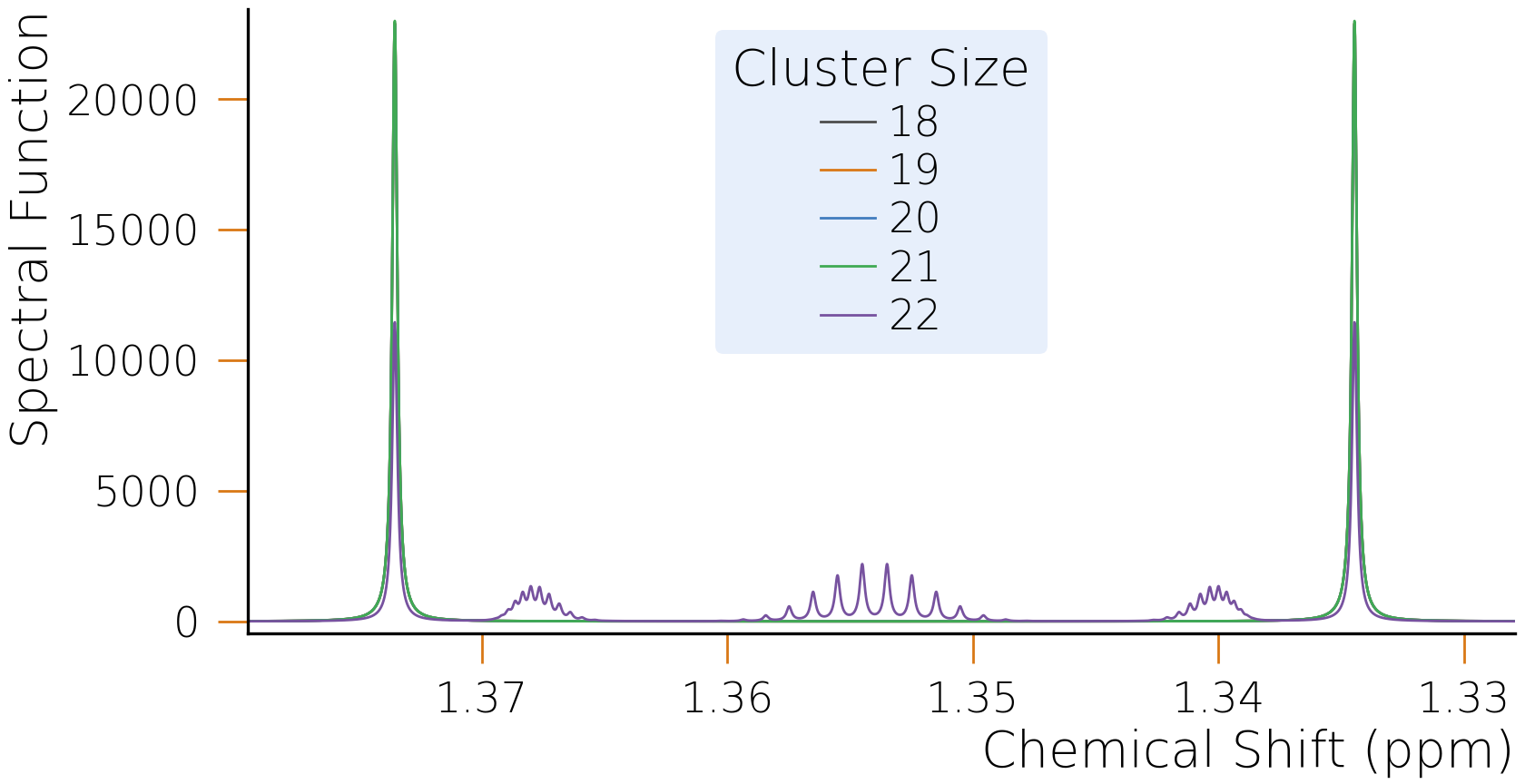}
   \includegraphics[width=0.49\textwidth]{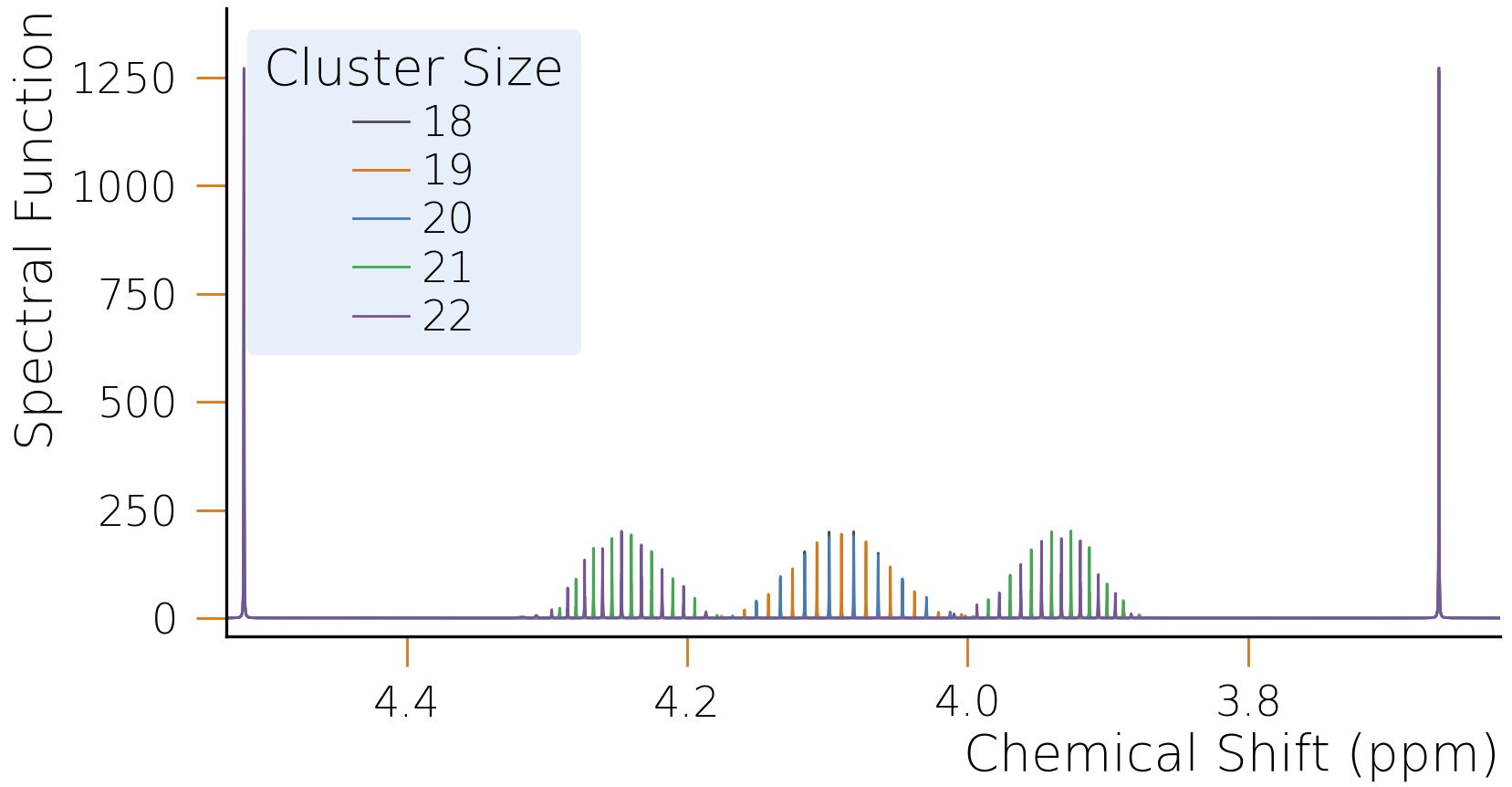}
   \caption{The spectral function for Diphosphane, for the case of high field and low broadening, at larger cluster sizes, which we can obtain due to the symmetry of the molecule using the simple clustering method. For the improved clustering see Fig.~\ref{fig:diphosphane_spectra_new_clustering}.}
   \label{fig:exact_diph}
\end{figure*}

To investigate this possibility, we can in fact compare these results against the exact solution for Diphospane, due to its previously mentioned high level of symmetry. The two phosphorous nuclei have the same chemical shift as each other, which is also true for the two central hydrogen nuclei. These two central hydrogen nuclei couple with the same strength to their respective phosphorous nuclei. However, much more significant is the fact that all of the eighteen hydrogen nuclei contained in the Methyl groups have identical chemical shifts, and also couple identically to their respective phosphorous nuclei. Thus, all of the ``magnetically equivalent'' spins can be combined into a single (higher representation) spin object, by exploiting the local SU(2) symmetry. Specifically, this reduction leads to an effective system of two spin-9/2 particles coupled to four spin-1/2 particles, which can be solved in approximately 10 minutes of computational time on a laptop with 32 Gigabytes of memory and an Intel(R) Core(TM) Ultra 7 165U processor.

In Figure \ref{fig:exact_diph} we again display the spectral function of Diphosphane at the two relevant portions of the spectrum, this time for larger cluster sizes. We continue to see the oscillating behaviour at higher chemical shifts, right up to the exact solution. In addition, we see a dramatic qualitative change in the spectrum at lower chemical shift. As we increase the cluster size from 21 to 22 spins (the full size of the system), a significant fraction of the weight in the two large peaks gets transferred to many other intermediate peaks which appear for the first time. This extreme behaviour verifies our suspicion that our basic clustering method is inadequate for this molecule.

These results we have seen for Diphosphane and TPPO clearly indicate that for molecules which exhibit a high level of symmetry and a small number of direct spin-spin interactions, a refinement to our basic clustering method must be introduced.

%% file: Sections/improve.tex
\section{Improving the Clustering Scheme}
\label{sec:improve}

Here we discuss a relatively simple extension to our basic clustering scheme which enables us to obtain the correct convergence behaviour for both Diphosphane and TPPO. We emphasize that our intent here is not to develop the best possible clustering method for these molecules. Rather, we simply wish to demonstrate that very simple improvements to the basic clustering scheme can already lead to a noticeable improvement for these molecules which possess very few direct couplings in the presence of high symmetry. An ideal clustering scheme might, for example, utilize higher order terms in the J couplings, or possibly invoke ideas from graph theory, and it should certainly include a careful consideration of all of the symmetries present in the structure of a given molecule. Such a study would, however, be outside the scope of this work.

With this caveat in mind, our extended scheme will construct the cluster for a given nucleus as follows. If the chosen cluster size does not exceed the number of other nuclei which couple directly to the given nucleus, then the clustering proceeds as usual. Otherwise, since we cannot fill the cluster with directly coupled nuclei, we turn to the nucleus which is most strongly coupled to the original given nucleus (according to the metric in eq. (\ref{cluster_metric})), and designate it as the secondary nucleus. We now add further nuclei to the cluster, except that the coupling is now between the secondary nucleus and these additional nuclei. If the cluster is still not filled after considering all of the nuclei which are directly coupled to the secondary nucleus, then we find the nucleus which is most strongly coupled to either the first or second nuclei, and designate it as the tertiary nucleus, repeating the procedure with this third nucleus. This process repeats until the cluster has been filled.

\begin{figure}[t]
   \centering
   \includegraphics[width=0.45\textwidth]{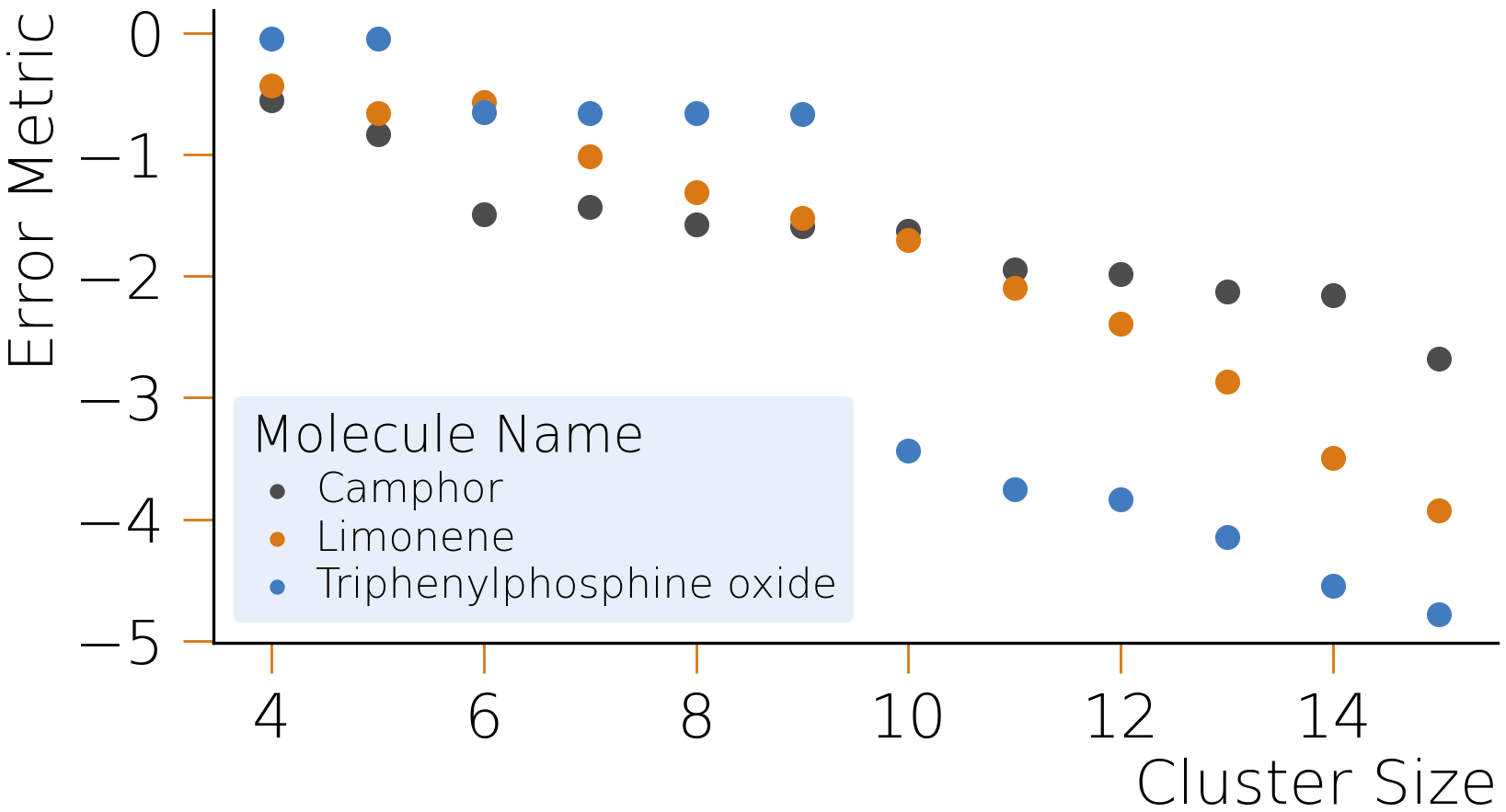}
   \caption{The convergence for selected molecules with 16 nuclear spins for the case of low broadening and very low field, using the extended clustering.}
   \label{fig:new_clustering_16_convergence}
\end{figure}

Figure \ref{fig:new_clustering_16_convergence} displays the convergence behaviour of the three exactly solvable molecules we have previously considered, however now with the spectral functions obtained using this extended clustering. We choose to examine the most extreme case, at very low field and low broadening, in order to highlight the success of this extension. As we can see, the spectral function for TPPO now converges well, while the convergence behaviour of the other two molecules is not significantly altered.

\begin{figure}[t]
   \centering
   \includegraphics[width=0.45\textwidth]{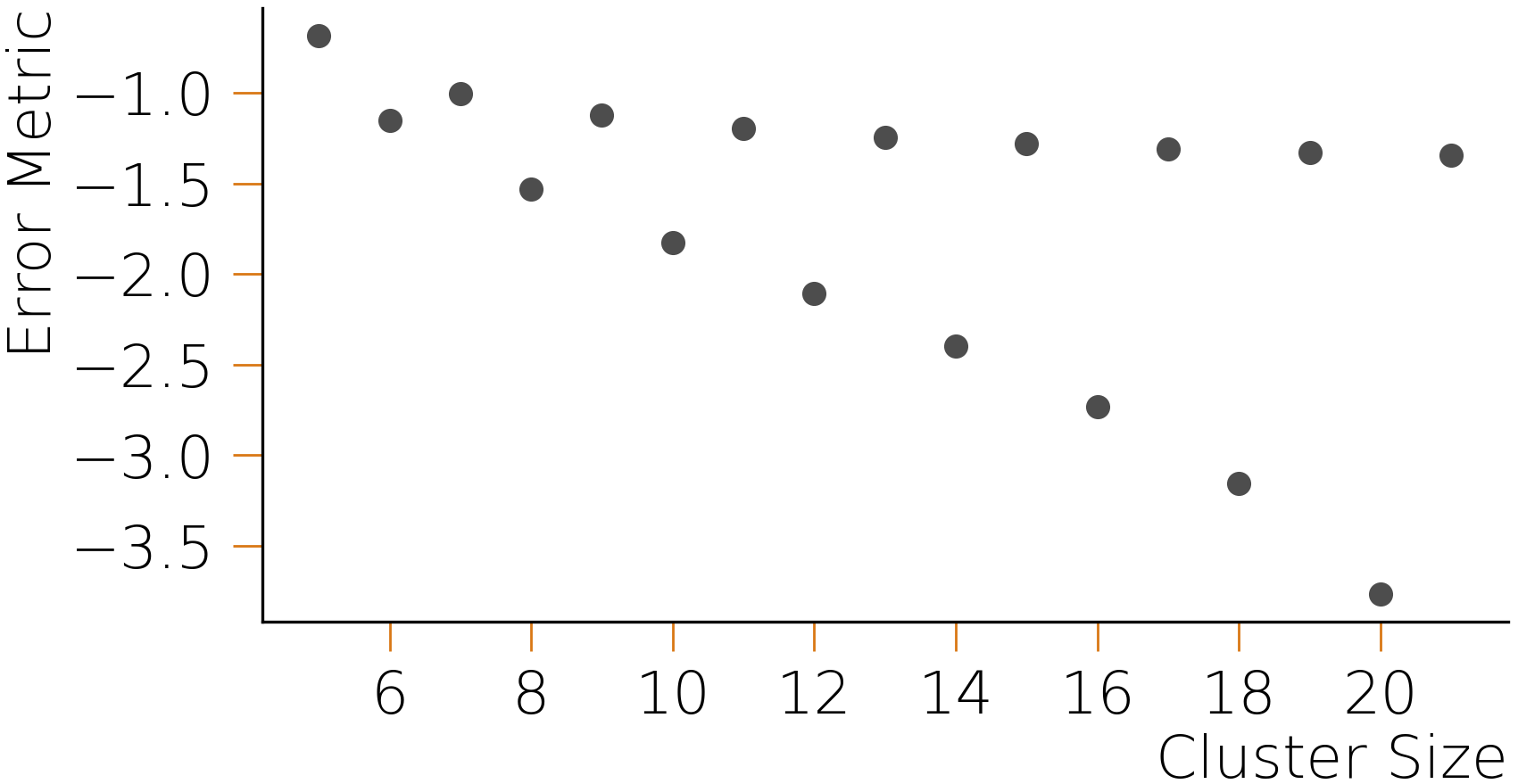}
   \caption{The convergence for Diphosphane for the case of high field and low broadening, using the extended clustering scheme.}
   \label{fig:new_clustering_convergence_diphosphane}
\end{figure}

\begin{figure*}[t]
   \centering
   \includegraphics[width=0.49\textwidth]{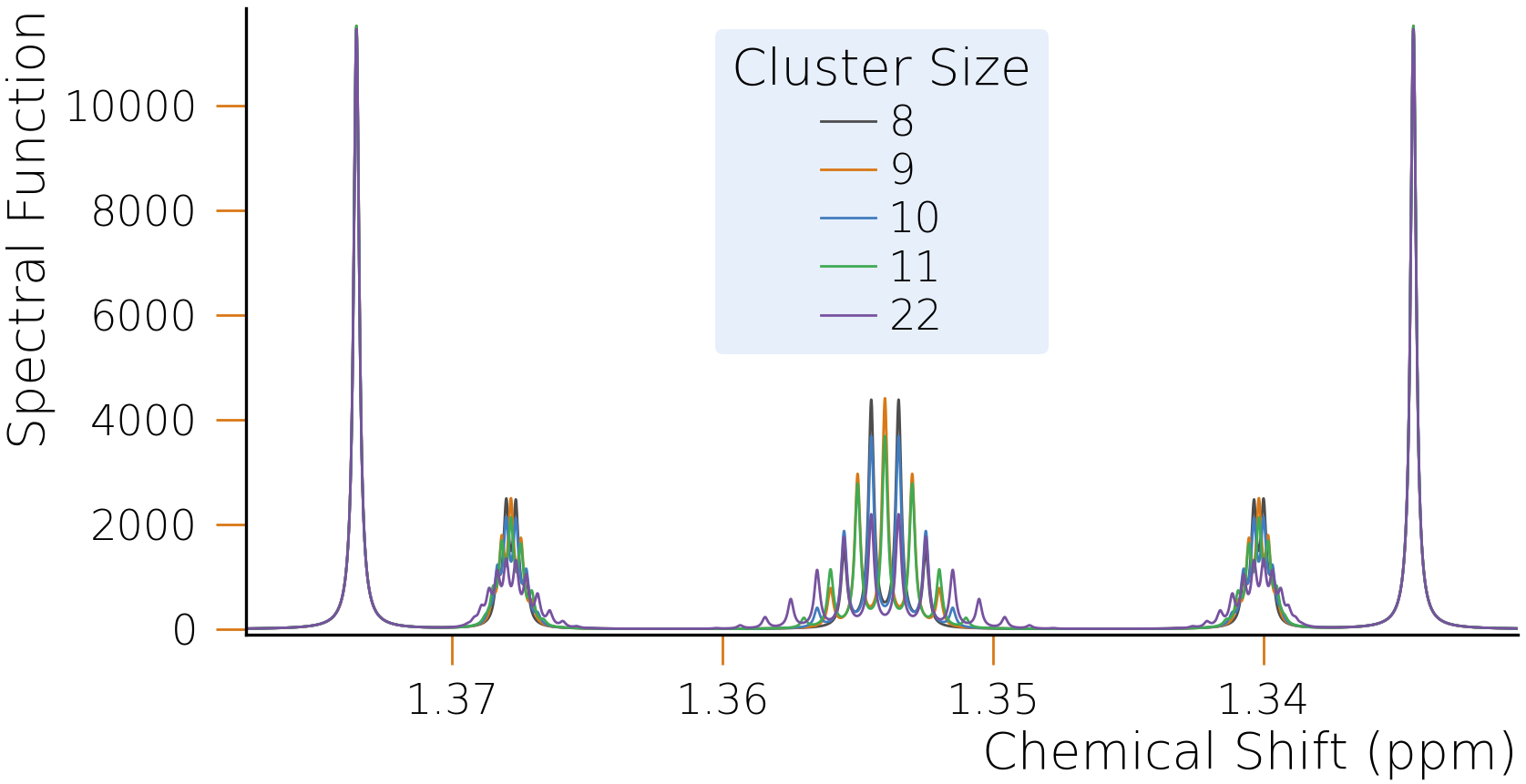}
   \includegraphics[width=0.49\textwidth]{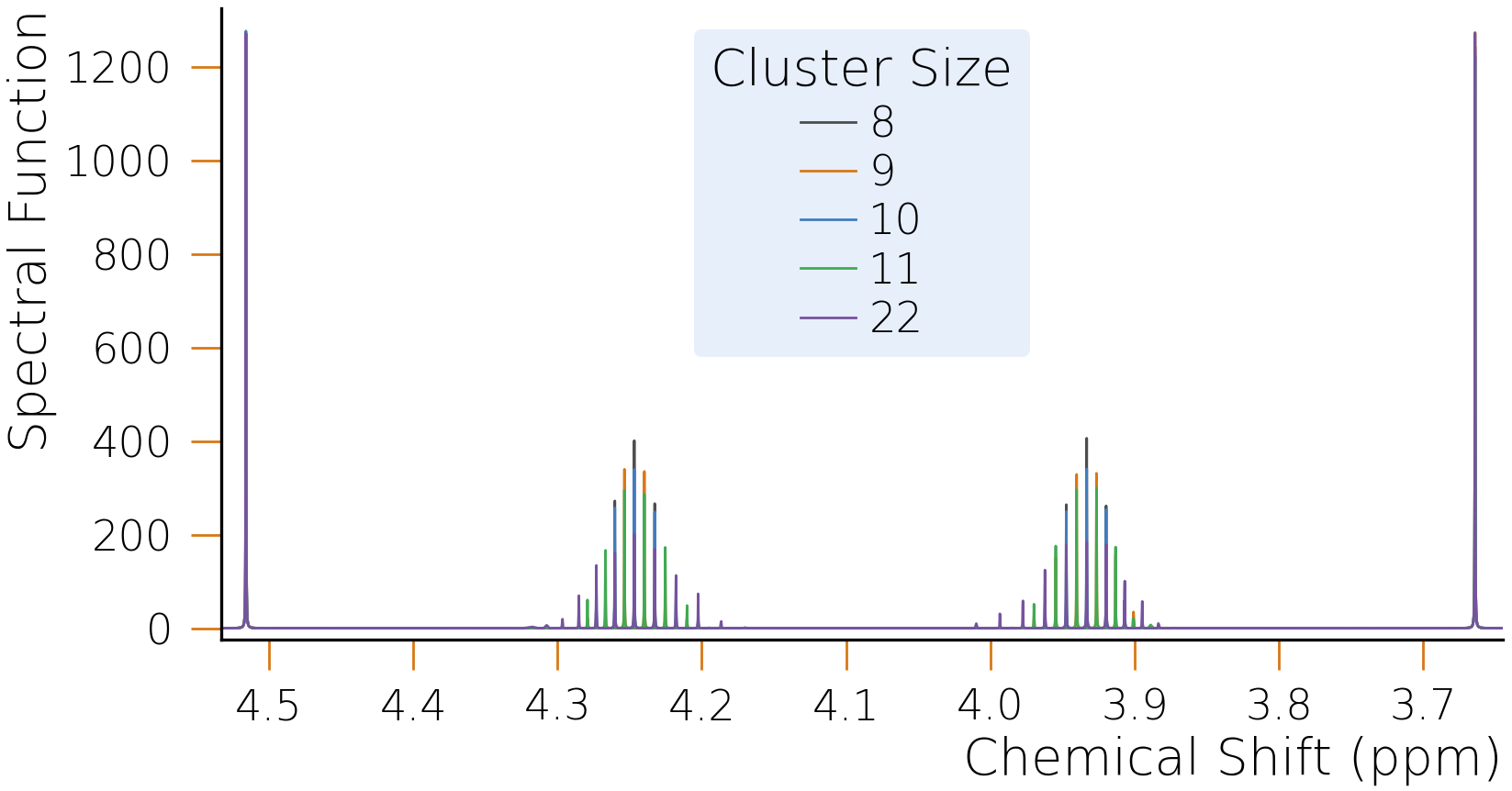}
   \caption{The spectral function of Diphosphane, for the case of high field and low broadening, using the extended clustering scheme. The spectral functions for cluster sizes of eight through eleven already demonstrate good qualitative agreement with the exact solution at 22 spins.}
   \label{fig:diphosphane_spectra_new_clustering}
\end{figure*}

A similar behaviour is reflected in the convergence properties of the larger molecules. Diphosphane is in fact the only one of these molecules which has its convergence behaviour impacted in any meaningful way by the extended clustering, and so we limit our discussion here to this molecule. Figure \ref{fig:new_clustering_convergence_diphosphane} displays the convergence for this molecule, which we note does indeed still demonstrate an unusual oscillating pattern. However, an examination of the spectral functions obtained with this new clustering, which we display in Figure \ref{fig:diphosphane_spectra_new_clustering}, demonstrates that the proper qualitative features of the spectrum are now visible in the approximate spectral functions, with reasonably accurate results being seen when using clusters as small as eight spins. Moving beyond eight spins, the only changes we see in the spectrum concern the precise number of peaks in the individual multiplets, along with their exact position. However, these changes occur in a predictable fashion, as their positions alternate between even and odd sized clusters, with each additional spin leading to an additional peak splitting. As we have seen that this behaviour indeed continues throughout all larger cluster sizes, we come to the conclusion that a knowledge of the total number of spins in the molecule, together with the approximate results obtained using cluster sizes eight through eleven, would be sufficient to identify the correct spectrum of this molecule.

Taken together, these results demonstrate that even for the most pernicious cases we have studied, only a minimal extension to our clustering scheme is necessary to obtain good convergence.

%% file: Sections/conclusion.tex
\section{Conclusion}
\label{sec:conclusion}

In this work we have investigated the performance of our NMR solver when applied to a collection of selected molecules. Across all parameter regimes of practical relevance, we have found that our basic clustering scheme performs very well for almost every molecule, with the performance increasing with higher magnetic fields and higher broadening. The only exceptions take the form of highly symmetric molecules which exhibit very few direct spin couplings. However, a simple extension to our clustering scheme is sufficient to tackle even this case. This demonstrates the utility of our clustering-based solver as a powerful tool for understanding molecular NMR spectra, even for molecules that we can not solve exactly.

With this conclusion in mind, we turn to the question of potential quantum advantage in the context of NMR. At the relatively high magnetic fields which are utilized in conventional NMR experiments, the performance of our solver casts doubt on the possibility of demonstrating quantum advantage in the calculation of traditional proton NMR spectra. While we cannot rule out the possibility that there may exist sufficiently large, highly symmetric molecules of industrial relevance which may pose a challenge for our solver, one must also consider whether the number of such molecules would be sufficiently high as to justify the wide scale application of quantum computing in this context.

At lower magnetic fields, however, the situation may be different. The decreasing performance of our solver with lower broadening and lower magnetic fields leads us to the domain of zero field NMR spectroscopy \cite{BARSKIY2025101558}. In such experiments, hyperpolarized samples are used to measure the NMR spectra at magnetic field strengths of only a few nanotesla, such that the Zeemann term can essentially be neglected. Furthermore, the broadening in these experiments can reach as low as 0.01 Hz, leading us to believe that classical solvers (including our own) should struggle in this regime. This therefore suggests a potential path towards quantum advantage \cite{Seetharam2023}.

The possibility of demonstrating quantum advantage in the context of zero field NMR is a question which we are currently investigating and hope to report on in more depth in the future.

%% file: Sections/acknowledgment.tex
\acknowledgments{

This work was supported by the German Federal Ministry of Education and Research, through projects QSolid (13N16155) and Q-Exa (13N16065),
as well as the European Innovation Council (EIC), through the Horizon Europe EIC Transition project HQS-NextNMR (grant agreement 101214526).
We also want to thank Alexander Zech, Julius Kleine Büning, Peter Pinski, and Ilya Kuprov for helpful discussions.

}

%% file: Sections/appSF.tex
\section{Derivation of the Spectral Function}
\label{sec:appSF}

Here we demonstrate how the spectral function
\begin{equation}
C\left ( \omega \right ) \propto \eta \sum_{n,m} \frac{\langle E_{n}| \hat{M}^{-} | E_{m} \rangle  \langle E_{m} | \hat{M}^{+} |  E_{n} \rangle  }{\eta^{2} + [\omega - (E_{n} - E_{m})]^{2}}
\label{eqn:spectralfunction}
\end{equation}
is connected to measurements made in an NMR experiment, specifically the so-called pulse-acquire experiment. To this end, we will first discuss what properties are measured in such an experiment, and then derive from them the spectral function.

\subsection{The pulse acquire experiment}

In an NMR experiment, the magnetic field resulting from the magnetization of the sample is measured. The average magnetic moment of a single molecule in the sample along a specified direction is given by the expectation value of the appropriate spin operator,
\begin{equation}
m^{\alpha} = \hbar \sum_{i} \gamma_{i} \langle \hat{I}^\alpha \rangle = \hbar \langle \hat{M}^{\alpha} \rangle
\end{equation}
where we have ignored the small correction to the magnetic moment due to the chemical shifts, and have used the notation 
\begin{equation}
\hat{M}^{\alpha} \equiv \sum_{i}\gamma_{i} \hat{I}_{i}^{\alpha},
\end{equation}
with $\alpha \in \{x, y, z\}$. To connect this object to the magnetic field measured in an NMR experiment, we note three important facts. First, the induced magnetic field due to a single molecule is proportional to its magnetic moment. Second, a measurement of the total magnetic field is essentially an ensemble average over macroscopically many quantum measurements, one for each molecule in the sample, and is thus proportional to the expectation value above. Third, since we are not worried about the overall normalization of the spectral function, any overall proportionality factors, such as the permeability or the total number of molecules, are irrelevant. For these reasons, we can simply study the expectation value of the magnetic moment of a single molecule, and our task is therefore to understand how this quantity evolves in time in an NMR experiment.

After a sample is placed in an NMR spectrometer, it is allowed to come to thermal equilibrium, a state described by the density matrix
\begin{equation}
\rho_{\beta} = \frac{1}{Z}\e^{-\beta \hbar \hat{\cal{H}}}.
\end{equation}
with $Z$ the partition sum.
For a spectrometer approaching the limits of current technology, with a natural frequency $\nu$ of 1 GHz, the characteristic energy scale of the molecular NMR Hamiltonian is
\begin{equation}
h \nu \approx 6.626 \times 10^{-25} ~\text{J},
\end{equation}
while the thermal energy scale at a temperature of only one Kelvin is
\begin{equation}
k_{B}T \approx 1.381 \times 10^{-23} ~\text{J}.
\end{equation}
Since the thermal energy scale is several orders of magnitude larger than the characteristic energy scale of the molecule in any conceivable scenario, it is an incredibly good approximation to take
\begin{equation}
\begin{split}
Z \rho_{\beta} & = \mathds{1} - \beta \hbar \hat{\cal{H}}   \\ & = \mathds{1} - \beta \hbar \left [- \sum_l \gamma_l B^z  \hat{I}^z_l + 2 \pi  \sum_{k < l} J_{kl}  \mathbf{\hat I}_k \cdot \mathbf{\hat I}_l \right],
\end{split}
\end{equation}
where we have again ignored the correction due to the chemical shifts. Furthermore, even at magnetic fields as low as the Earth's magnetic field, the coupling terms are several orders of magnitude smaller than the Zeeman terms, and thus these operators can be ignored in the thermal density matrix above. We thus take
\begin{equation}
Z \rho_{\beta} = \mathds{1} + \beta \hbar B^z \sum_l \gamma_l \hat{I}^z_l = \mathds{1} + \beta \hbar B^z \hat{M}^{z}
\end{equation}

Once the system has reached its equilibrium state, an electromagnetic pulse is applied, which is implemented by sending radiofrequency current through a coil wrapped around the sample. Through proper control of the phase, frequency, amplitude, and duration of this pulse, it is possible to manipulate the nuclear spins in the sample, and depending on the frequency band where this pulse is effective, different isotopes can be excited. In the standard proton NMR pulse-acquire experiment which we discuss here, a single pulse close to the Larmor frequency of TMS is used to excite the hydrogen nuclei, and we assume that it is possible to engineer this pulse such that all hydrogen nuclei are affected equally. If we view the system from a frame of reference that rotates with the Larmor frequency of TMS, we can describe the effect of the pulse as inducing a $\pi/2$ rotation of the net magnetization of these nuclei into the XY plane (a rotation which we take to be around the X axis). The hydrogen portion of the density matrix in this frame is therefore transformed as
\begin{equation}
\rho_{\text{Hy}} \to \exp \left( -i \frac{\pi}{2} \hat{I}^x \right ) \rho_{\text{Hy}} \exp \left ( +i \frac{\pi}{2} \hat{I}^x \right).
\end{equation}
If only hydrogen nuclei are present in the molecule, then applying this transformation to the thermal density matrix (which takes the same form in the rotating frame) yields
\begin{equation}
Z \rho_{\beta} \to Z \rho_{0} = \mathds{1} - \beta \hbar B^z \hat{M}^{y}
\end{equation}
as the initial density matrix after the pulse. Contributions from other nuclei in the molecule, which are not flipped by the initial pulse, will eventually drop out of the formula for the net magnetization (for symmetry reasons similar to the one that allows us to ignore the identity operator later on). As a result, we can simply omit any contributions from other nuclei from this point on.

After the pulse has been applied, the system is allowed to evolve freely in time, and the resulting magnetization is measured. It is convenient to continue working in the rotating frame of reference, with the density matrix of the rotating frame and the density matrix of the laboratory frame being related by
\begin{equation}
\rho_{L} = \e^{-i \hat{\cal{H}}_{0} t}~\rho_{R} ~\e^{+i \hat{\cal{H}}_{0} t},
\end{equation}
with 
\begin{equation}
\hat{\cal{H}}_{0} = - \omega_{\text{ref}} \sum_l \hat{I}^z_l \equiv - \omega_{\text{ref}} \hat{I}^z
\end{equation}
corresponding to uniform precession at the Larmor frequency $\omega_{\text{ref}}$ of the hydrogen nuclei in the reference compound (TMS for proton NMR). The rotating frame density matrix evolves according to the shifted Hamiltonian
\begin{equation}
\hat{\cal{H}}_{R} = \hat{\cal{H}} - \hat{\cal{H}}_{0} = - \sum_l \left( \omega_l - \omega_{\text{ref}} \right) \hat{I}^z_l + 2 \pi  \sum_{k < l} J_{kl}  \mathbf{\hat I}_k \cdot \mathbf{\hat I}_l.
\end{equation}
We note that for a hydrogen nucleus in particular, the definition of the chemical shifts leads to
\begin{equation}
\omega_{l} - \omega_{\text{ref}} = \gamma_{\text{ref}} (1 + \delta_l) B^z - \gamma_{\text{ref}} B^z = \gamma_{\text{ref}} \delta_l B^z,
\end{equation}
where $\gamma_{\text{ref}}$ is the gyromagnetic ratio with respect to which the chemical shifts have been defined. The density matrix of the molecule in the rotating frame therefore evolves in time according to
\begin{equation}
\rho \left ( t \right ) = \e^{-i \hat{\cal{H}}_{R} t} \rho_{0} \e^{+i \hat{\cal{H}}_{R} t} \propto \mathds{1} - \beta \hbar B^z \e^{-i \hat{\cal{H}}_{R} t} \hat{M}^{y} \e^{+i \hat{\cal{H}}_{R} t}.
\end{equation}
The magnetic moment is given according to
\begin{equation}
m^{\alpha} = \hbar \langle \hat{M}^{\alpha} \rangle =  \hbar \text{Tr} \left [ \hat{M}^{\alpha} \rho \left ( t \right ) \right ].
\end{equation}
We can, if we like, omit nuclei other than hydrogen from $\hat{M}^{x/y}$ in the above expression, since these do not yield a significant signal in a proton NMR spectrum (this can be shown via a perturbative analysis).
Since $\hat{M}^{\alpha}$ is traceless, only the second term in the density matrix contributes to the trace, and we have
\begin{equation}
m^{x} = \hbar \langle \hat{M}^{x} \rangle \propto \text{Tr} \left [ \hat{M}^{x} \e^{-i \hat{\cal{H}}_{R} t} \hat{M}^{y} \e^{+i \hat{\cal{H}}_{R} t} \right ],
\label{eqn:magneticmomentx}
\end{equation}
along with
\begin{equation}
m^{y} = \hbar \langle \hat{M}^{x} \rangle \propto \text{Tr} \left [ \hat{M}^{y} \e^{-i \hat{\cal{H}}_{R} t} \hat{M}^{y} \e^{+i \hat{\cal{H}}_{R} t} \right ],
\label{eqn:magneticmomenty}
\end{equation}
Furthermore, if we assume that this magnetization decays over time due to some internal decoherence processes, we should in fact send
\begin{equation}
m^{\alpha} \to \e^{-\eta t}m^{\alpha} \,.
\end{equation}

\subsection{Computing the Spectral Function from the Correlation Function}

We can interpret these formulas (\ref{eqn:magneticmomentx}) and (\ref{eqn:magneticmomenty}) for the magnetic moments as correlation functions, respectively denoted $C_{XY}$ and $C_{YY}$, making use of the notation
\begin{equation}
C_{\alpha \beta} \left ( t \right ) = \text{Tr} \left [ \left ( \hat{M}^{\alpha} \right )^{\dagger} \e^{-i\hat{\cal{H}}_{R}t} \hat{M}^{\beta} \e^{+i\hat{\cal{H}}_{R}t} \right ]
\end{equation}
If we extend this notation, and thereby define
\begin{equation}
\begin{split}
C_{++} \left ( t \right ) & = \text{Tr} \left [ \left ( \hat{M}^{+} \right )^{\dagger} \e^{-i\hat{\cal{H}}_{R}t} \hat{M}^{+} \e^{+i\hat{\cal{H}}_{R}t} \right ] \\ & = \text{Tr} \left [ \hat{M}^{-} \e^{-i\hat{\cal{H}}_{R}t} \hat{M}^{+} \e^{+i\hat{\cal{H}}_{R}t} \right ],
\end{split}
\end{equation}
then using the linearity of the trace, we can write
\begin{equation}
C_{++}( t) = \left [ C_{XX}(t) + C_{YY}(t) \right ]  + i \left [ C_{XY}(t) - C_{YX}(t) \right ].
\end{equation}
If we further make use of the cyclic nature of the trace, together with the rotational symmetry of the Hamiltonian under rotations around the Z axis,
\begin{equation}
R \hat{M}^{x} R^{\dagger} = \hat{M}^{y} ;~ R \hat{M}^{y} R^{\dagger} = -\hat{M}^{x} ;~ R \e^{\pm i\hat{\cal{H}}_{R}t} R^{\dagger} = \e^{\pm i\hat{\cal{H}}_{R}t}
\end{equation}
we can rewrite this as
\begin{equation}
C_{++}(t) = 2 \left ( C_{YY}(t) + i C_{XY}(t) \right ).
\end{equation}
Inserting two complete sets of states in the definition of $C_{++}$, we find
\begin{equation}
\begin{split}
C_{++} \left ( t \right ) & = \sum_{n,m} \langle E_{n} |  \hat{M}^{-}  \e^{-i\hat{\cal{H}}_{R}t} |E_{m} \rangle \langle E_{m} | \hat{M}^{+} \e^{+i\hat{\cal{H}}_{R}t} | E_{n} \rangle \\ & = \sum_{n,m} \e^{-i\left(E_{m} - E_{n}\right)t} \langle E_{n} |  \hat{M}^{-} |E_{m} \rangle \langle E_{m} | \hat{M}^{+}  | E_{n} \rangle
\end{split}
\end{equation}
We now consider the half-sided transform of this object,
\begin{equation}
C_{++} \left ( w \right ) = \int_{0}^{+\infty} C_{++} \left ( t \right ) \e^{-i\omega t}\e^{-\eta t} dt
\end{equation}
Note that the quantity $\eta$ serves here as a convergence factor, although its physical motivation has been discussed previously. Using the fact
\begin{equation}
\int_{0}^{+\infty} \e^{-i \left ( \omega + E_{m} - E_{n} \right ) t}\e^{-\eta t} dt = \frac{-i}{\omega + E_{m} - E_{n} - i \eta}
\end{equation}
We find that
\begin{equation}
C_{++} \left ( w \right ) = -i \sum_{n,m} \frac{\langle E_{n} |  \hat{M}^{-}  |E_{m} \rangle \langle E_{m} | \hat{M}^{+}  | E_{n} \rangle}{\omega + E_{m} - E_{n} - i \eta}.
\end{equation}
Since $\hat{M}^{-}$ and $\hat{M}^{+}$ are conjugates of each other, the numerator of this expression is real, and so taking the real part of the full expression, we obtain
\begin{equation}
\text{Re}\left[C_{++} \left ( w \right )\right] = \eta \sum_{n,m} \frac{\langle E_{n} | \hat{M}^{-} |E_{m} \rangle \langle E_{m} | \hat{M}^{+} | E_{n} \rangle}{\eta^{2} + [\omega + E_{m} - E_{n}]^{2}},
\end{equation}
which is the spectral function (\ref{eqn:spectralfunction}), as desired.

We note that the eigenstates above are those of $\hat{\cal{H}}_{R}$, the Hamiltonian in the rotating frame (indeed, our solver performs the diagonalization on the rotating frame Hamiltonian). However, the only effect this has on the spectral function is to shift all of the frequencies by the Larmor frequency (as one would intuitively expect), and since we ultimately plot everything in terms of the chemical shift anyways, this is ultimately a relatively unimportant distinction. We also emphasize that while it is not necessary to include nuclei other than hydrogen in the $\hat{M}^{\pm}$ operators, it is still crucial to include them in the Hamiltonian, since this will generate important splittings in the hydrogen spectrum.

%% file: Sections/appCalc.tex
\section{Calculation of Cosine Similarity}
\label{sec:appCalc}

Here we comment briefly on how we use the output of the solver to compute the cosine similarity. 

While the user of our NMR solver can in principle specify the points at which to evaluate the spectral function, the default behaviour of the solver is to determine the most relevant frequency points automatically. This is done using an equal-area partitioning of the spectral function at different levels of approximation, resulting in a nonlinear frequency grid. This is advantageous numerically, since computational effort is not wasted on portions of the spectrum where there is no signal. By default, the solver will return the spectral function at 2,000 points, although for the atypically low line widths which we simulate in the low broadening regime, we find it necessary to increase this to 20,000 points. It is worth noting that this automatic identification of the relevant frequency points may result in different frequency values for the same molecule, depending on the specified parameters of the problem (magnetic field, maximum cluster size, etc.).

In order to compute the cosine similarity, we must compute the quantity
\begin{equation}
\cos \theta_{ab} = \frac{\int_{-\infty}^{+\infty}C_{a}\left ( \omega \right )C_{b}\left ( \omega \right )d\omega}{\sqrt{\int_{-\infty}^{+\infty}C^{2}_{a}\left ( \omega \right )d\omega}~\sqrt{\int_{-\infty}^{+\infty}C^{2}_{b}\left ( \omega \right )d\omega}}.
\end{equation}
Each of the integrals in the above expression can be approximated with a Riemann sum as
\begin{equation}
\int_{-\infty}^{+\infty} f \left ( \omega \right ) d\omega \approx \delta \sum_{i} f  \left ( \omega_{i} \right ) 
\end{equation}
where $\delta$ (not to be confused with the chemical shift) is the spacing between some equally-spaced points $\omega_{i}$ which cover the entire range in which at least one of the functions $C_{a}$ and $C_{b}$ is non-zero. This reduces the expression for the cosine similarity to
\begin{equation}
\cos \theta_{ab} = \frac{\sum_{i} C_{a}\left ( \omega_{i} \right )C_{b}\left ( \omega_{i} \right )}{\sqrt{\sum_{i} C^{2}_{a}\left ( \omega_{i} \right )}~\sqrt{\sum_{i} C^{2}_{b}\left ( \omega_{i} \right )}}.
\end{equation}
For a sufficiently large number of points, this finite sum will converge to the value of the continuous integral.

In practice, we perform the integration with respect to the shifted and scaled variable,
\begin{equation}
\Delta = \frac{\omega - \omega_{\text{ref}}}{\omega_{\text{ref}}},
\end{equation}
which does not alter the value of the cosine similarity, since the Jacobian of such a transformation is a simple constant factor, which cancels in the definition of the cosine similarity. We find that 100,000 frequency points is sufficient for the convergence of the cosine similarity in all of the cases we consider. Explicitly computing the value of the spectral function at all of these points is however unnecessary, as the values of the spectral function at the 20,000 points returned by the solver are sufficiently dense that a good interpolation of the data can be performed. In particular, we make use of the PCHIP shape-preserving interpolator in scipy, as this appears to be more robust against Runge phenomena at the edges of the spectrum, as opposed to other possible options. Any large gaps in the spectrum are also filled in explicitly with zero values, in order to aide the interpolation process.

%% file: Sections/appParam.tex
\section{Parameter Source}
\label{sec:appParam}

In Table \ref{tab:param}, we list the molecules which we study in our analysis, and indicate the source of the chemical shifts and couplings which appear in the NMR Hamiltonian. All of the active nuclei for the molecules mentioned below are $^{1}$H nuclei, except for the case of a.) Diphosphane, which in addition to containing 20 $^{1}$H nuclei, also possesses two $^{31}$P nuclei, as well as the case of b.) Triphenylphosphine oxide, which in addition to containing 15 $^{1}$H nuclei, also possesses one $^{31}$P nucleus.

We divide the molecules into six groups, depending on how we obtain their chemical shifts and couplings. For the molecules in Group A, we compute the couplings internally, while we obtain the chemical shifts from the Biological Magnetic Resonance Data Bank [\url{https://bmrb.io/}]. For the molecules in Group B, we compute the couplings internally, while we obtain the chemical shifts from the publication \cite{doi:10.1021/cr200106v}. For the molecules in Group C, the parameters are obtained from the collection of Hans Reich [\url{https://organicchemistrydata.org}]. For the molecules in Group D, we compute all of the parameters internally. We obtain the parameters for the Diphosphane molecule (Group E) from the \texttt{Spinach} library \cite{Hogben2011}. Lastly, for the molecule Friedelin (Group F), we compute the NMR parameters internally, although we have manually adjusted some of the shifts to match experimental data provided to us by PhytoLab GmbH \& Co. KG.

\begin{table*}
   \caption{The list of molecules studied in this work, along with their size and parameter source.}
   \label{tab:param}
   \small
   \centering
   \begin{tabular}{lccr}
   \toprule\toprule
   \textbf{Molecule Name} & \textbf{Nuclei Count} & \textbf{Molecule Group} \\ 
   \midrule
   Cyclobutane & 8 & B \\
   THF & 8 & B \\
   Cyclopentene & 8 & B \\
   Cyclohexenone & 8 & B \\
   Toluene & 8 & B \\
   Norbornadiene & 8 & B \\
   methylenecyclobutane & 8 & B \\
   quadricyclane & 8 & B \\
   Naphthalene & 8 & C \\
   Propane & 8 & C \\
   Cinnamaldehyde & 8 & D \\
   t-Butyl chloride & 9 & B \\
   2-Methyl-2-cyanopropane & 9 & B \\
   Cyclopentane & 10 & B \\
   t-Butylacetylene & 10 & B \\
   MTBE (methyl t-butyl ether) & 12 & B \\
   Anethole & 12 & D \\
   Camphor & 16 & A \\
   Limonene & 16 & D \\
   Triphenylphosphine oxide & 16 & D \\
   (+)-Neomenthol (SSR) & 20 & D \\
   (-)-Menthol (RSR) & 20 & D \\
   (+)-Isomenthol (SRR) & 20 & D \\
   (-)-Neoisomenthol (SSS) & 20 & D \\
   Artemisinin & 22 & A \\
   1,2-di-tert-butyl-diphosphane (Diphosphane) & 22 & E \\
   Androstenedione & 26 & D \\
   Patchouli alcohol & 26 & D \\
   4,10,11,11-Tetramethyltricyclo[5.3.1.01,5]undecan-10-ol & 26 & D \\
   Friedelin & 50 & F \\
   \bottomrule
   \end{tabular}
\end{table*}

%% file: Sections/appExtraData.tex
\section{Additional Convergence Data}
\label{sec:appExtraData}

\begin{figure*}
   \centering
   \includegraphics[width=0.32\textwidth]{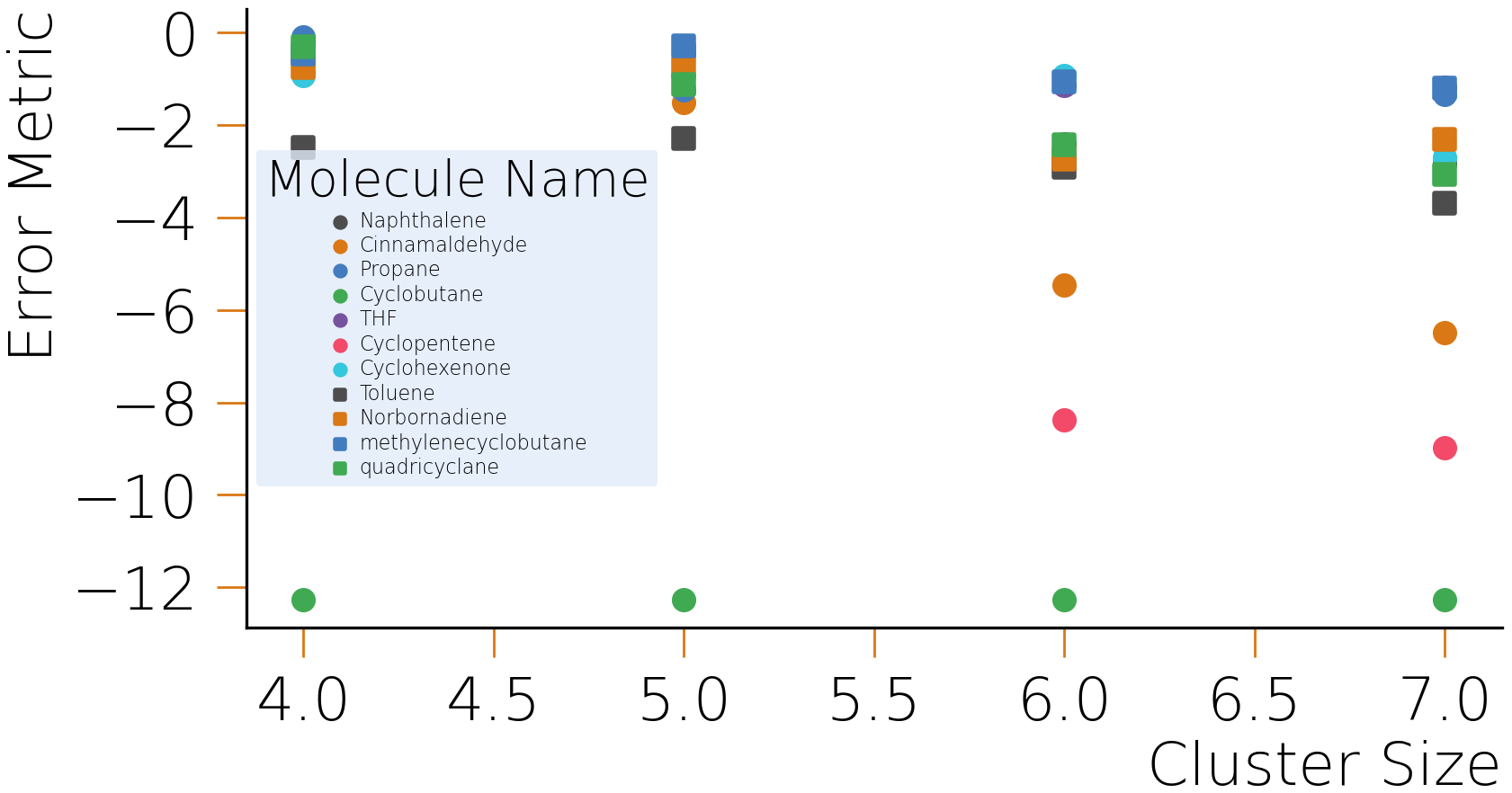}
   \includegraphics[width=0.32\textwidth]{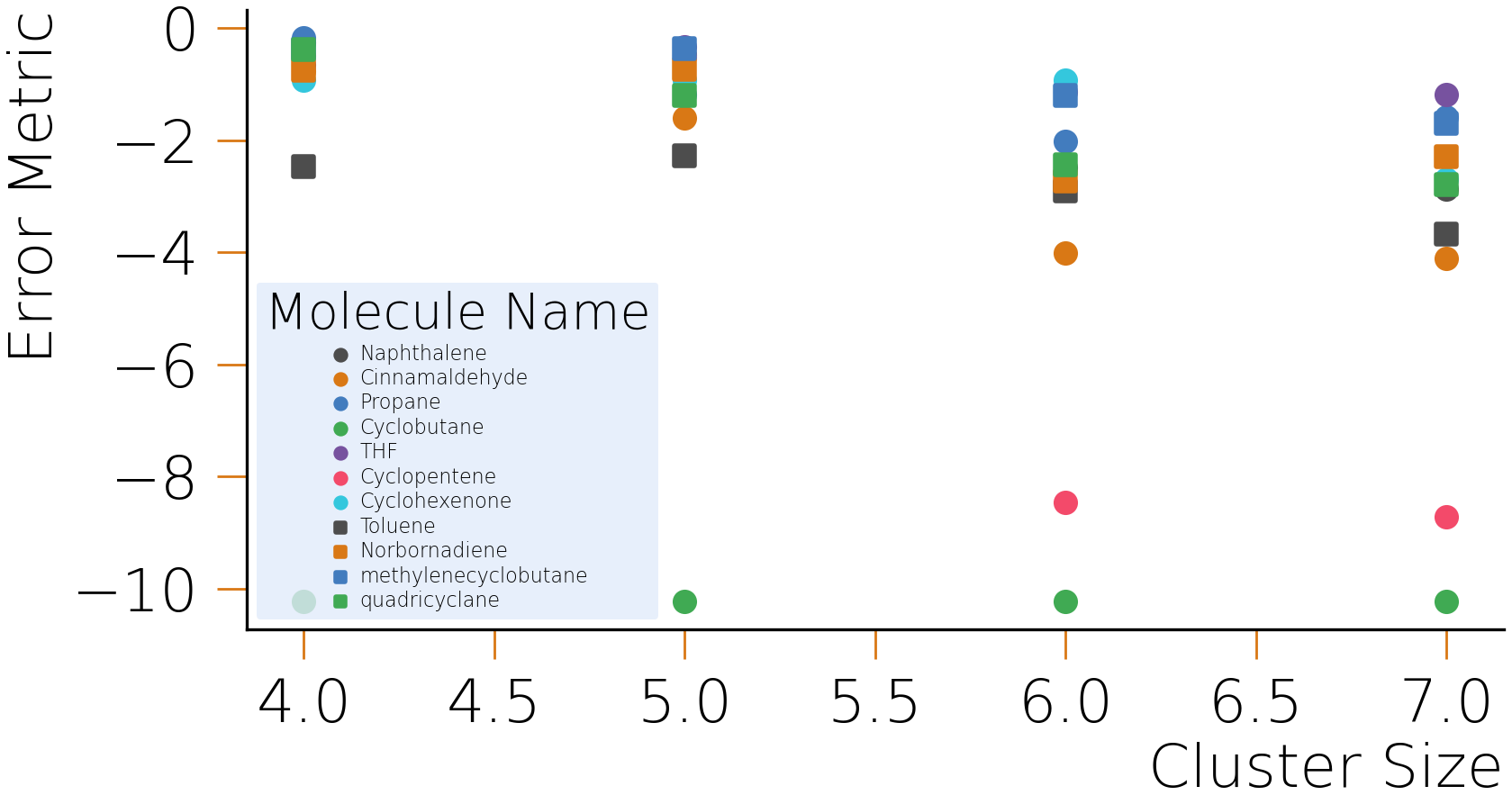}
   \includegraphics[width=0.32\textwidth]{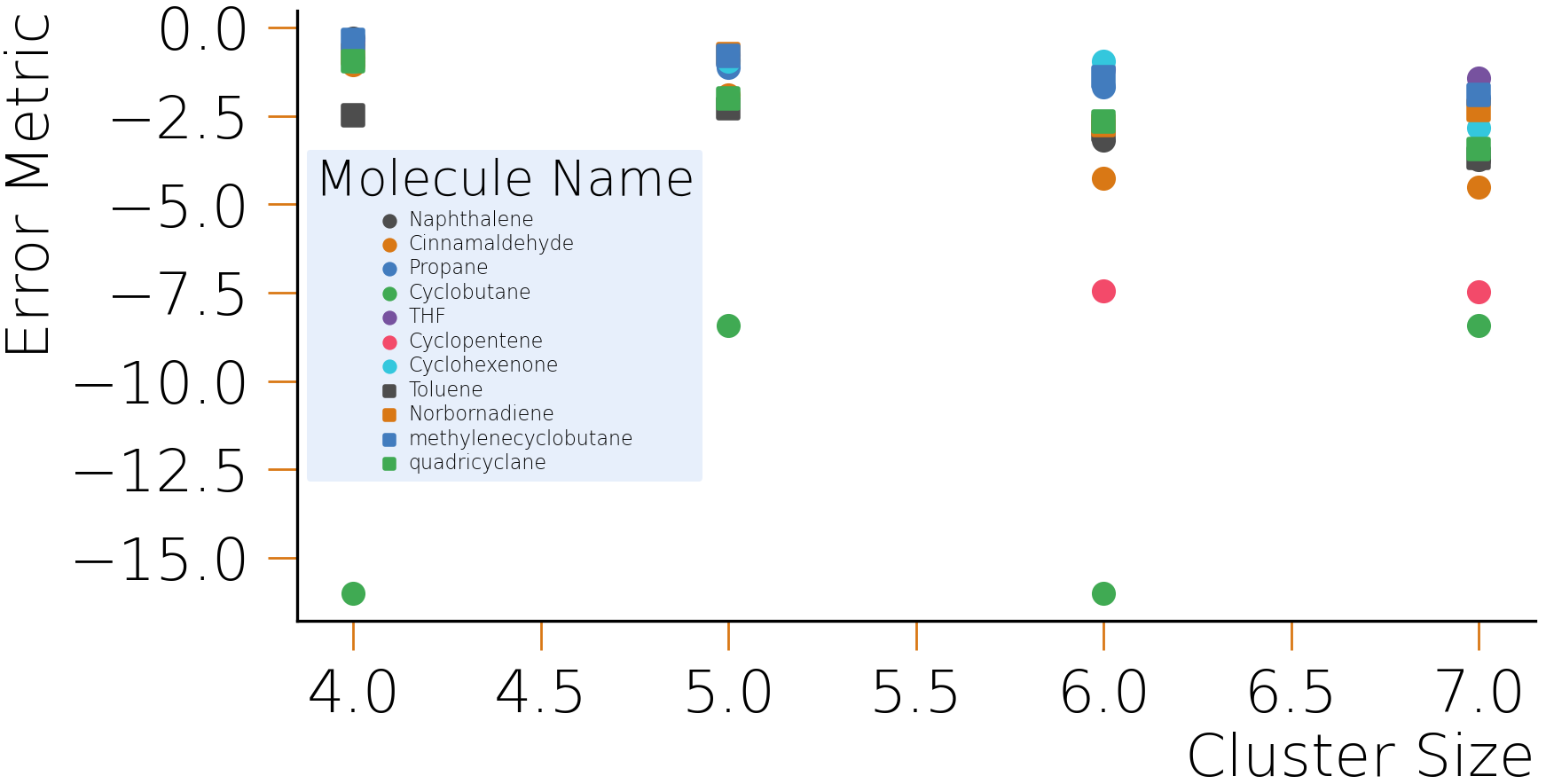}
   \caption{The convergence for selected molecules with 8 nuclear spins for the case of high broadening, for high, low, and very low field.}
   \label{fig:8high}
\end{figure*}

\begin{figure*}
   \centering
   \includegraphics[width=0.32\textwidth]{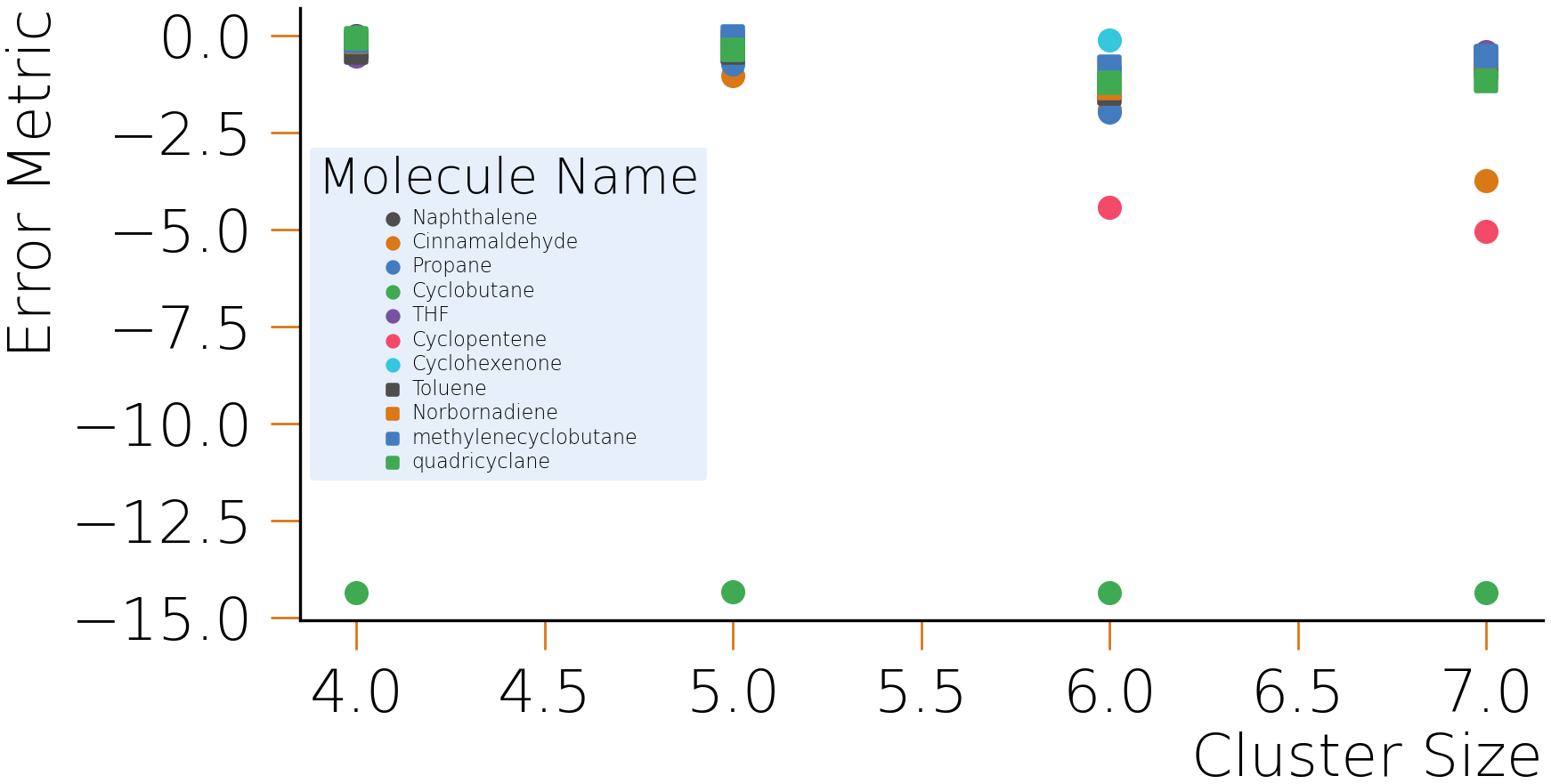}
   \includegraphics[width=0.32\textwidth]{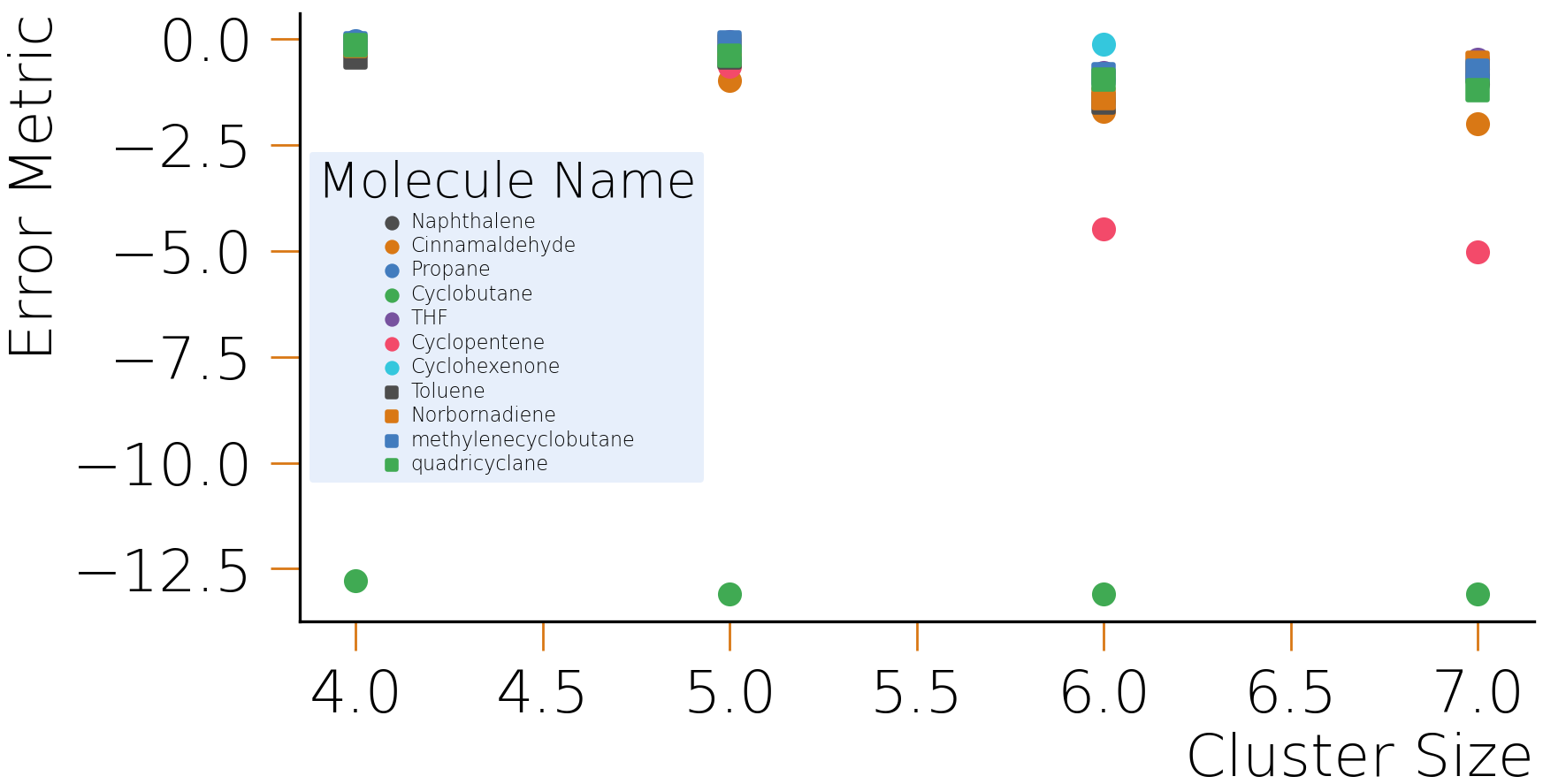}
   \includegraphics[width=0.32\textwidth]{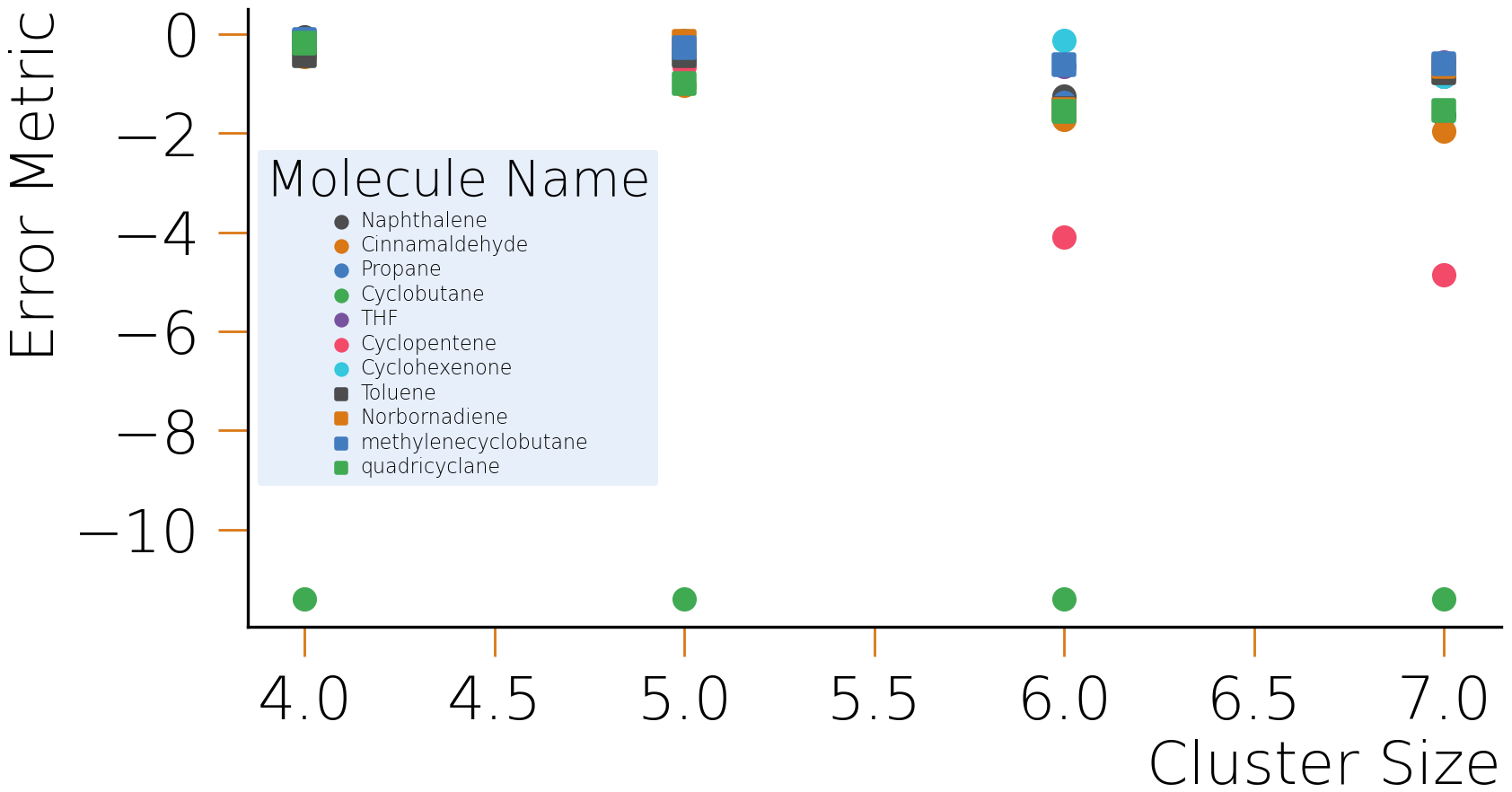}
   \caption{The convergence for selected molecules with 8 nuclear spins for the case of low broadening, for high, low, and very low field.}
   \label{fig:8low}
\end{figure*}

\begin{figure*}
   \centering
   \includegraphics[width=0.32\textwidth]{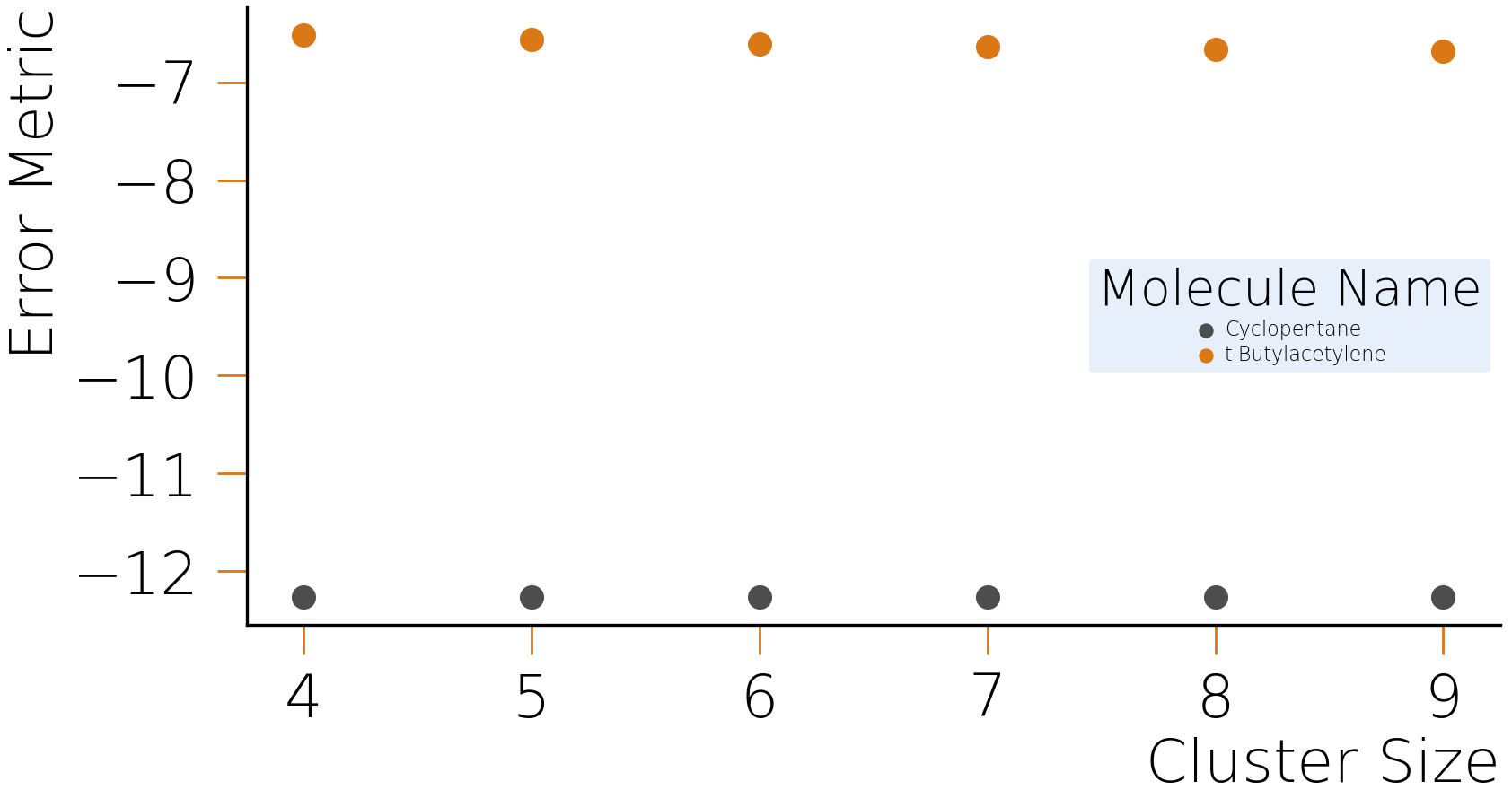}
   \includegraphics[width=0.32\textwidth]{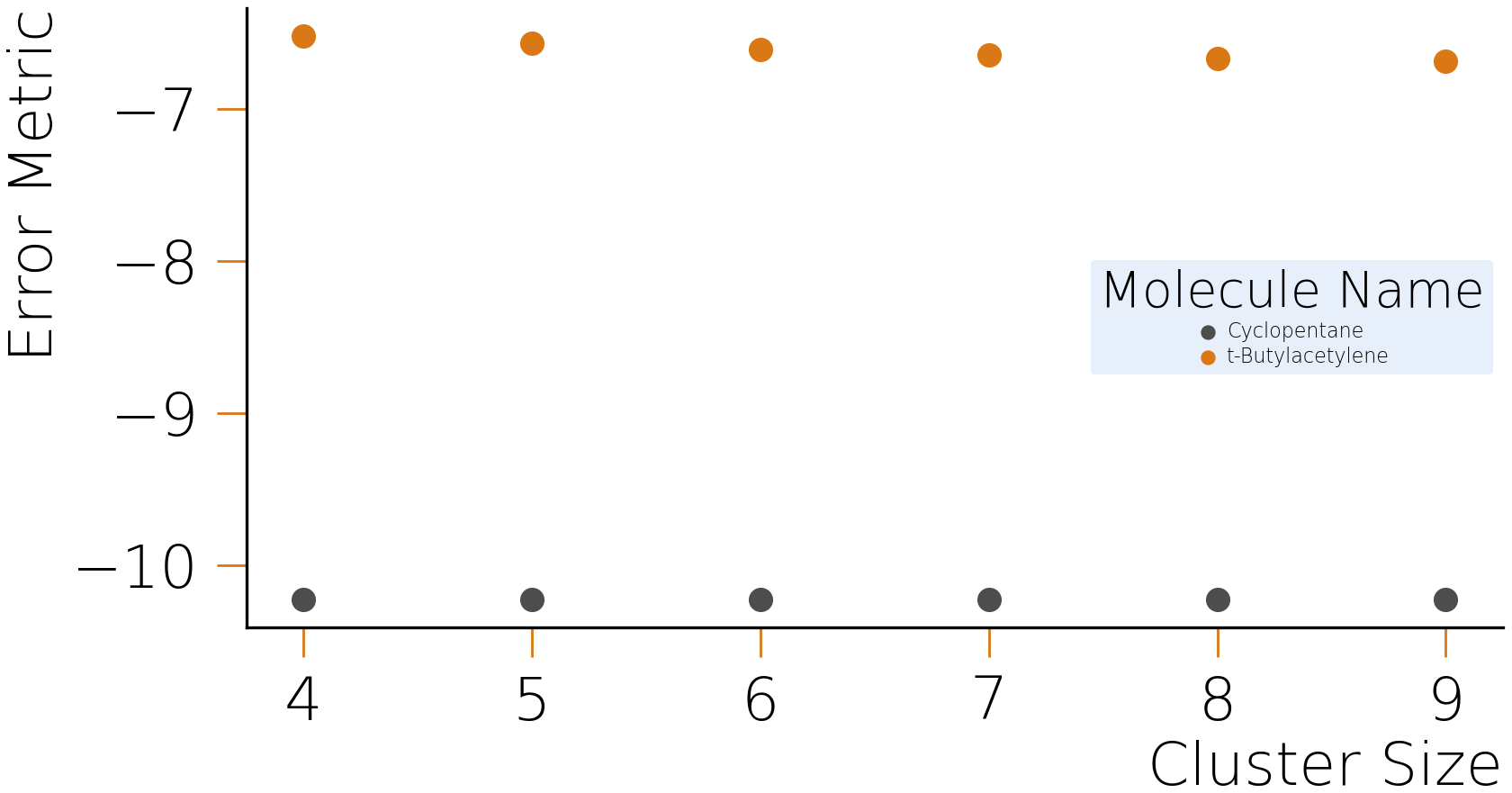}
   \includegraphics[width=0.32\textwidth]{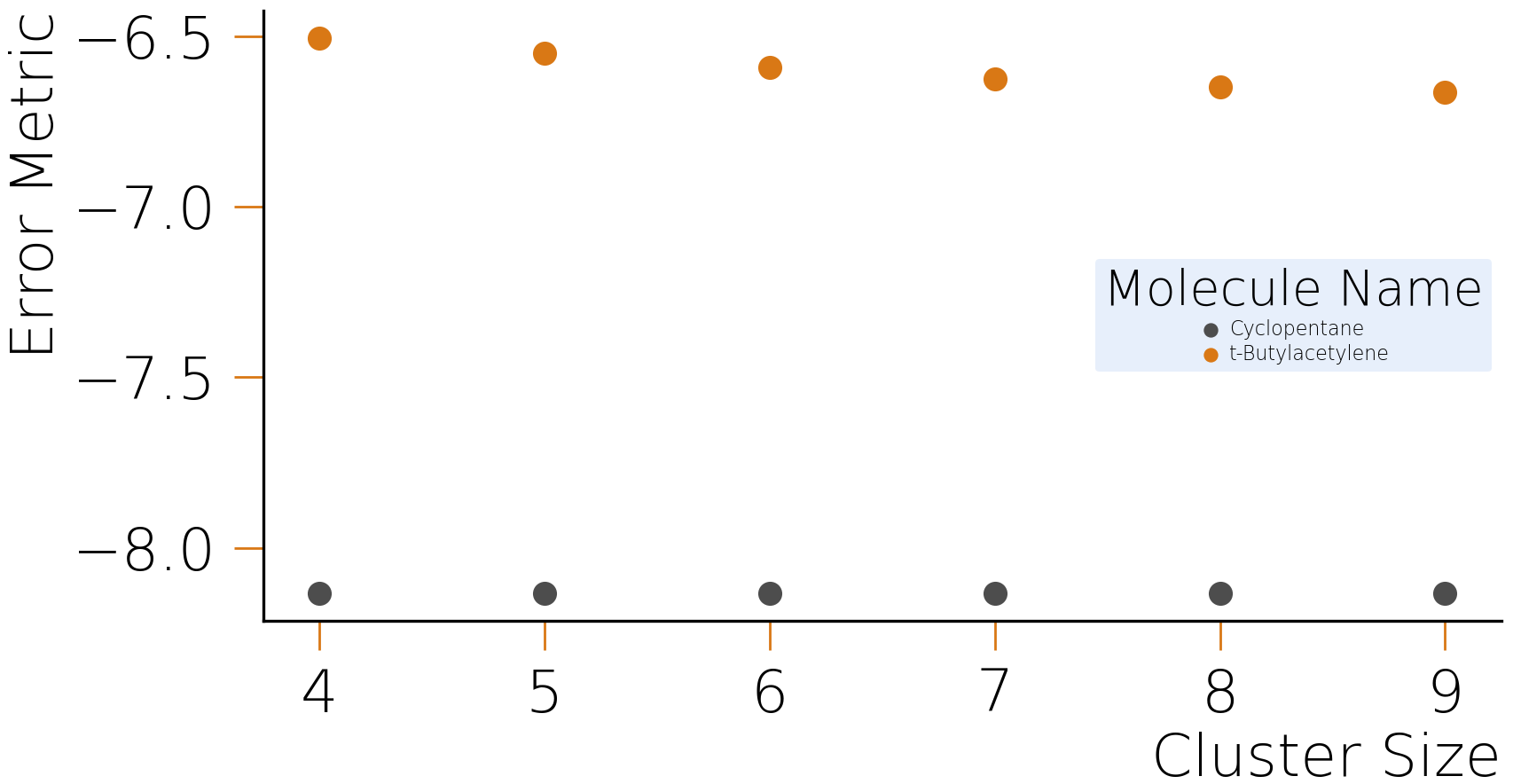}
   \caption{The convergence for selected molecules with 10 nuclear spins for the case of high broadening, for high, low, and very low field.}
   \label{fig:10high}
\end{figure*}

\begin{figure*}
   \centering
   \includegraphics[width=0.32\textwidth]{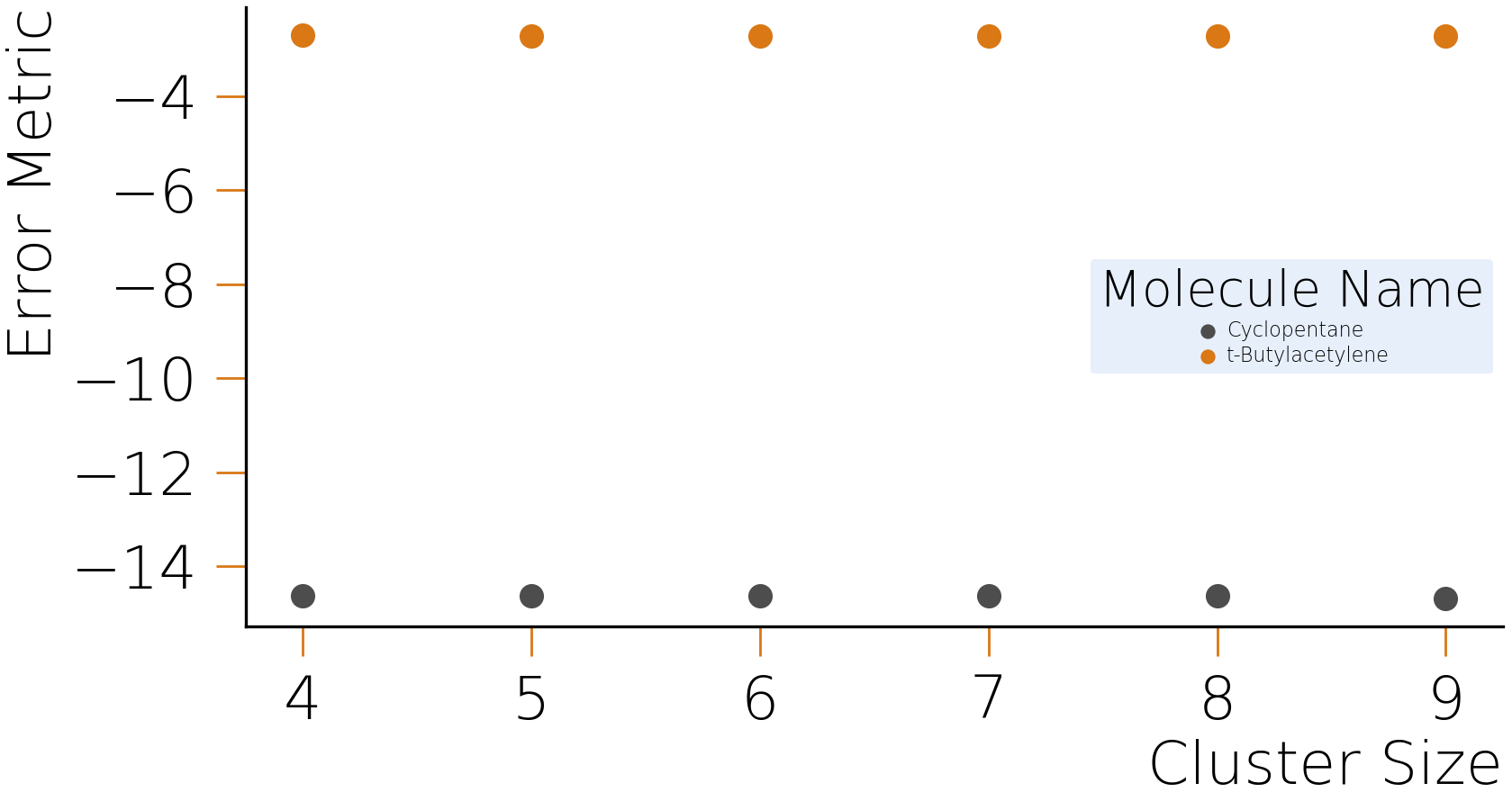}
   \includegraphics[width=0.32\textwidth]{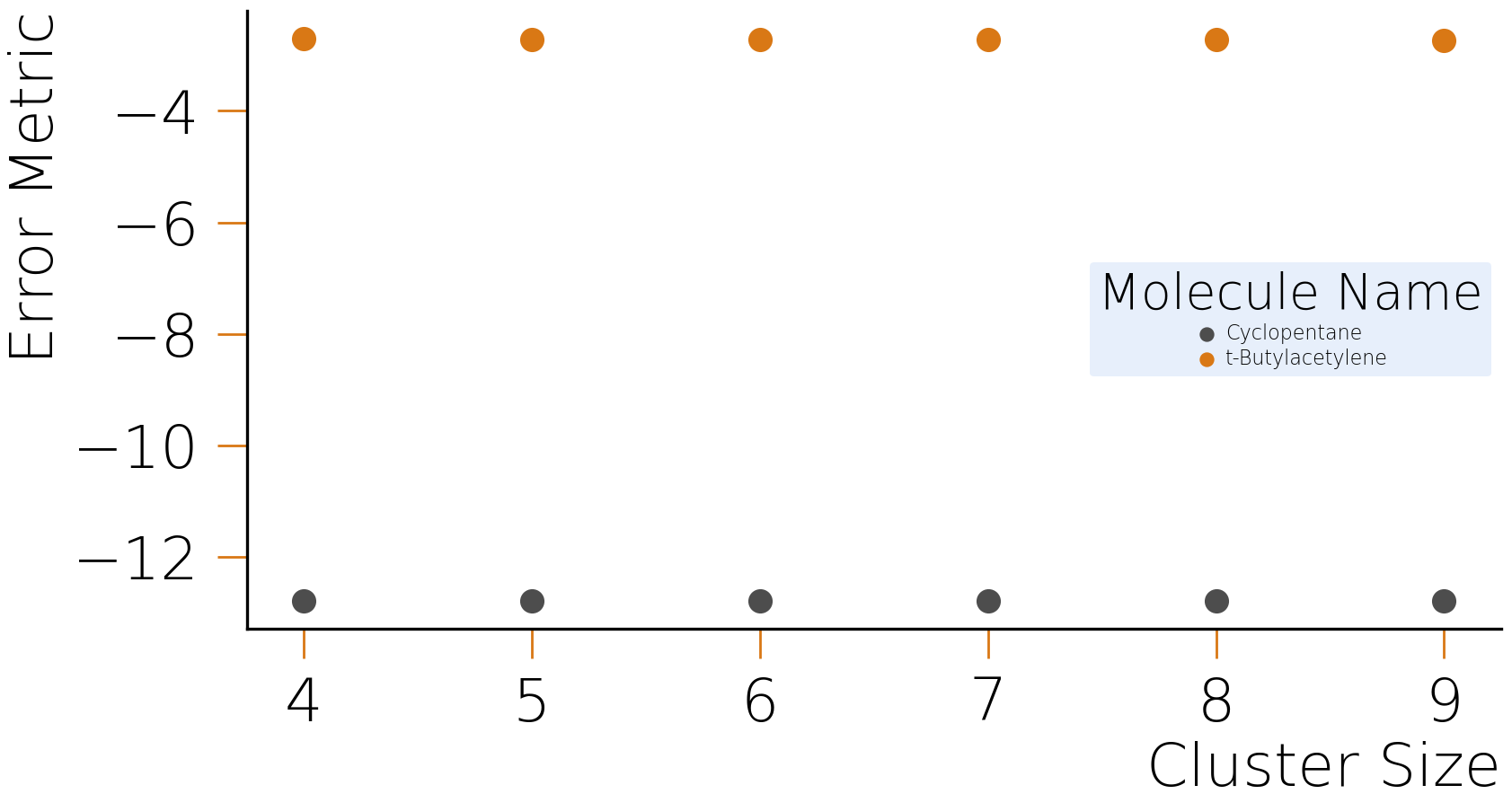}
   \includegraphics[width=0.32\textwidth]{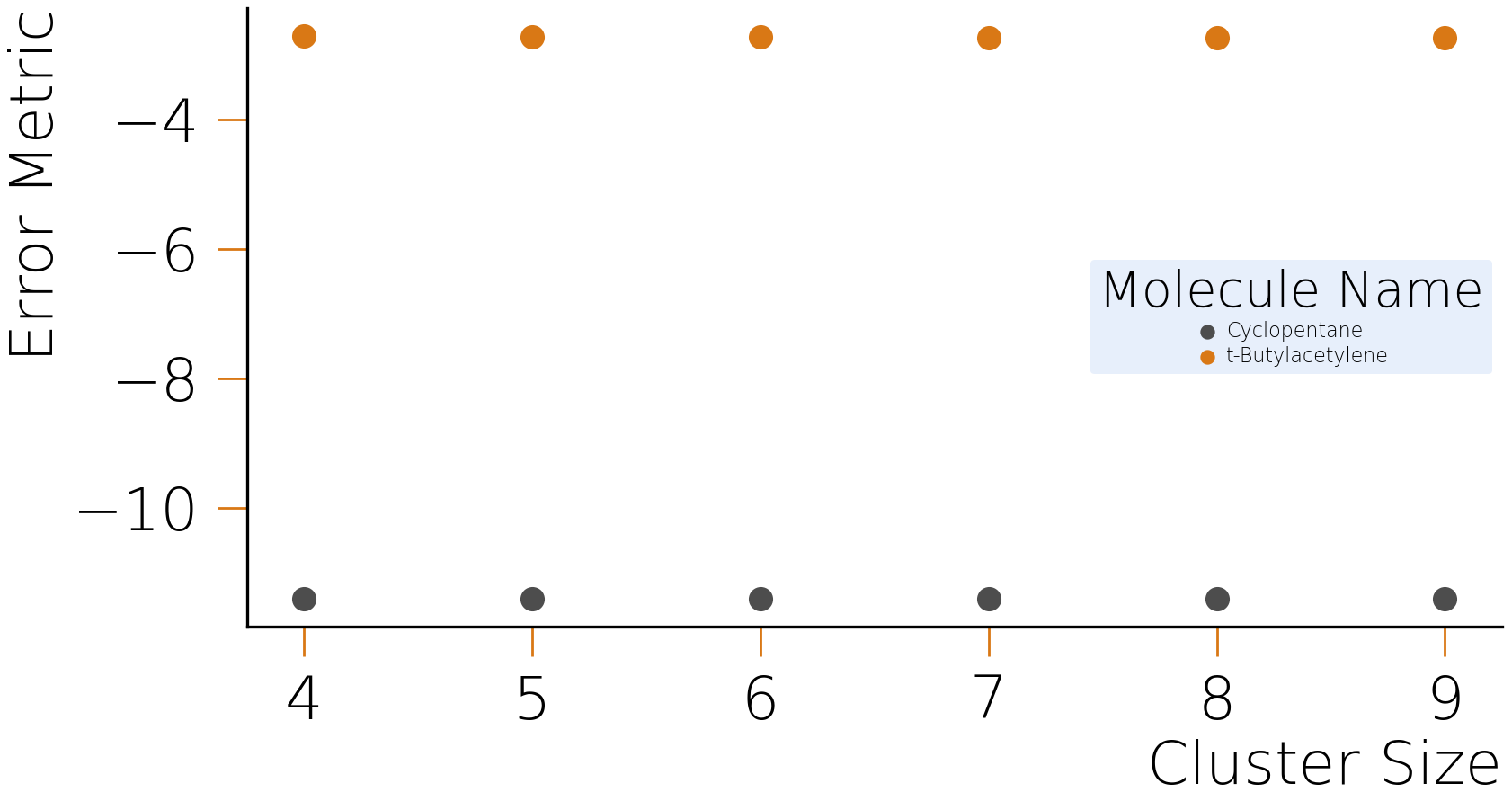}
   \caption{The convergence for selected molecules with 10 nuclear spins for the case of low broadening, for high, low, and very low field.}
   \label{fig:10low}
\end{figure*}

\begin{figure*}
   \centering
   \includegraphics[width=0.32\textwidth]{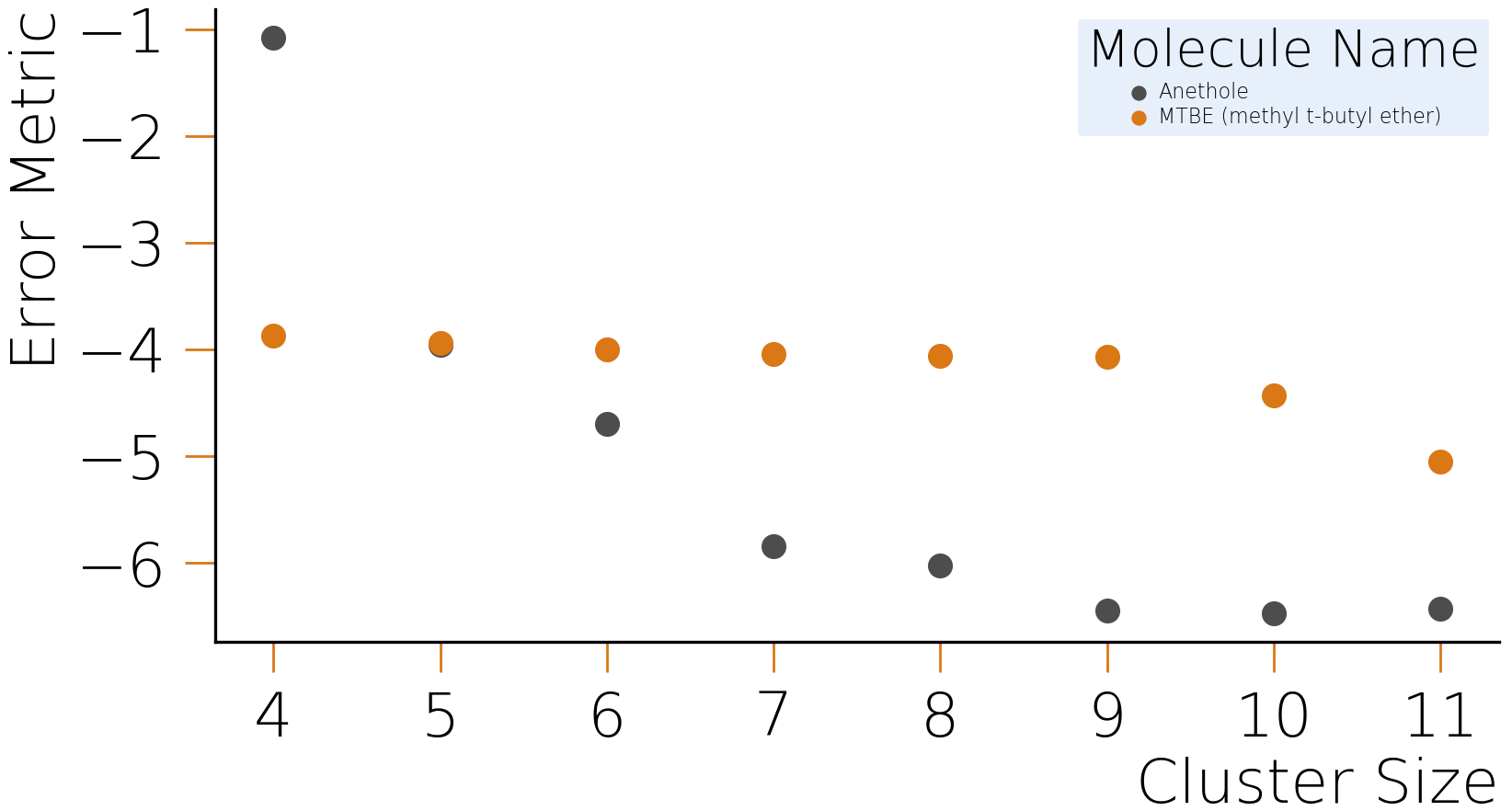}
   \includegraphics[width=0.32\textwidth]{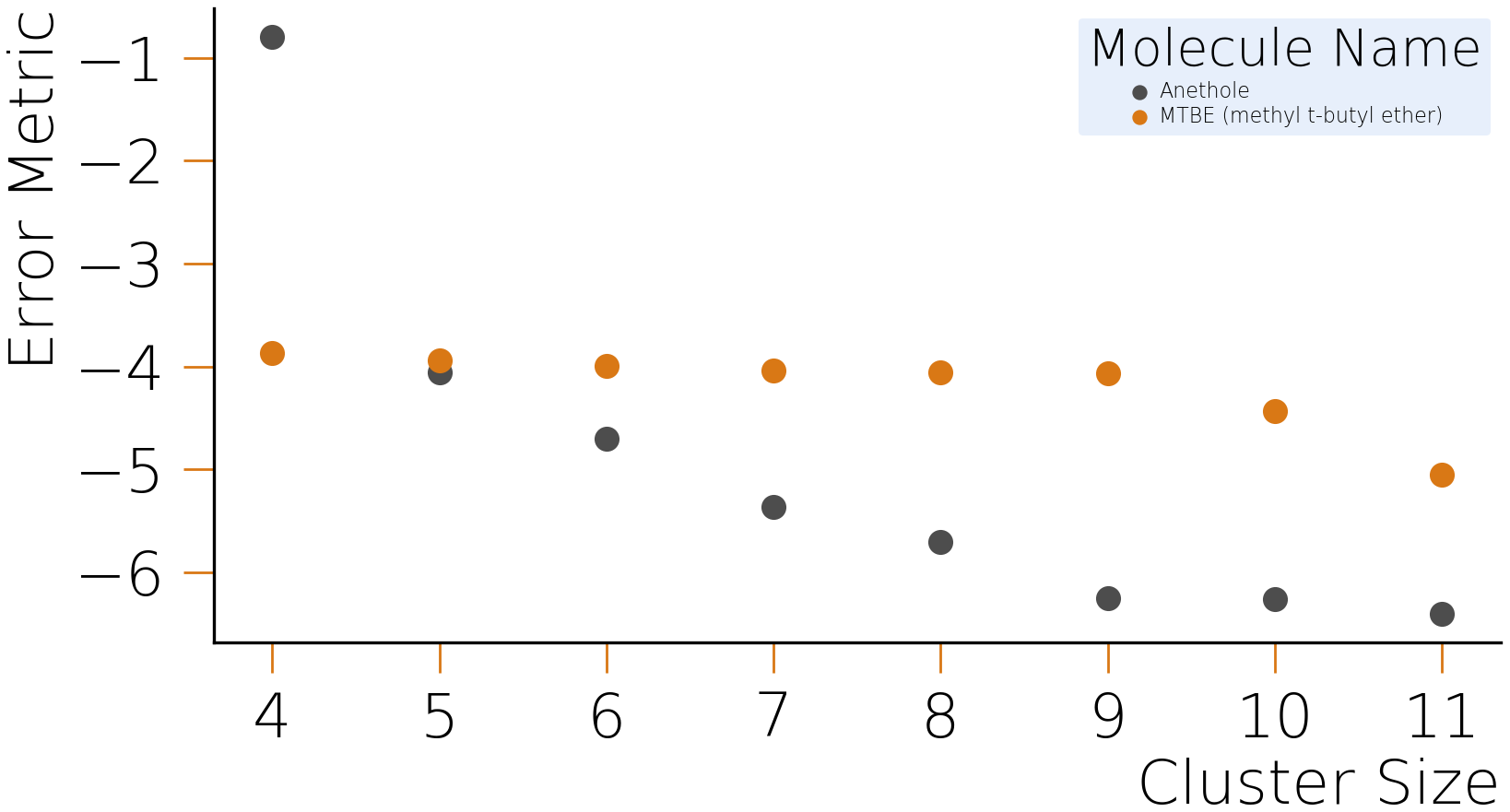}
   \includegraphics[width=0.32\textwidth]{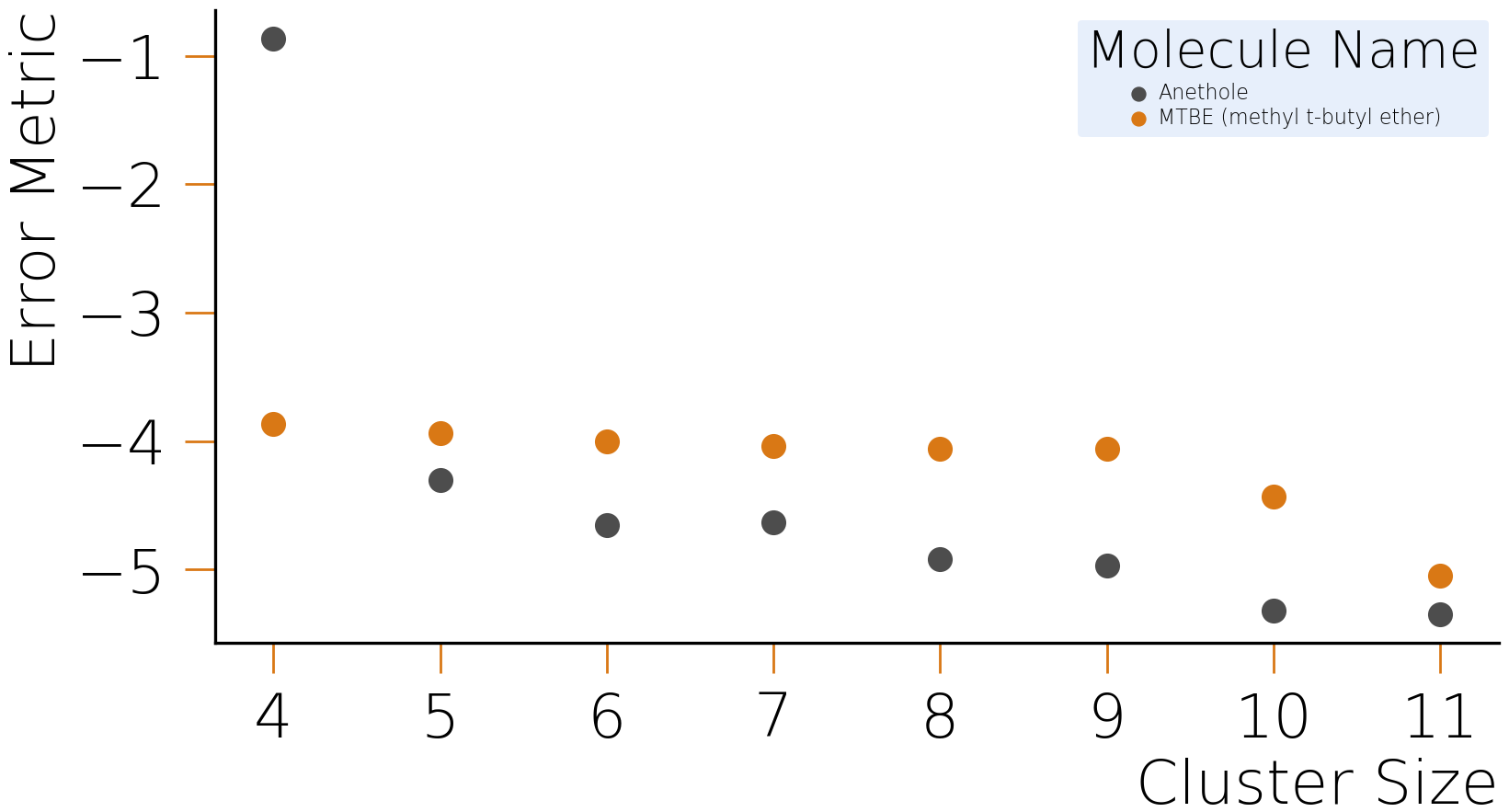}
   \caption{The convergence for selected molecules with 12 nuclear spins for the case of high broadening, for high, low, and very low field.}
   \label{fig:12high}
\end{figure*}

\begin{figure*}
   \centering
   \includegraphics[width=0.32\textwidth]{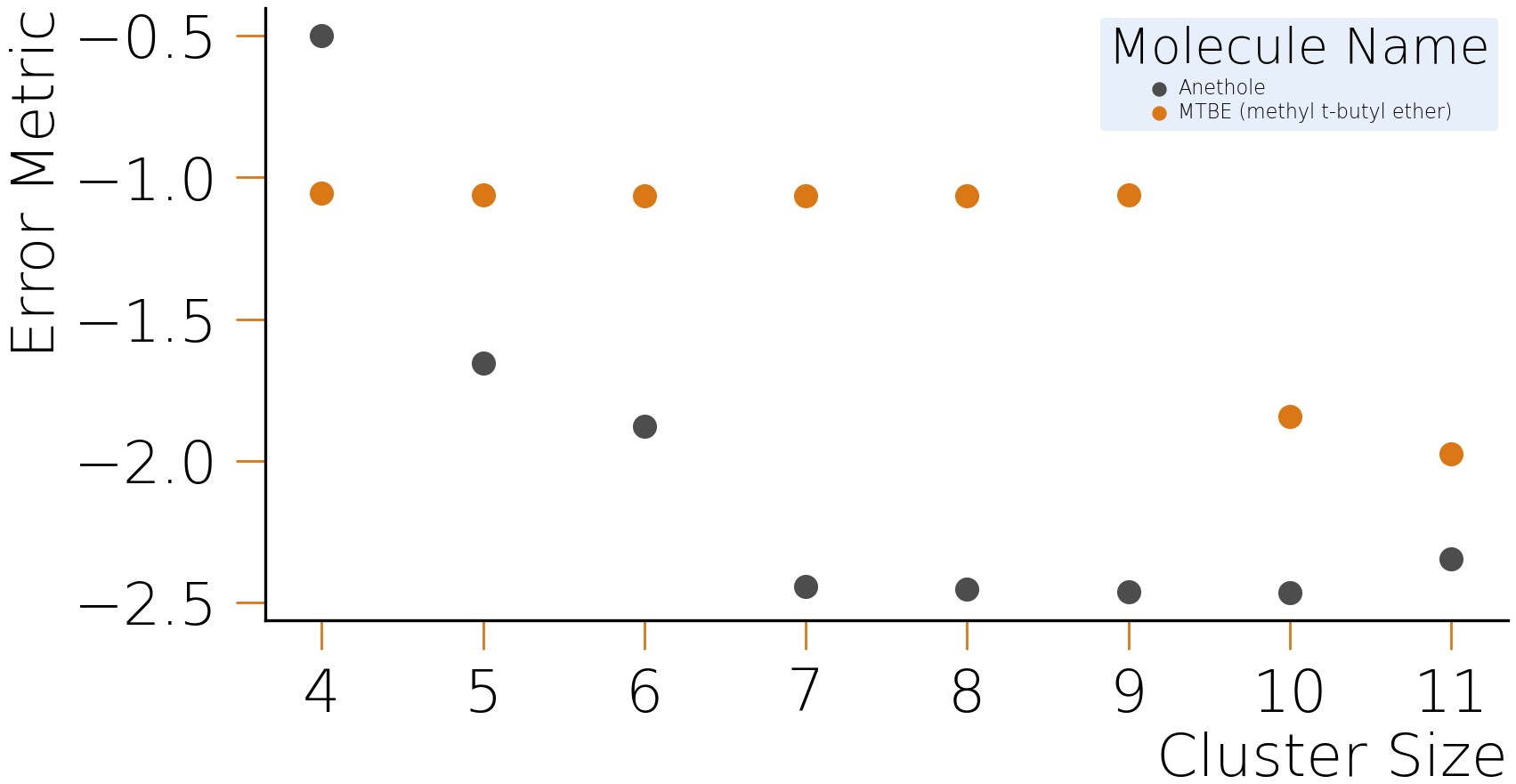}
   \includegraphics[width=0.32\textwidth]{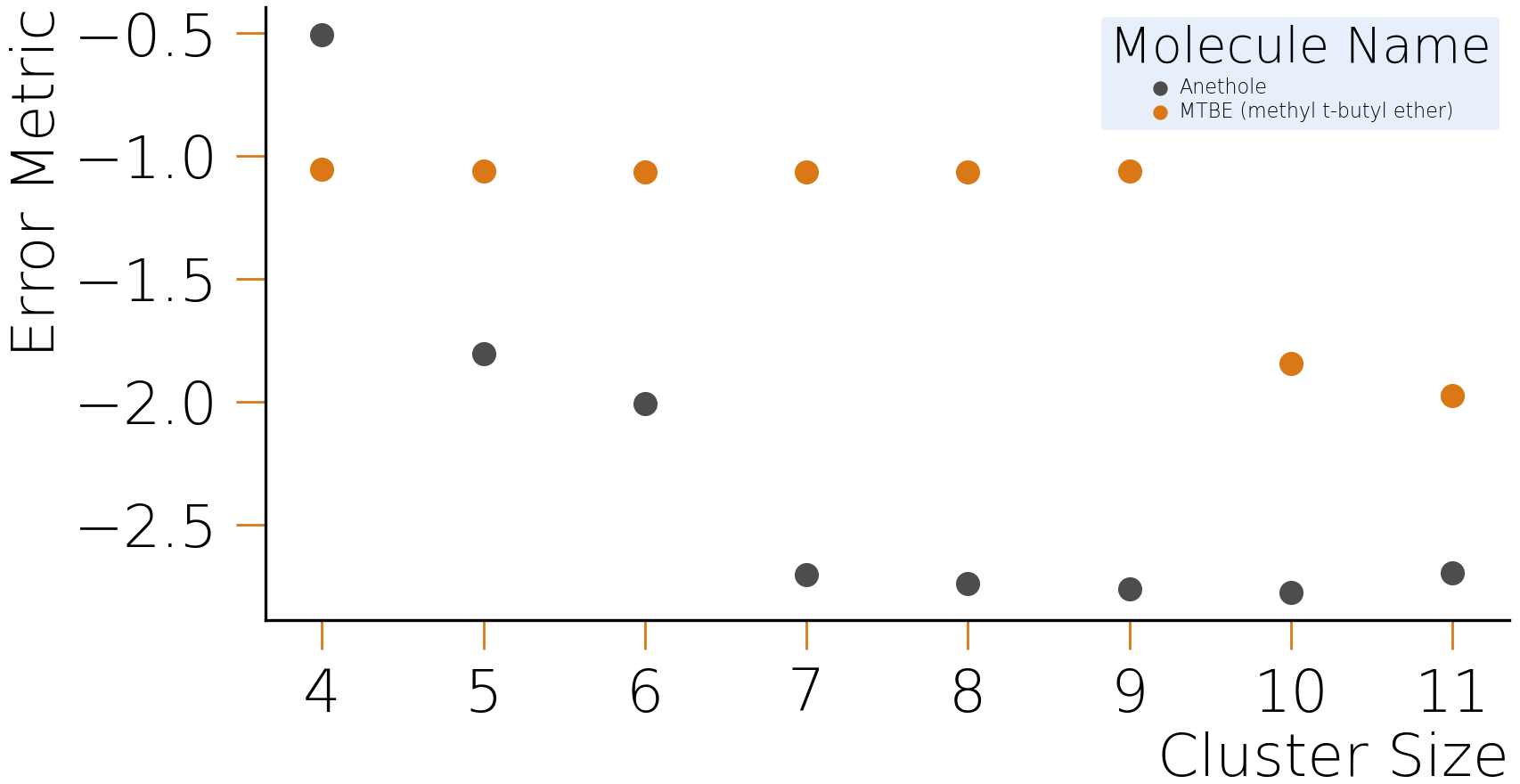}
   \includegraphics[width=0.32\textwidth]{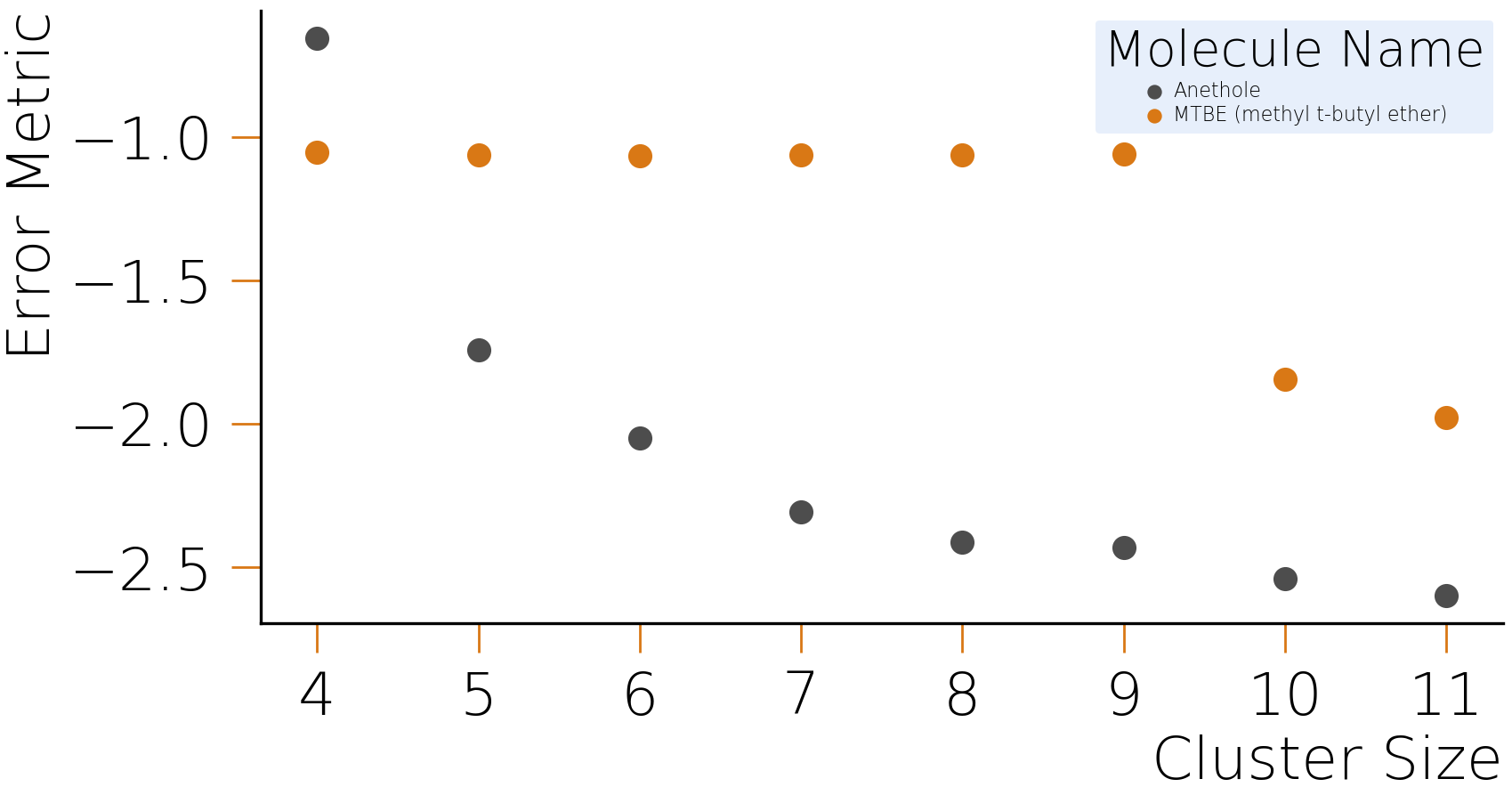}
   \caption{The convergence for selected molecules with 12 nuclear spins for the case of low broadening, for high, low, and very low field.}
   \label{fig:12low}
\end{figure*}

In this Appendix we present some additional convergence data for exactly solvable systems. Figures \ref{fig:8high} through \ref{fig:12low} display the results for selected molecules of size 8, 10, and 12, for the usual parameter regimes. The convergence data is again generally what we would expect, although there are perhaps some stronger finite-size effects. We also have data for two 9-site molecules, t-Butyl chloride and 2-Methyl-2-cyanopropane, but both of these immediately converge to within machine precision for even the smallest cluster sizes, for every possible parameter regime.